\documentclass[12pt]{amsart}
\usepackage{amssymb,amsbsy,latexsym,amsmath,mathrsfs,amsthm}
\usepackage{amsfonts,euscript}
\usepackage{tabularx}
\usepackage{latexsym}
\usepackage{array}
\usepackage{booktabs}
\usepackage{ctable}
\usepackage{caption}
\usepackage{subcaption}
\usepackage{verbatim}
\usepackage{graphicx}
\usepackage{epstopdf}
\usepackage{pdflscape} % rotate the page
\graphicspath{{figs/}}
\numberwithin{equation}{section}
\numberwithin{figure}{section}
\numberwithin{table}{section}

\usepackage{graphicx}
\usepackage[rightcaption]{sidecap}
\usepackage{epstopdf}
\usepackage{epsf}
\usepackage{epsfig}

\newtheorem{theorem}{Theorem}[]

\theoremstyle{definition}

\newcommand{\D}{\displaystyle}

\newcommand{\bfa}[1]{\boldsymbol{#1}}
\newcommand{\R}{\mathcal{R}}
\newcommand{\p}{\bfa{p}}

\newif\ifnotesw \noteswtrue

\begin{document}
\bibliographystyle{abbrv}
\title{\small{Sensitivity analysis in an Immuno-Epidemiological Vector-Host Model}}% with Application to Rift Valley Fever}}

\author{Hayriye Gulbudak$^{\dagger}$}
\address{Department of Mathematics,
University of Louisiana at Lafayette
217 Maxim Doucet Hall,
P.O. Box 43568,
Lafayette, LA }
\email{hayriye.gulbudak@louisiana.edu}

\author{Zhuolin Qu}
\address{Department of Mathematics,
University of Texas at San Antonio,
One UTSA Circle,
San Antonio, TX 78249 }
\email{zhuolin.qu@utsa.edu}

\author{Fabio Milner}
\address{School of Mathematical and Statistical Sciences, 
Arizona State University,
825 Wexler Hall,
PO Box 871804,
Tempe, AZ 85287}
\email{fmilner@asu.edu} 

\author{Necibe Tuncer}
\address{Department of Mathematical Sciences,  
Florida Atlantic University,
Science Building, Room 234
777 Glades Road
Boca Raton, FL 33431}
\email{ntuncer@fau.edu}

\thanks{$^\dagger$author for correspondence.}
%%%%%%%%%%%%%%%%%%%%%%%%%%%%%%%%%%%%%%%%%%%%%%%%%%%%%%%%%%%%%%%%%%%%%%%%%%%%%%
\begin{abstract} Sensitivity Analysis (SA) %has proven to be 
is a useful tool to measure the impact of changes in model parameters on the infection dynamics, particularly to quantify the expected efficacy of disease control strategies. SA has only been applied to epidemic models at the population level, ignoring the effect of within-host virus-with-immune-system interactions on the disease spread. Connecting the scales from individual to population can help inform drug and vaccine development. Thus the value of understanding the impact of \emph{immunological} parameters on epidemiological quantities. %such as \emph{basic reproduction number, final size of the epidemic} or \emph{the infectiousness at different phases of an outbreak} %is an important development that can benefit from the use of SA in multi-scale models. 
Here we consider an age-since-infection structured vector-host model, in which epidemiological parameters are formulated as functions of within-host virus and antibody densities, governed by an ODE system. 
We then use SA for these immuno-epidemiological models to investigate the impact of immunological parameters on population-level disease dynamics such as \emph{basic reproduction number, final size of the epidemic} or \emph{the infectiousness at different phases of an outbreak}. As a case study, we consider Rift Valley Fever Disease (RFVD) utilizing parameter estimations from prior studies. SA indicates that $1\%$ increase in within-host pathogen growth rate can lead up to $8\%$ increase in $\mathcal R_0,$ up to $1 \%$ increase in steady-state infected host abundance, and up to $4\%$ increase in infectiousness of hosts when the reproduction number $\mathcal R_0$ is larger than one. These significant increases in population-scale disease quantities suggest that control strategies that reduce the within-host pathogen growth can be important in reducing disease prevalence.%abundance population-scale impact of the disease. 

\bigskip\noindent

\end{abstract}
\date{\today}
\maketitle
\pagestyle{myheadings}
\markboth{\sc SA in a Nested Immuno-Epidemiological Vector$-$Host Model}
{\sc }

\baselineskip14pt
%\begin{runninglinenumbers}
%\modulolinenumbers[5]

%%%%%%%%%%%%%%%%%%%%%%%%%%%%%%%%%%%%%%%%%%%%%%%%%%%%%%%%%%%%%%%%%%%%%%%%%
%%%%%%%%%%%%%%%%%%%%%%%%%%%%%%%%%%%%%%%%%%%%%%%%%%%%%%%%%%%%%%%%%%%%%%%%%
\setcounter{tocdepth}{3}
%\tableofcontents

\section{Introduction} 

Sensitivity Analysis (SA) is a common methodological approach for determining the expected impact of control strategies on common disease outbreak quantities such as final size and basic reproduction number \cite{mpeshe2011mathematical}. SA has been used extensively for ODE models but has not been extended to age-structured models.  In this study, we develop a method that extends SA to immuno-epidemiological models \cite{tuncer2016structural, gulbudak2020infection,gandolfi2015epidemic,angulo2013sir}. We couple a time-since-infection-structured epidemiological system with an ODE immunological model. The main motivation for this modeling approach is that many recent vector-borne epidemic models share the limitation of exploring only between-host transmission while ignoring the impact of within-host virus-with-immune-response interactions, which may be important to guide drug and vaccine development, for example. 
%Some work has been done towards extending classical epidemic models into immune-epidemiological ones by formulating epidemiological parameters such as transmission and recovery rates dependent on pathogen and immunological state variables \cite{Angulo et al. (1) and (2)}.
 Here we demonstrate the approach through a multi-scale model, first introduced in \cite{gulbudak2017vector}. The parameters for this model are studied in  \cite{tuncer2016structural} based on Rift Valley Fever Disease (RVFD) immunological data and human epidemiological data from the $2006-2007$ Kenya Outbreak \cite{CDCdata,munyua2010rift}. Using the SA approach developed here, we quantify the impact of within-host parameters on the Rift Valley Fever Disease (RVFD) dynamics. The same multi-scale modeling framework can be adapted to other arbovirus diseases such as Dengue and West Nile Virus (WNV). 

RFVD is a viral disease transmitted by mosquitoes, mainly from the {\it Aedes}\/ and {\it Culex}\/ genera, and causes illness and death in several different mammal species, including livestock (e.g. cattle, buffalo, sheep, goat, and camel), as well as in humans. RVFD has resulted in significant negative socio-economic impacts, for example, due to abortion among RVF-infected livestock and high mortality among younger ones. In 2018, a panel of experts convened by the World Health Organization (WHO) listed RVF among diseases that pose big public-health risks, yet few or no intervention strategies have been developed. SA can provide helpful insights on the impact of possible pharmaceutical interventions for RVFD control.

Sensitivity analysis has been utilized in several ODE models to assess the impact of epidemic parameters on epidemic quantities. For example, Gaff et al. \cite{gaff2007epidemiological,gaff2011mathematical} considered an ODE vector-host RVF model to assess the effectiveness of some control interventions on RVFD. Fischer et al.\;\cite{fischer2013transmission} utilized SA to investigate the effect of temperate climate on the RVFD dynamics.  Mpeshe et al.\;\cite{mpeshe2011mathematical} formulated an ODE model of RVF incorporating parameters dependent of human behavior to investigate disease dynamics and explore sensitivity of the model to variation in those parameters.  Xiao et al. \cite{xiao2015modelling} recently studied the effect of both seasonality and socioeconomic status in a multi-patch model. To the best of our knowledge, the SA of immuno-epidemiological models has never been carried out, despite the value and usefulness of SA of the underlying immunological model parameters on epidemic variables related quantities related to them.% to provide accurate information on the impact of control strategies such as vaccination, or drug treatment. 

In this study, we develop a novel approach for SA in immuno-epidemiological models to investigate the impact of immunological parameters on the disease dynamics. In particular, we consider a time-since-infection-structured vector-host model in which epidemiological model parameters are described as functions of within-host virus-antibody densities that are governed by an ODE system. We first define the basic reproduction number, $\R_0$ that serves as a threshold between extinction and persistence of RVFD. Then we use this SA approach to investigate the impact of changes in immunological parameters on epidemic quantities such as basic reproduction number, and final disease abundance when $\mathcal R_0>1.$ Interestingly, our analytical and numerical results suggest that immunological parameters such as viral growth rate and immune activation rate can have a large impact on disease outcomes, underscoring the importance of pharmacological intervention strategies.

This paper is organized as follows. In Section \ref{intro_multi_scale_mdls}, we present an immuno-epidemiological model, first introduced in \cite{gulbudak2017vector, tuncer2016structural}, and summarize the stability and persistence conditions for the disease. In Section \ref{sec:SA},  we develop a novel approach for SA in immuno-epidemiological models to assess the impact of the within-host parameters on epidemic quantities. 
Furthermore in subsection \ref{SA_outbreaks}, we consider three distinct stages of an outbreak ---initial, peak and die-out--- and show how infectiousness of hosts at these different stages of infection is altered by slight changes in the immunological parameters through the phases of an outbreak. In the last section, we summarize our results and draw some conclusions.

%%%%%%%%%%%%%%%%%%%%%%%%%%%%%%%%%%%%%%%%%%%%%%%%%%%%%%%%%%%

\section{An immuno-epidemiological vector-borne disease model} \label{intro_multi_scale_mdls}
\begin{figure}[htbp]
\centering
\includegraphics[width=0.48\textwidth]{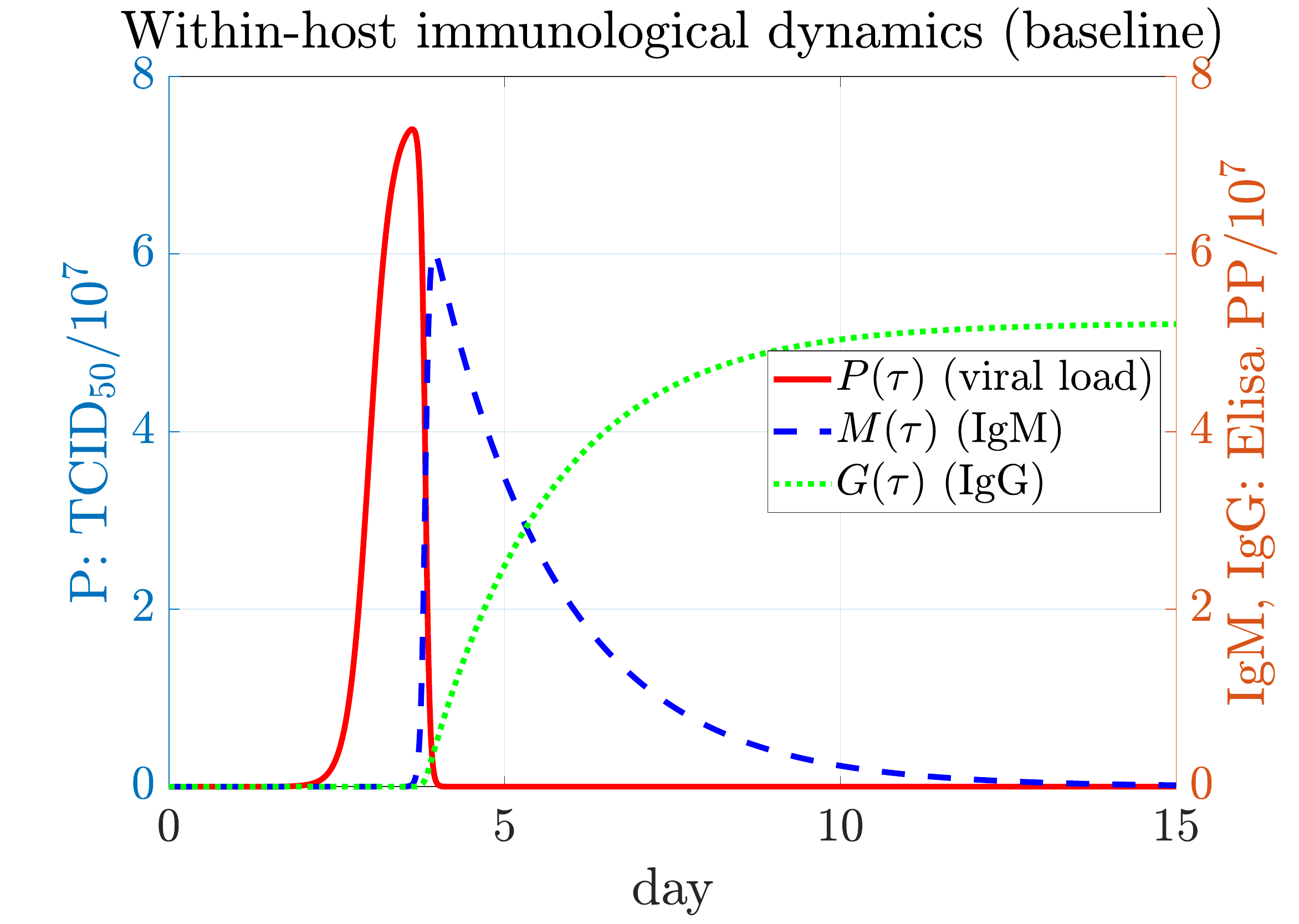}\hfill\includegraphics[width=0.48\textwidth]{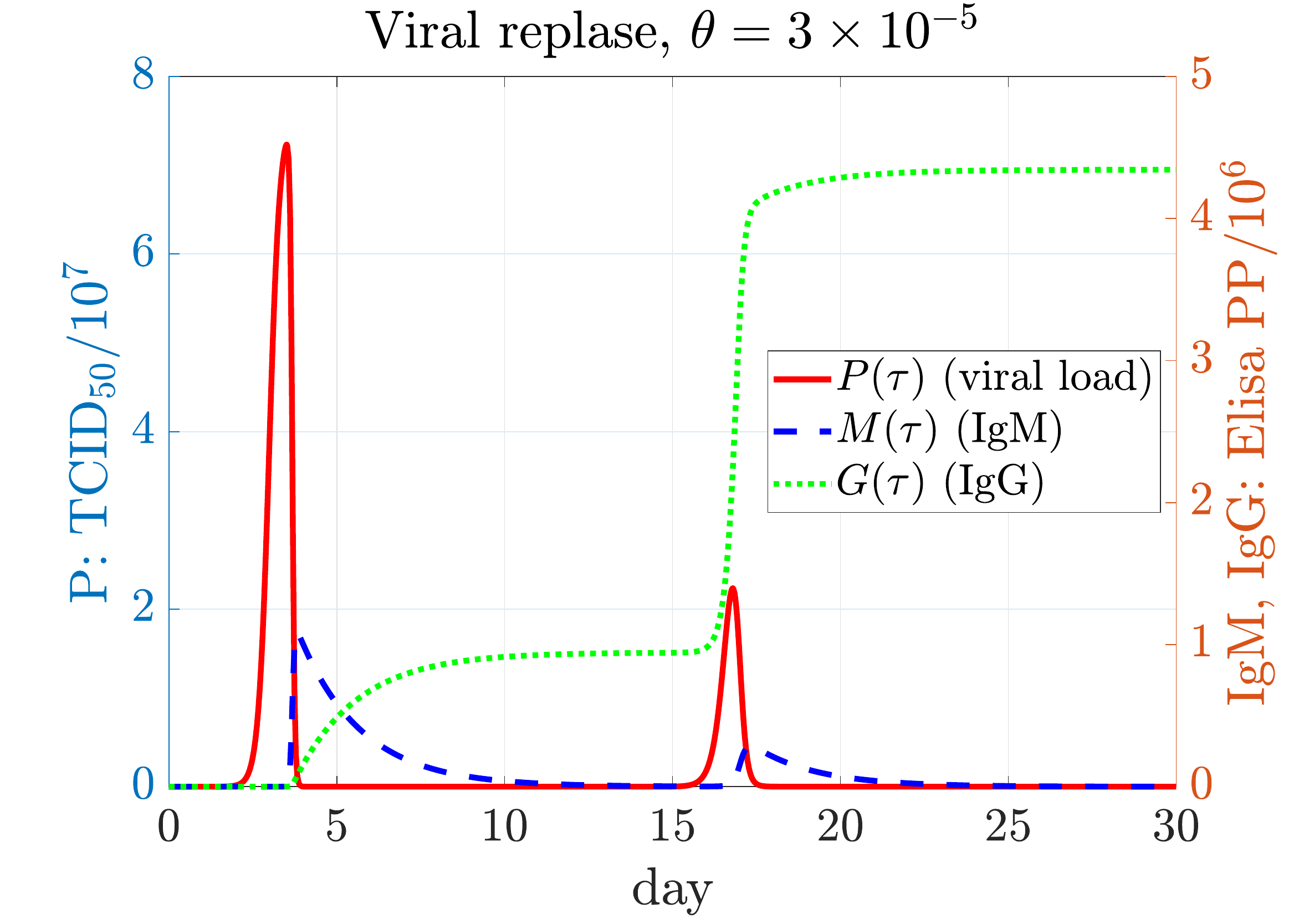}
\caption{\emph{Different within-host pathogen-immune response antibody dynamics} \textbf{Left.} General within-host dynamics. \textbf{Right.}  \emph{Resurgence of virus.}
The experiments done on monkeys suggest that in some cases, viral load appears to be controlled after an initial infection but makes a resurgence after disseminating into new tissue \cite{pepin2010rift}.}  
\label{fig:within-hostdynamics}
\end{figure}

 A simple model that captures vector-borne disease spread on two-scales (namely immunological and the epidemiological) was first introduced in \cite{gulbudak2017vector}.  At the individual scale, we consider the following immune response model with %age-of-infection dependent 
 three state variables representing serum density of the pathogen and of two specific antibodies released by B-cell lymphocytes ---IgM and IgG--- with densities given, respectively, by $P=P(\tau),$ $M=M(\tau)$ and $G=G(\tau),$ where $\tau$ is the time elapsed since infection:
 \begin{equation}\label{Immune_model}
\frac{dP}{d \tau} =\left(f(P)-\theta M-\delta G\right)P,\
\frac{dM}{d \tau} =\left(aP-(q+c)\right)M,\
\frac{dG}{d \tau} =qM + bGP.\
\end{equation}
The initial values are $P(0)=P_0>0, M(0)=M_0\ge0,G(0)=G_0\ge0$, and $M_0+G_0>0$ to ensure that there is a pathogen and an immune response. It is assumed that the pathogen replicates with a logistic per capita growth rate $f(P)=r\left( 1-\frac PK\right)$ and that, upon exposure to the virus, the IgM and IgG antibodies get activated at unit rates $a$ and $b$, respectively.  The IgM antibodies are responsible for a \emph{a quick immunological response (innate)}: they kill virus at a unit rate $\theta$ and decay at a unit rate $c$. The IgG antibodies kill the pathogen at a unit rate $\delta$, and they are mainly responsible for \emph{long-term immunity (adaptive)} \cite{gulbudak2017vector,tuncer2016structural}.
Mature B-cells activated by antigen stimulation proliferate quickly in lymphoid follicles and undergo genetic alterations resulting in a switch of the immunoglobulin isotype produced from IgM to either IgG, IgE, or IgA. To keep the immune response model lower-dimensional, we model the B-cell population indirectly through the IgM and IgG antibodies it produces. We incorporate this switch in antibody production by the B-cells by ``converting" IgM antibodies to IgG antibodies at a unit rate $q$ \cite{honjo2002molecular}. 
%Mature B lymphocytes, which have completed functional VDJ recombination of both H and L chain genes, express IgM on the surface and migrate to the secondary lymphoid organs such as spleen and lymph nodes where they encounter antigens. B lymphocytes activated by antigen stimulation proliferate vigorously in lymphoid follicles and often form special microenvironments called germinal centers, where the second wave of genetic alterations, namely class switch recombination (CSR) and somatic hypermutation (SHM), takes place in the immunoglobulin gene loci. CSR replaces the immunoglobulin CH gene to be expressed from C  to C , C" or C , resulting in switching of immunoglobulin isotype from IgM to either IgG, IgE, or IgA, respectively, without changing the antigen specificity.
All parameters and state variables of this within-host model and their definitions are given in Table \ref{table:parameters} and Table \ref{table:variables}. Different dynamics of this immune response model can be found in Figure \ref{fig:within-hostdynamics}.

A detailed analysis of the immunological model extended to include vector-to-host inoculum size dependent on age-of-infection of vectors was presented in \cite{gulbudak2020immuno}. In general upon viral progression within an infected mosquito midgut, the amount of pathogen in an infected mosquito's saliva dynamically changes with respect to its infection-age, determining the vector-to-host inoculum size $P_0$ that is the amount of pathogen injected to a susceptible host by an infectious vector during the infectious contact.
 Assuming that $P_0$ is constant,
all analytical results therein also hold for our present immunological model.
In particular, the following result is established \cite{gulbudak2020immuno}:
 \begin{theorem}\label{withhost} If $M_0>0$ (or $G_0>0$), then the pathogen eventually clears ($\displaystyle\lim_{\tau \rightarrow \infty}P(\tau)=0$), the IgM antibodies decay to zero after viral clearance and, subsequently, the IgG memory antibodies reach a steady-state; i.e.\ 
$\displaystyle\lim_{\tau \rightarrow \infty}M(\tau)=0 \ \text{and} \ \displaystyle\lim_{\tau \rightarrow \infty}G(\tau)=G^+,$ where $G^+>0$ depends on the initial condition $(P_0,M_0,G_0)$.
\end{theorem}

The innate and adaptive immune responses vary from organism to organism and depend on several factors, including the initial viral dose, the strain of the virus, the age of the host, and the species of the host. However, infections like RVFV in the laboratory generally follow three qualitative scenarios:
\begin{itemize}
\item{The viral load within the host grows rapidly to high levels within the host, and the host dies.} 
\item{There is a delayed onset of complications of infections: e.g. viral load appears to be controlled after an initial infection but makes a resurgence after disseminating into new tissue, such as the central nervous system. This often leads to long-term consequences such as blindness and may also be lethal.} 
\item{The viral load within the host is brought under control and annihilated by a robust immune response.  
}
\end{itemize}
Our immunological model captures the general dynamics of these different infection and immune-response scenarios as seen in Figure \ref{fig:within-hostdynamics} and as suggested by the analytical result above (Theorem \ref{withhost}). In general, within-host dynamics in arbovirus diseases behave as depicted in Figure \ref{fig:within-hostdynamics} (left), corresponding to the third scenario mentioned above. Also, in a lab experiment with monkeys infected with RVFV, it was observed that monkeys died when the viral concentration rebounded as seen in Figure \ref{fig:within-hostdynamics}(right). In surviving monkeys, the within-host dynamics imitated the third scenario %, though an infected monkey still experienced death under the same within-host dynamics 
\cite{Morrill}.
In an ideal immune response, the immune system releases virus-specific antibodies, targeting the virus and bringing it to extinction. These dynamics are captured in Figure \ref{fig:within-hostdynamics} (left).
However, sometimes there is a resurgence of virus within the host as the virus spreads into new organs and tissues, such as the central nervous system. This is often associated with delayed onset of symptoms, including blindness, and may result in fatalities \cite{pepin2010rift}. Our model can capture these dynamics as well, as can be seen in Figure \ref{fig:within-hostdynamics} (right).

\medskip
At the \emph{population scale}, we consider a time-since-infection structured vector-host model, \eqref{Epi_model}, where
$S_H(t)$ and $R_H(t)$ are the numbers of susceptible and recovered individuals in the host population and $i_H(\tau,t)$ is the infected-host density, structured by age-of-infection $\tau$. The total number of infected individuals is $ I_H(t) = \int_0^\infty i_H(\tau,t) d\tau$. The vector compartments $S_V(t)$ and $I_V(t)$ represent, respectively, the size of susceptible and infected vector populations at time $t$. The full model is as follows:

{\small 
\begin{equation}\label{Epi_model}
\left\{
\begin{aligned}
\frac{d S_H}{dt} & = \Lambda - \beta_V S_H(t)I_V(t) -d S_H(t),\\
\frac{\partial i_H}{\partial t}+\D\frac{\partial i_H}{\partial \tau} & =  
 -(\alpha (\tau)+ \kappa (\tau)+ \gamma(\tau)+ d)\; i_H(\tau,t),\quad
i_H(0,t)  = \beta_V S_H(t) I_V(t),\\
\frac{d R_H}{dt}& = \D\D\int _0^\infty {\gamma(\tau)i_H(\tau,t)d\tau}-dR_H(t),\\
%\end{aligned}
%\right.
%\end{equation} }
%\begin{equation}\label{Epi_model}
%\left\{
%\begin{aligned}
\frac{d S_V}{dt} &= \eta -  \left( \int_0^\infty {\beta_{H}(\tau)i_H(\tau,t) d \tau}+\mu \right)S_V(t),\\
\frac{d I_V}{dt} &= S_V(t) \int_0^\infty {\beta_{H}(\tau)i_H(\tau,t) d \tau}-\mu I_V(t).
\end{aligned}
\right.
\end{equation} }

To \emph{bridge the scales}\/ from individual to population, the unit infectiousness and the disease-induced death and recovery rates are formulated as functions of immunological variables as suggested by \emph{data} in \cite{handel2015crossing, fraser2014virulence}, as follows:
\begin{equation}\left\{
\begin{aligned}
\label{linkfns}
\beta_H(\tau)=\frac{C_\beta P(\tau)}{C_0+ P(\tau)},\\[1em]
\alpha(\tau)= \zeta P(\tau),\\[1em]
\kappa(\tau)=\xi M(\tau),%\text{ and }
\\[1em]
\gamma(\tau)=C_\gamma\frac{G(\tau)}{P(\tau)+\epsilon_0}.
\end{aligned}\right.
\end{equation}

\medskip

% The general within-host dynamics is displayed in Fig.\ref{fig:within-hostdynamics} (left panel). 
Fig.\;\ref{fig:linking_baseline} in the Appendix shows how some of the corresponding epidemiological parameters evolve over the course of host infection.
Note that Tuncer et al.\cite{tuncer2016structural} considers mass action term, satisfying \textit{balance equation} for describing the interactions between vectors and hosts with following underlying assumptions: 
\begin{itemize}
\item $c_v N_v=$total number of contacts  (bites received) per host. Then total number of contacts (bites received) that hosts' have: $(c_v N_v) \times (N_H)$.
\item $c_H N_H=$ total number of contacts (given bites) per vector. Then total number of contacts (given bites) that vectors' have: $(c_H N_H)\times (N_v)$. Therefore it makes sense when the host population size is small (or not changing much). 
\item To satisfy balance equation (conservation law), we assume that $c_v=c_H$.
\item Here c is proportionality constant. Mass action assumes that per vector (or per host) contact is proportional to total host (or total vector) population. Therefore it has different meaning than the one in standard incidence.
\end{itemize}
Changes to the functional form of the interaction terms such as considering \textit{frequency depending force of infection rate} (standard incidence) will be further explored in later work. 
\medskip

\begin{figure}[ht]
\centering
\includegraphics[width=0.85\textwidth]{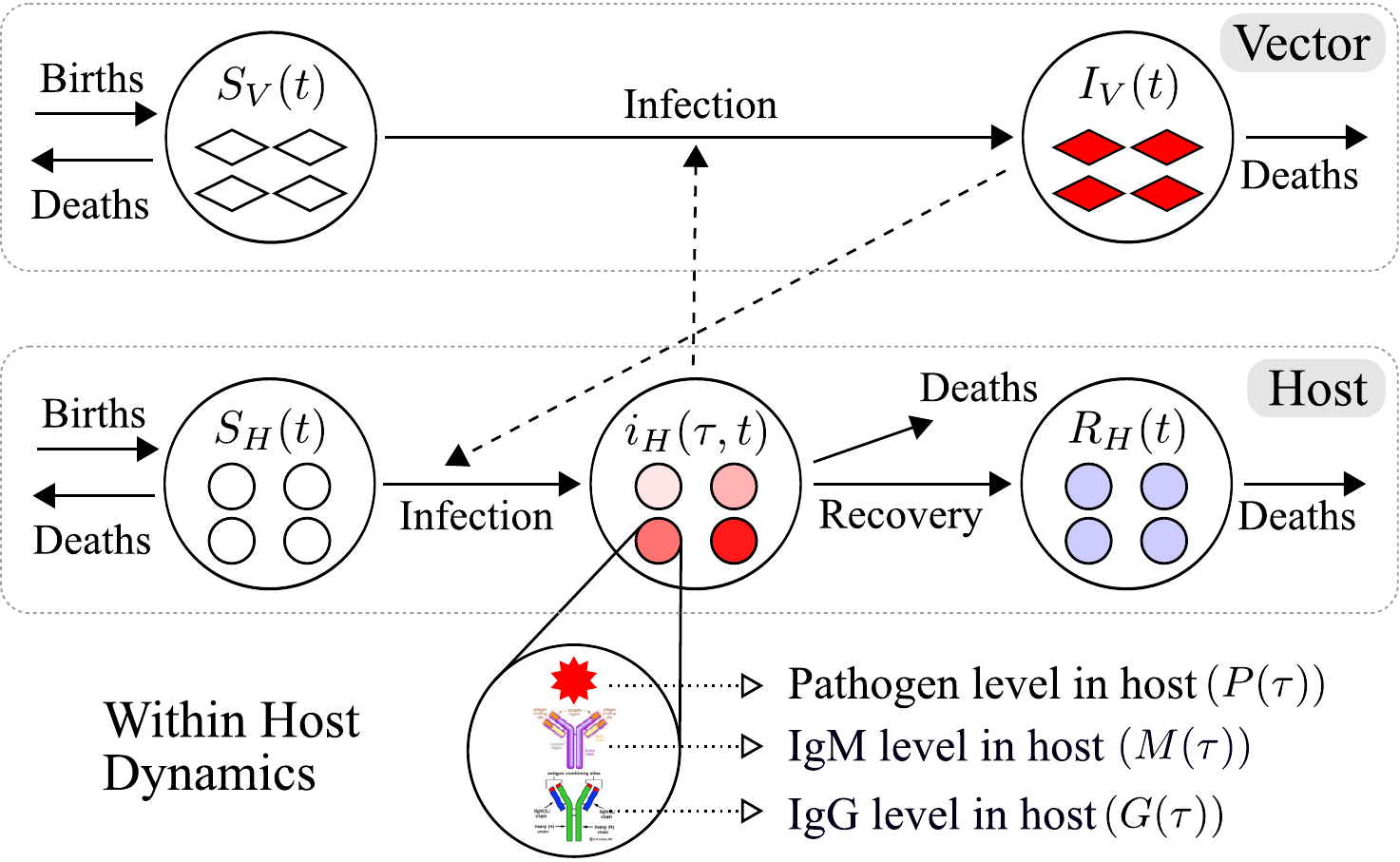}
\caption{Schematic illustration of the immuno-epidemiological vector-host model\eqref{Immune_model}-\eqref{linkfns}.}
\label{fig:schematic}
\end{figure}

A schematic diagram for the full system is presented next in Fig.\;\ref{fig:schematic}.

{\small \begin{landscape}
\begin{table}[htbp]
\caption{Model parameters: Parameter estimates of the immuno-epidemiological model \eqref{Immune_model}-\eqref{linkfns} fitted \cite{tuncer2016structural}. $\theta$ and $\delta$ are the killing efficiency parameters that re-scale the population immune cells. *$q$ value is chosen differently (fit separately) \label{table:parameters}}
\begin{tabular}{lllll}
\toprule
& immunological parameters (within-host model \eqref{Immune_model}) & dimension & baseline & range \\
\midrule
$r$ & Unit pathogen growth rate & $ \mbox{day}^{-1}$  & $6.3033$ & $1 \sim 10$ \\
$K$ & Carrying capacity of pathogens & TCID$_{50}$  & $7.57\times 10^7$ & $1.5\times 10^7 \sim 10^8$ \\
$a$ & Unit IgM-activation rate per pathogen&  $(\mbox{TCID}_{50} \times \mbox{day})^{-1}$ &$3.93\times 10^{-7}$ & $ 10^{-8} \sim 5\times 10^{-7}$ \\
$b$ &  Unit IgG-activation rate per pathogen & $(\mbox{TCID}_{50} \times \mbox{day})^{-1}$ & $9.58 \times 10^{-8}$ &$ 5\times 10^{-9} \sim 10^{-6}$ \\
$\theta$ & Unit IgM pathogen-destruction rate per-pathogen & $(\mbox{Elisa PP} \times \mbox{day})^{-1}$ & $5\times 10^{-7}$ & $10^{-9} \sim 10^{-4}$ \\ 
$\delta$ & Unit IgG pathogen-destruction rate per pathogen &$(\mbox{Elisa PP} \times \mbox{day})^{-1}$  &$8.3918\times 10^{-8}$ &$1.7 \times 10^{-10}\sim 1.7\times 10^{-5}$  \\
$q$ & Unit IgM-to-IgG production switch rate & $\mbox{day}^{-1}$ &$0.4232^*$ &$  2 \sim  20 $ \\  
$c$ & Unit IgM decay rate &  $\mbox{day}^{-1}$ & $0.1155$ &  $ 0.01 \sim 1$ \\
\midrule
& epidemic parameters (vector-host model \eqref{Epi_model}) & & & \\
\midrule
$\Lambda=d$ & Host recruitment rate& host$\times$day$^{-1}$ & ${1}/{(365\times10)}$ & \\
$\eta=\mu$ & Vector recruitment rate  & vector$\times$day$^{-1}$ & ${1}/{40}$ & \\
$\beta_V$ &  Per-capita-per-infected vector transmission rate & (vector$\times$ day)$^{-1}$ & $0.2$ & \\
$\beta_H(\tau)$ & Unit age-of-infection-dependent per host transmission rate  & (host$\times$ day)$^{-1}$ & - & \\
$\alpha (\tau)$ & Unit pathogen-induced death rate & day$^{-1}$ & - & \\
$\kappa (\tau)$ & Unit immune-response-induced death rate & day$^{-1}$ & - & \\
$\gamma (\tau)$ & Unit recovery rate $\tau$ days post-infection & day$^{-1}$ & - & \\
$d$ & Unit natural death rate of hosts  & day$^{-1}$  & $ {1}/{(365\times10)}$ & \\
$\mu$ & Unit natural death rate of vectors  & day$^{-1}$  & ${1}/{40}$ & \\
\midrule
& linking function parameters \eqref{linkfns} & & & \\
\midrule
$C_{\beta}$ & transmission efficiency of pathogen infection & (host$\times$ day)$^{-1}$ & $0.5365$ & \\
$C_0$ &  half-saturation constant in transmission rate & $\mbox{TCID}_{50}$ & $3.03 \times 10^3$ & \\
$\zeta$ & unit per host pathogen-lethality rate &$(\mbox{TCID}_{50}\times\mbox{day})^{-1}$ &$6.21\times 10^{-8}$ & \\
$\xi$ &  unit per host immune-response-lethality rate & $(\mbox{Elisa PP}\times\mbox{day})^{-1}$ & $8.51\times 10^{-5}$& \\
$C_{\gamma}$ &  recovery coefficient &  $\frac{\mbox{TCID}_{50}}{\mbox{Elisa PP}\times\mbox{day}}$ &  $0.7212$ & \\
$\epsilon_0$ &  half-saturation constant in recovery rate & \mbox{TCID}$_{50}$ & $ 7.43\times 10^{-4}$ & \\
\bottomrule
\end{tabular}
\end{table}
\end{landscape}}

\begin{table}[htbp]
\caption{State Variables \label{table:variables}}
\centering 
\begin{tabular}{ll}
\toprule
& Meaning \\
\midrule
$P(\tau)$ & Pathogen concentration at infection-age $\tau$  \\
$M(\tau)$ & IgM  concentration at infection-age $\tau$ \\  
$G(\tau)$ &  IgG concentration at infection-age $\tau$ \\
$S_H(t)$ & Number of susceptible hosts at time $t$ \\
$ i_H(\cdot,t)$ & Age-of-infection density of  infected hosts at time $t$\\
$ R_H(t)$ & Cumulative number of recovered hosts at time $t$\\
$S_V(t)$ & Number of susceptible vectors at time $t$\\
$I_V(t)$ & Number of infected vectors at time $t$\\
\bottomrule
\end{tabular}
\end{table}

Next, we highlight the threshold dynamics of the system, including long-term behavior of the solutions, explicit expressions of equilibria and basic reproduction number.  

Let us define the \textit{immune-response-dependent reproduction number of the epidemic} as
\begin{equation}\label{R0}
\mathcal R_0 = \D\frac{\beta_V S^0_H }{\mu}\; \D\int_0^\infty{S_V^0 \beta_H(\tau) e^{-\D\int_0^\tau{(\alpha(s)+\kappa(s)+\gamma(s)+d)ds}}d\tau}, %:= \mathcal{R}_{VH}\times\mathcal{R}_{HV},
\end{equation}
where $S^0_H =\D\frac{\Lambda}{d}$ and $S_V^0=\D\frac{\eta}{\mu}.$   A detailed analysis of our model \eqref{Immune_model}-\eqref{linkfns} including also vector age-of-infection was presented in \cite{gulbudak2020immuno}. The vector compartments here correspond to the ones in that paper integrated over vector-infection-age and, therefore,
all analytical results therein (e.g. threshold conditions) also hold for our present model. In particular, the following results are established for our model:
\begin{theorem}\label{Appendix_thm1}
The disease-free equilibrium (DFE) $\mathcal E_0=(S^0_H,0,0,S_V^0,0)$ is locally asymptotically stable if $\R_0<1$ and unstable if $\R_0>1.$
\end{theorem}
This result is actually global for $\mathcal E_0$:
\begin{theorem}\label{Appendix_thm2}
$\mathcal E_0$ is globally asymptotically stable if $\R_0<1$ and unstable if $\R_0>1.$
\end{theorem}
Furthermore, when $\R_0>1$, the system \eqref{Epi_model} has a unique endemic equilibrium (EE)
$\mathcal E^* =(S^*_H,\  i^*_H(\tau),\ R^*_H,\  S^*_V, \ I^*_V)$ (also presented in \cite{gulbudak2017vector}), where

\begin{equation}\left\{
\begin{aligned}
S^*_H &=  \left(\beta_V \D\frac{S^*_V}{\mu}\D\int_0^\infty{\beta_H(\tau)\pi_H(\tau)d\tau}\right)^{-1},\\ 
i^*_H(\tau)&= i^*_H(0)\pi_H(\tau), \\ 
R^*_H&=\frac{i^*_H(0)}{d} \D\int_0^\infty{\gamma(\tau)\pi_H(\tau)d \tau}, \label{eq_host_e}
\end{aligned}\right.
\end{equation}

\begin{equation}
\left\{
\begin{aligned}
S^*_V& =\eta \left(i^*_H(0)\D\int_0^\infty{\beta_H(\tau)\pi_H(\tau)d\tau}+\mu \right),\\  I^*_V&=\frac{S^*_V}{\mu}\D\int_0^\infty{\beta_H(\tau)i^*_H(\tau)d\tau},\label{eq_vector_e}
\end{aligned}
\right.
\end{equation}
with 
\begin{equation}\label{prob_inf}
i^*_H(0)=S_0^H\left( 1-\D\frac{1}{\mathcal R_0} \right)/ \left( \D\frac{1}{d} + \D\frac{\mu}{\beta_V\eta}\right),\
\pi_H(\tau) = e^{-\D\int_0^\tau{(\alpha(s)+\kappa(s)+\gamma(s)+d)ds}},
\end{equation}
and the epidemiological parameters $\beta(\tau), \ \gamma(\tau),\ \alpha(\tau),$ and $\kappa(\tau)$ are given by \eqref{linkfns}.
For this equilibrium we have the following results.
\begin{theorem}\label{Appendix_thm3}
The endemic equilibrium\  $\mathcal E^* =(S^*_H, i^*_H(\tau),R^*_H, S^*_V, I^*_V)$ %where the expressions for the components are given in \eqref{eq_host_e} and \eqref{eq_vector_e} , 
is locally asymptotically stable whenever it exists (i.e. for $\R_0>1$).
\end{theorem}

% In the presence of a disease, one also would like to understand under what conditions the disease will remain endemic for large time. In \cite{gulbudak2020immuno}, we show that when $\R_0>1$, the disease is uniformly weakly endemic, i.e. eventually the prevalence is bounded away from zero with a positive distance $\epsilon>0.$ 

\begin{theorem}\label{Appendix_thm4}
If $\R_0>1,$ then the disease is uniformly weakly endemic.
\end{theorem}
In fact, the global stability of a unique endemic equilibrium is shown in \cite{Magaltwogroupinfection2013} under an equivalent threshold condition to $\mathcal R_0>1.$  These results show that $\mathcal R_0$ represents both a threshold between extinction and persistence and a {strict} threshold for disease eradication.

%%%%%%%%%%%%%%%%%%%%%%%%%%%%%%%%%%%%%%%%%%%%%%%%%%%%%%%%%%%%%%%%%%%%%%%%%%%%%%%%%%%%%%%%
\section{Impact of Immune parameters on Epidemic Quantities}\label{sec:SA}

SA of  immuno-epidemiological models not only provides a measure of the influence of epidemic parameters on the spread and abundance of the disease at the population scale, but also of those at within-host scale. In particular, SA of immunological parameters on epidemic quantities might provide valuable information on the expected impact of control strategies including those targeted toward:\emph{(i) the host population such as vaccination, and drug treatment, (ii) the vector population, such as Wolbachia-based control strategies, and (iii) curbing viral production by providing important comparisons of the efficacy of different immune variables.} 
To determine the impact of immunological parameters on the disease dynamics at population scale, we carry out the SA of the epidemiological quantities $\R_0, I_H^*$ and $I^*_V$ with respect to the immune model parameters in \eqref{Immune_model}.  Notice that the prior studies on SA only focus on the impact of \textit{epidemic} parameters on epidemic quantities. 

The normalized forward sensitivity index of a quantity of interest (QOI) $q$ to a parameter of interest (POI) $p$ can be defined as the ratio of relative change in the variable to the relative change in the parameter:
\begin{align*}
\gamma^ {q} _{p}=\frac{\partial q}{\partial p}\times \frac{p}{q}.
\end{align*}
The baseline parameter values and ranges for RVFD are presented in Table \ref{table:parameters}. The baseline values were fitted to multi-scale data in \cite{tuncer2016structural}: for immunological parameter estimations, immunological RVF time-series data was obtained from livestock (in laboratory experiments), and for the epidemiological model, incidence data was acquired from the $2006-2007$ Kenya Outbreak \cite{CDCdata,munyua2010rift}. This simultaneous fitting of immunological and epidemiological model parameters induces practical identifiability of %fitted multi-scale 
model parameters \cite{tuncer2016structural}.\\

%\noindent\textbf{Numerical method.} 
We first consider the normalized sensitivity index for the basic reproduction number $\gamma_p^{\R_0}$. Note that for consistency regarding parameter estimates in Tuncer et al. \cite{tuncer2016structural}, we assume that $\Lambda=d,$ and $\eta=\mu,$ which set the total number of susceptible hosts, $S_H^0,$ equal to 1 at the DFE, $\mathcal E_0$.

From \eqref{R0}, we have 
\begin{equation}\label{R0_p}
\frac{\partial {\R_0}}{\partial p} = \frac{\beta_V}{\mu} \int_0^\infty \left( \frac{\partial \beta_H(\tau ,p)}{\partial p}\pi_H(\tau,p)+\beta_H(\tau,p)\frac{\partial \pi_H(\tau,p)}{\partial p}\right) \,d\tau,
\end{equation}
where $\beta_H(\tau,p)$ and $\pi_H(\tau,p)$ are defined in \eqref{linkfns} and \eqref{prob_inf}, respectively, and
\begin{align}
\frac{\partial \beta_H(\tau, p)}{\partial p}&=\frac{C_{\beta} C_0}{(C_0 + P(\tau))^2} \frac{\partial P(\tau,p)}{\partial p},\nonumber\\ 
\frac{\partial \pi_H(\tau, p)}{\partial p} &=-\pi_H(\tau)\int_0^\tau \left( \zeta \frac{\partial P(s,p)}{\partial p}+\xi \frac{\partial M(s,p)}{\partial p}\right.\nonumber\\
&\left.\qquad\qquad + C_{\gamma}\frac{\frac{\partial G(s,p)}{\partial p}(P(s,p)+\epsilon_0)- G(s,p)\frac{\partial P(s,p)}{\partial p}}{(P(s,p)+\epsilon_0)^2}\right)\,ds.\label{pi_p}
\end{align}
Note that we need to compute $\D\frac{\partial \bfa x}{\partial p}(\tau,p)$, where $\bfa x(\tau,p) = (P(\tau, p), M(\tau, p), G(\tau, p))$, the state variables of the within-host model \eqref{Immune_model}. From Clairaut's Theorem, we have
\begin{equation}\label{partial_p}
\frac{\partial}{\partial \tau}\left(\frac{\partial \bfa x}{\partial p}\right)=\frac{\partial}{\partial p}\left(\frac{\partial \bfa x}{\partial \tau}\right),
\end{equation}
where $\partial \bfa x/\partial \tau$ is the right-hand side of the system \eqref{Immune_model}. Let $\bfa f(\bfa x, p) = \partial \bfa x/\partial \tau$, and consider the following system for the derivatives $\partial \bfa x/\partial p$:
\begin{equation}
\label{senseqn}
\left\{
\begin{aligned}
&\frac{\partial }{\partial \tau} \left(\frac{\partial \bfa x }{\partial p}\right) = \frac{\partial \bfa f }{\partial \bfa x} \frac{\partial \bfa x }{\partial p} + \frac{\partial \bfa f }{\partial p},\\
&\frac{\partial \bfa x }{\partial p} (0) = \bfa 0, \\
\end{aligned}
\right.
\end{equation}
where $\partial \bfa f /\partial \bfa x$ is the Jacobian matrix of the right-hand side of the system \eqref{Immune_model}. For example, taking parameter $ p = a$, the IgM immune response activation rate, we can derive the corresponding system \eqref{senseqn} as follows
\begin{equation*}
\underbrace{ 
\begin{pmatrix} 
\frac{\partial}{\partial t} \left(\frac{\partial P }{\partial a}\right)\\[0.7em]
\frac{\partial }{\partial t} \left(\frac{\partial M }{\partial a}\right)\\[0.7em]
\frac{\partial }{\partial t} \left(\frac{\partial G }{\partial a}\right)\\ 
\end{pmatrix}}_{\frac{\partial }{\partial t} \left(\frac{\partial \boldsymbol x }{\partial a}\right)}
=\underbrace{ 
\begin{pmatrix} 
r\left( 1- \frac{2P}{K}\right)-\theta M-\delta G   &  \ -\theta P &\ -\delta P   \\
a M &  \ a P- (q+c) &\ 0   \\
b G &  \ q &\ b P  
\end{pmatrix}}_{\mathrm{ \frac{\partial \bfa f }{\partial \bfa x}}}
\underbrace{ 
\begin{pmatrix}
\frac{\partial P }{\partial a}\\[0.7em]
\frac{\partial M }{\partial a}\\[0.7em]
\frac{\partial G }{\partial a}\\[0.7em]
\end{pmatrix}}_{\mathrm{ \frac{\partial \bfa x }{\partial a}}}
+
\underbrace{ 
\begin{pmatrix}
0\\
M P\\
0\\
\end{pmatrix}}_{\mathrm{ \frac{\partial \bfa f }{\partial a}}}.
\end{equation*}

\noindent By solving the extended system \eqref{senseqn}-\eqref{Immune_model}, we obtain the solutions for $\bfa x$ and the derivatives $\partial \bfa x/\partial p$ at all times $\tau$, which are then utilized to estimate the integrands in \eqref{R0_p} and \eqref{pi_p}. We then numerically approximate the integrals by using the trapezoidal rule. Note that \eqref{senseqn} is the standard derivation for the \textit{variational system} of an ODE.

\noindent Similarly, we assess the impact of within-host parameters on the final epidemic size, consider the endemic host population size $I^*_H$ and endemic vector population size $I_V^*$ as the QOIs, which are given in \eqref{eq_host_e} and \eqref{eq_vector_e}, respectively. The corresponding sensitivity matrices are derived as follows:
\begin{align*}
&\frac{\partial I^*_H}{\partial \p}=\displaystyle\int_0^\infty\!\bigg[\frac{\partial i^*_H(0)}{\partial \p}\pi_H(\tau)+ i^*_H(0)\frac{\partial \pi_H(\tau)}{\partial \p}\bigg]d\tau,\ \ \ \  \frac{\partial i^*_H(0)}{\partial \p}=\frac{1}{\mathcal{R}_0^2}\frac{\partial \R_0}{\partial \p}\bigg/ \bigg(\frac{1}{d} + \frac{1}{\beta_V}\bigg),\\
&\frac{\partial I^*_V}{\partial \p}=\frac{1}{(\R_0 S^*_H)^2}\bigg(\frac{\partial \R_0}{\partial \p} S^*_H -\frac{\R_0}{d}\frac{\partial i^*_H(0)}{\partial \p}\bigg).
\end{align*}

\begin{figure}[htbp]
\centering
\includegraphics[width=0.49\textwidth]{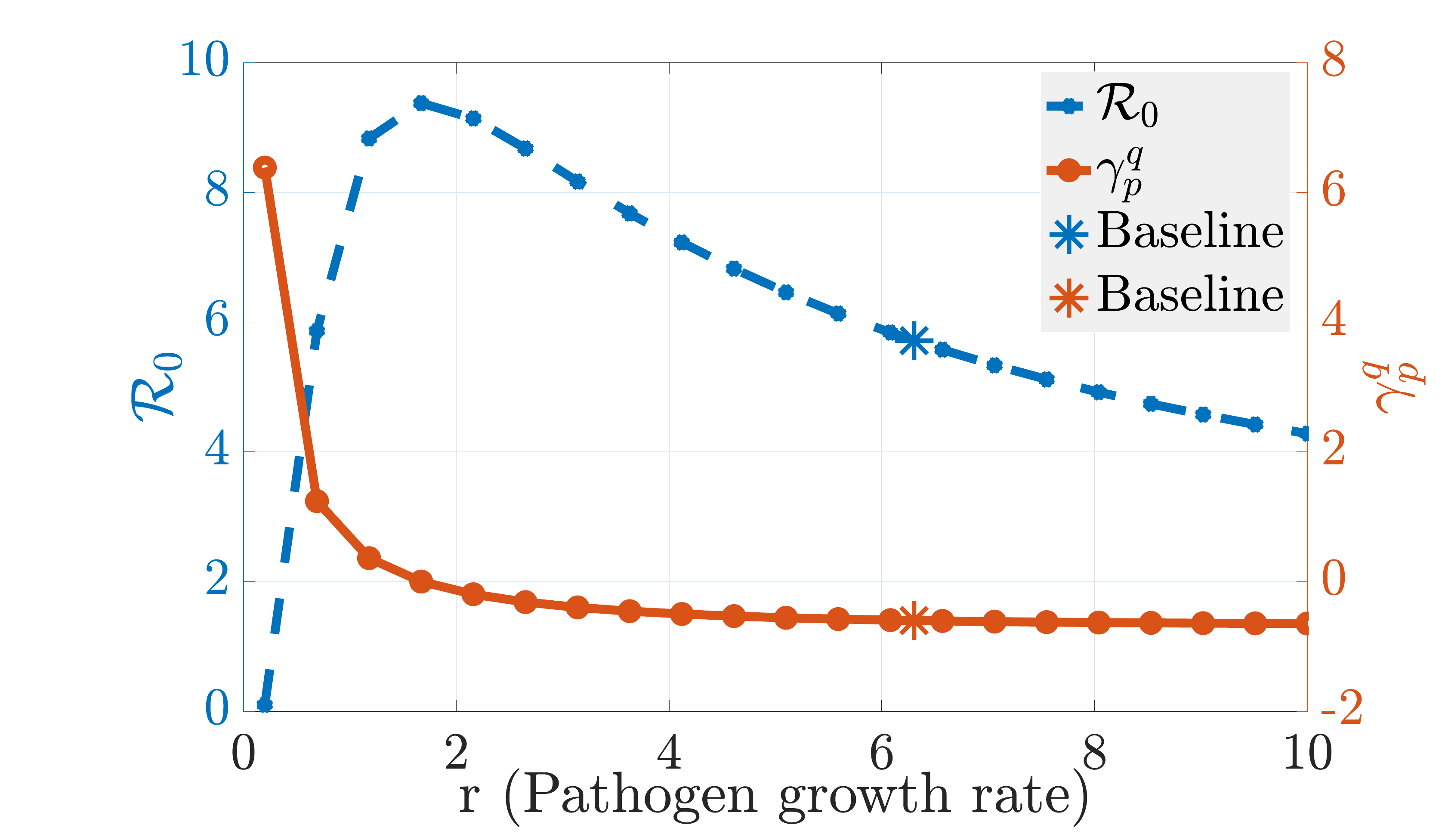}
\includegraphics[width=0.49\textwidth]{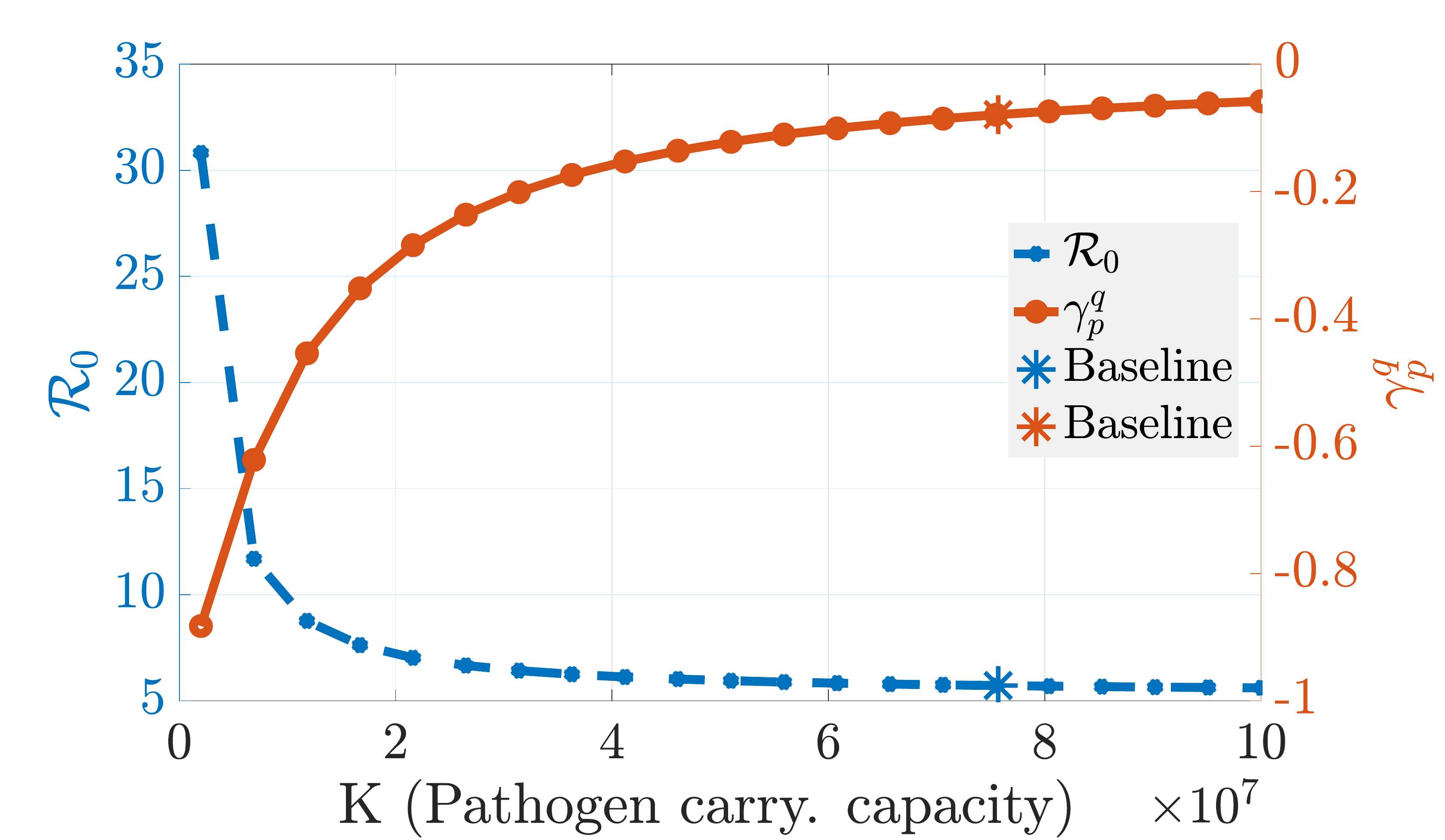}
\includegraphics[width=0.49\textwidth]{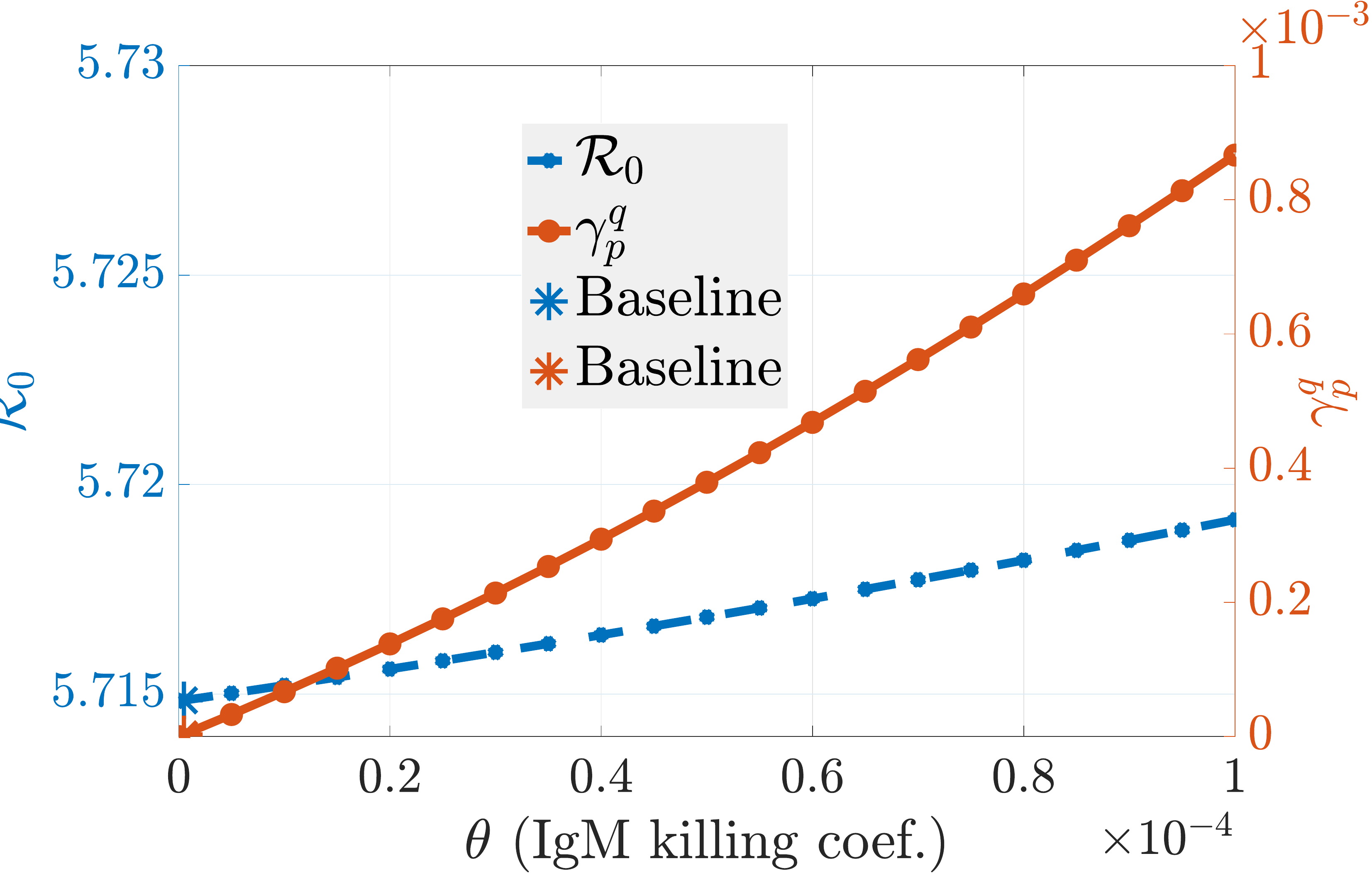}
\includegraphics[width=0.49\textwidth]{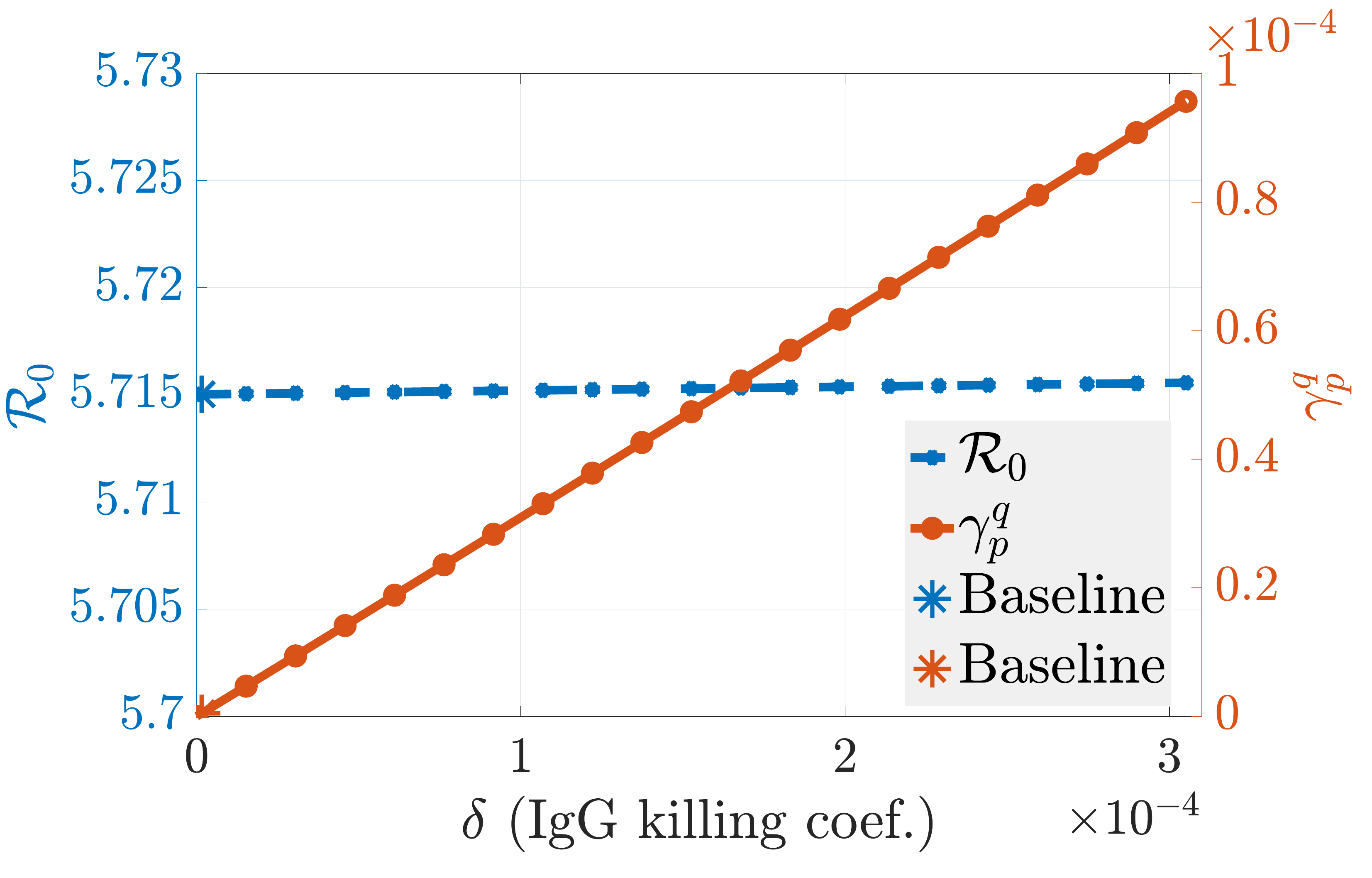}
\includegraphics[width=0.49\textwidth]{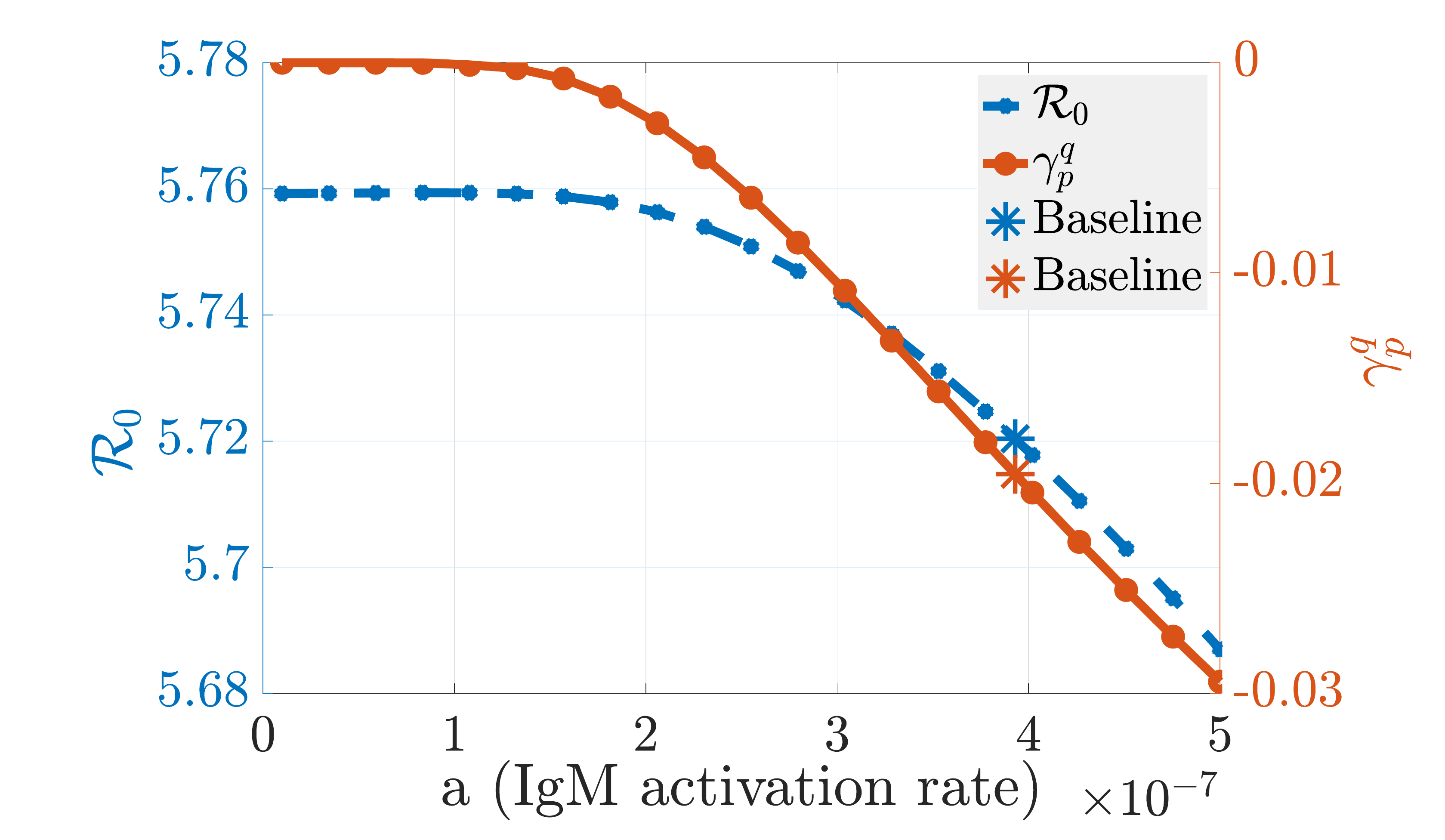}
\includegraphics[width=0.49\textwidth]{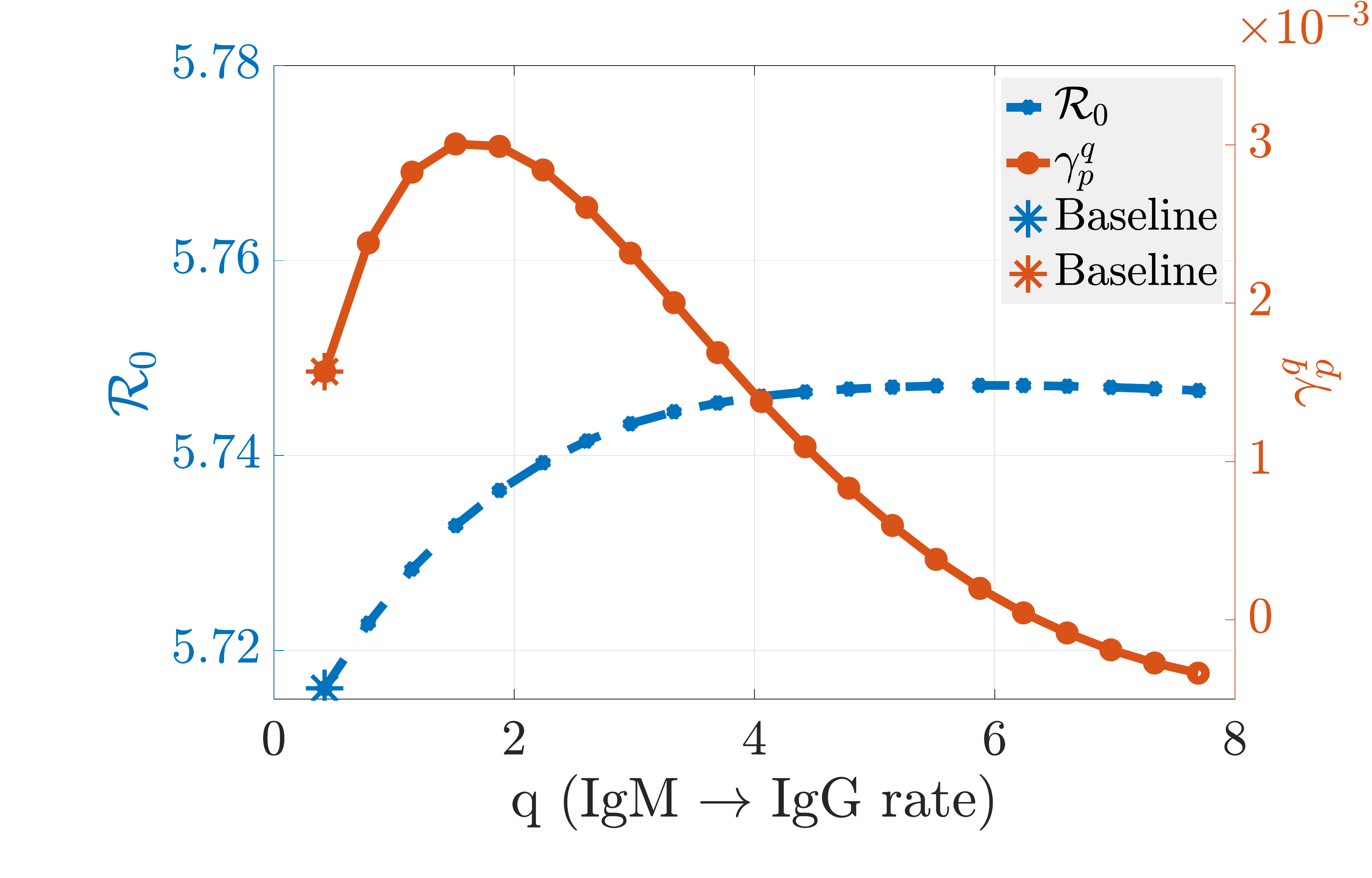}
\includegraphics[width=0.49\textwidth]{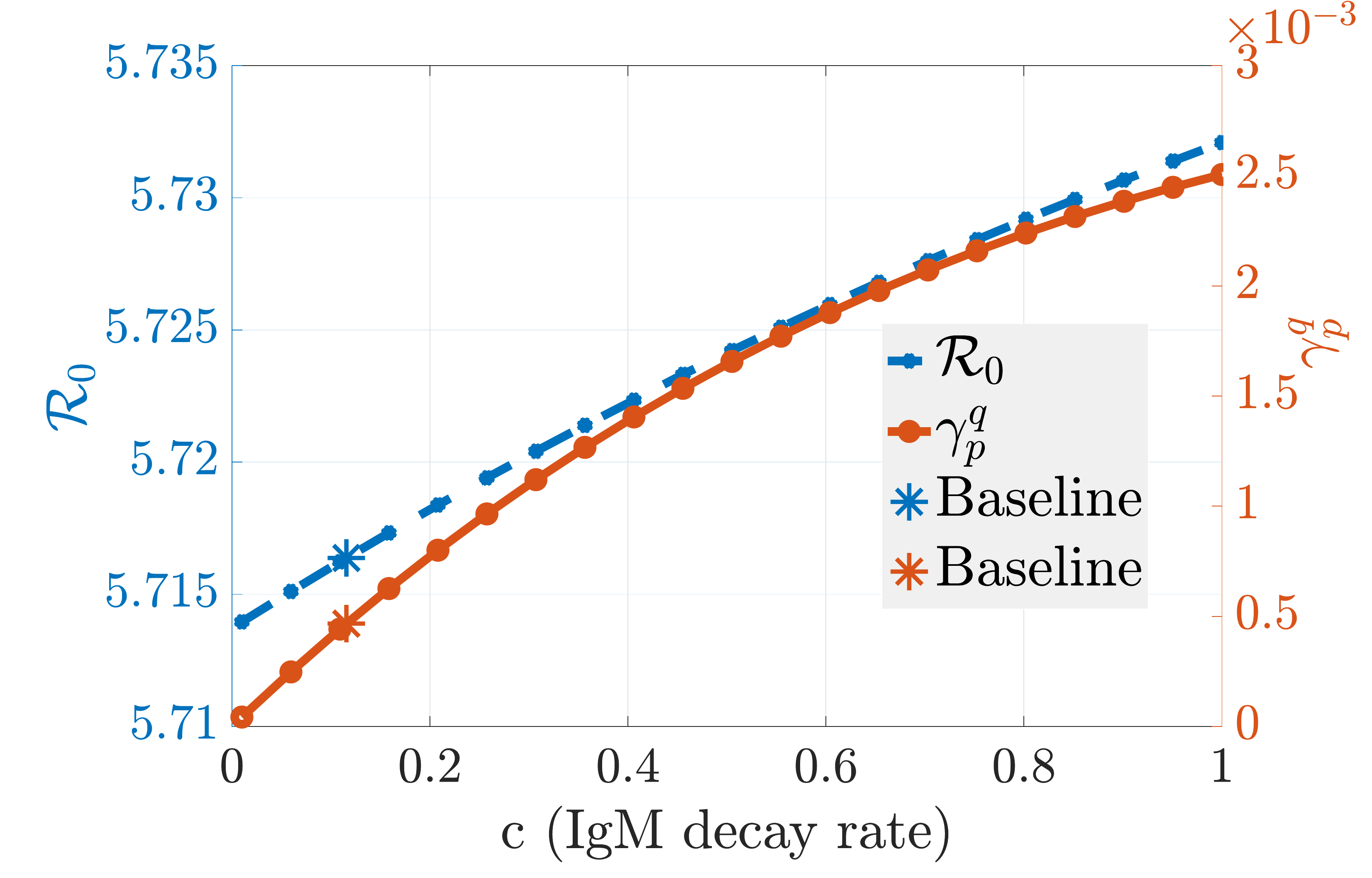}
\includegraphics[width=0.49\textwidth]{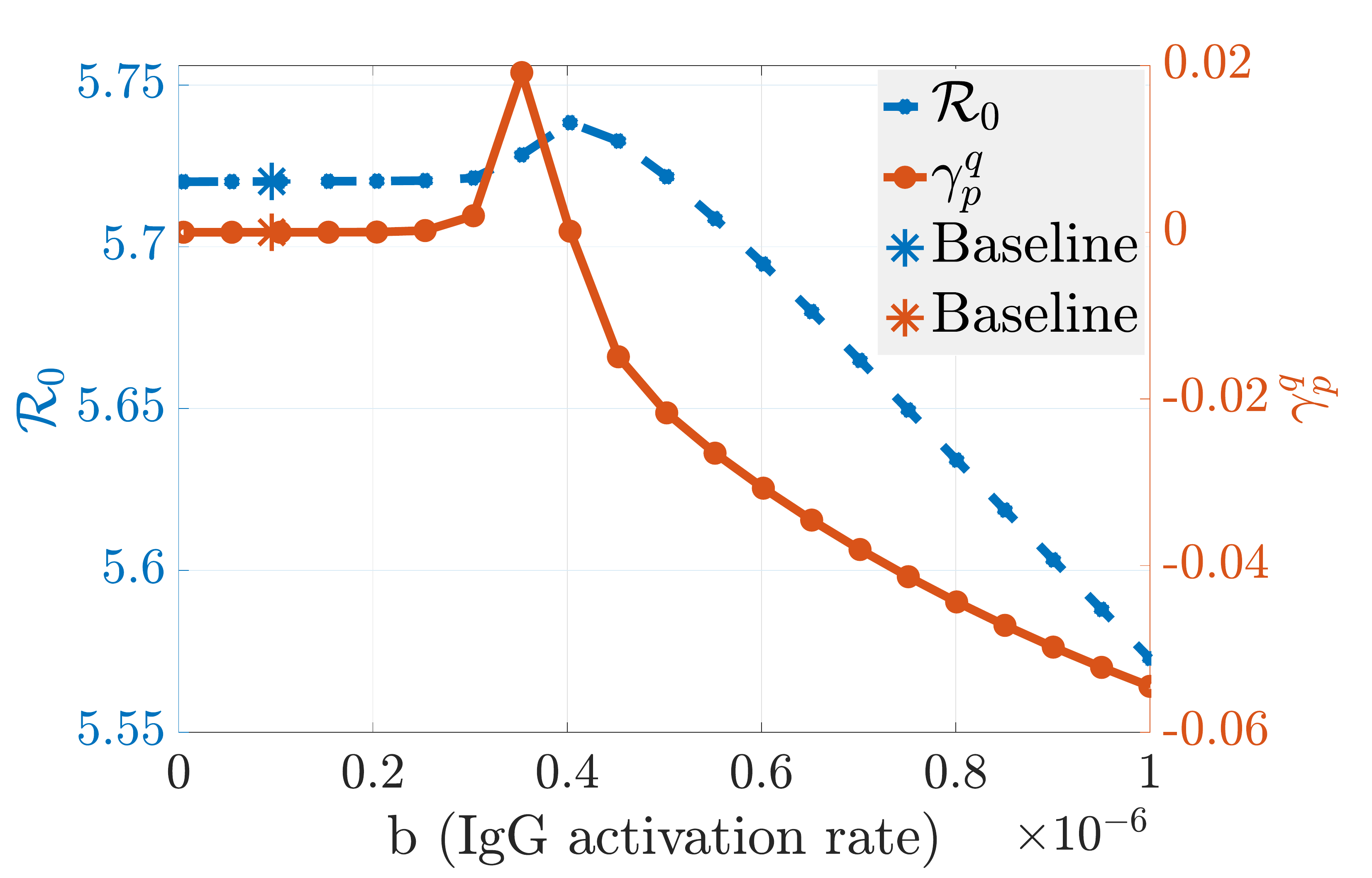}
\caption{\emph{Impact of immune parameters on the epidemic threshold, $\R_0.$} The \textit{immune-response-dependent epidemiological reproduction number,} $\R_0$ (left y-axis, blue dotted curve) along with the response curves for the extended SA, $\gamma^{\R_0}_{\p}$ (right y-axis, orange curve) are plotted against the immune parameter values $p$ over a range of values around the baseline estimates.}  
\label{fig:SA_R0}
\end{figure} 

\begin{figure}[htbp]
\centering
\includegraphics[width=0.49\textwidth]{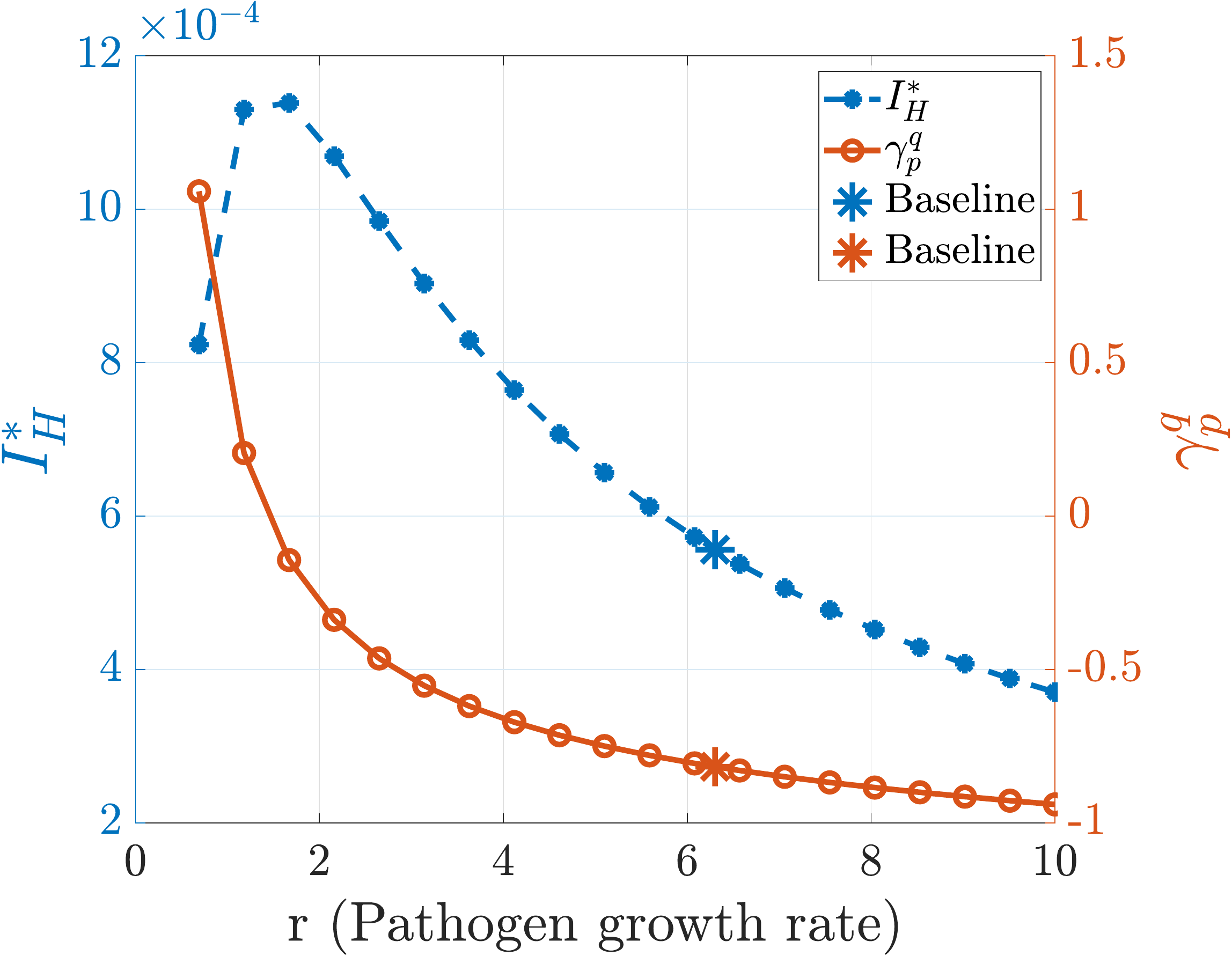} 
\includegraphics[width=0.49\textwidth]{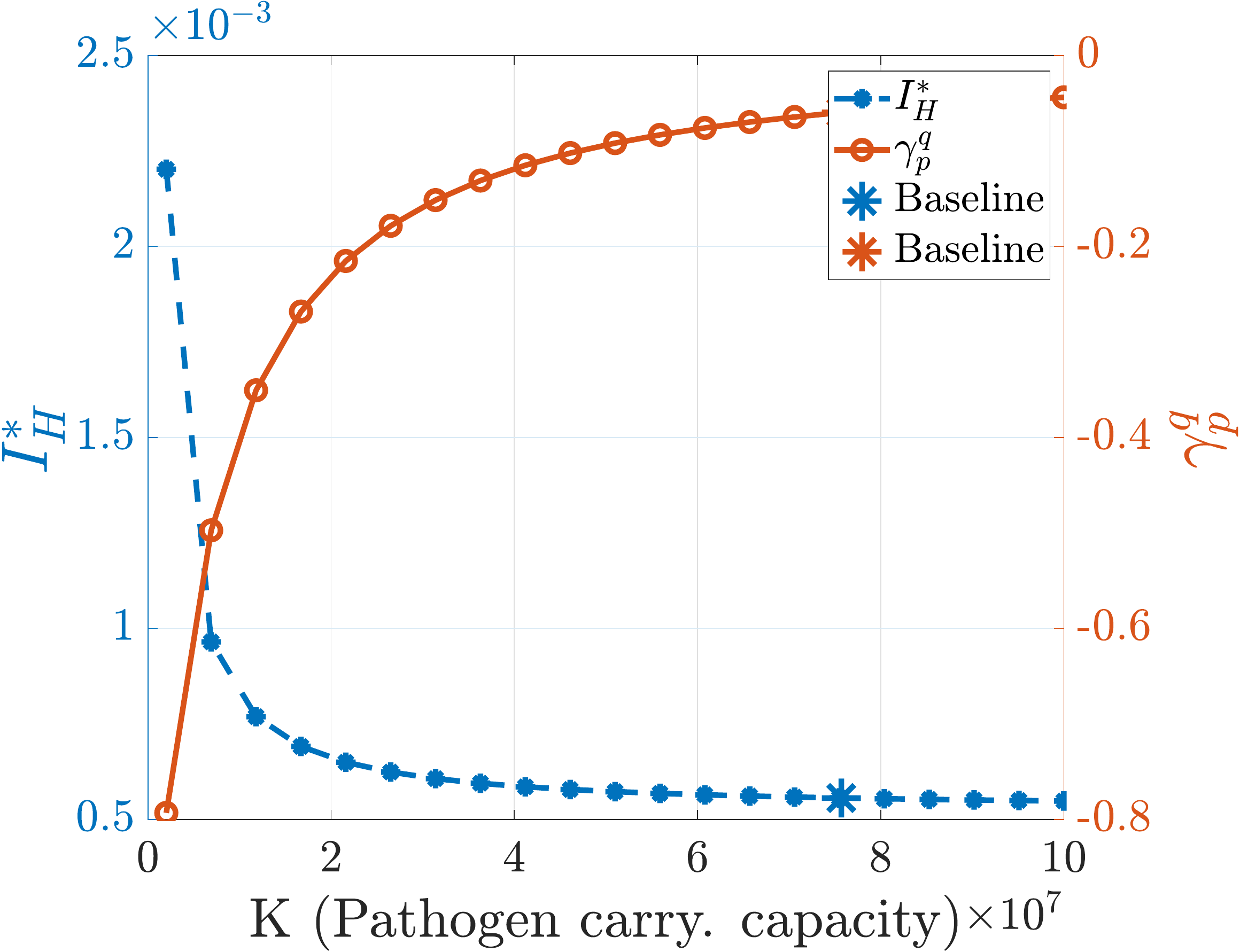}
\includegraphics[width=0.49\textwidth]{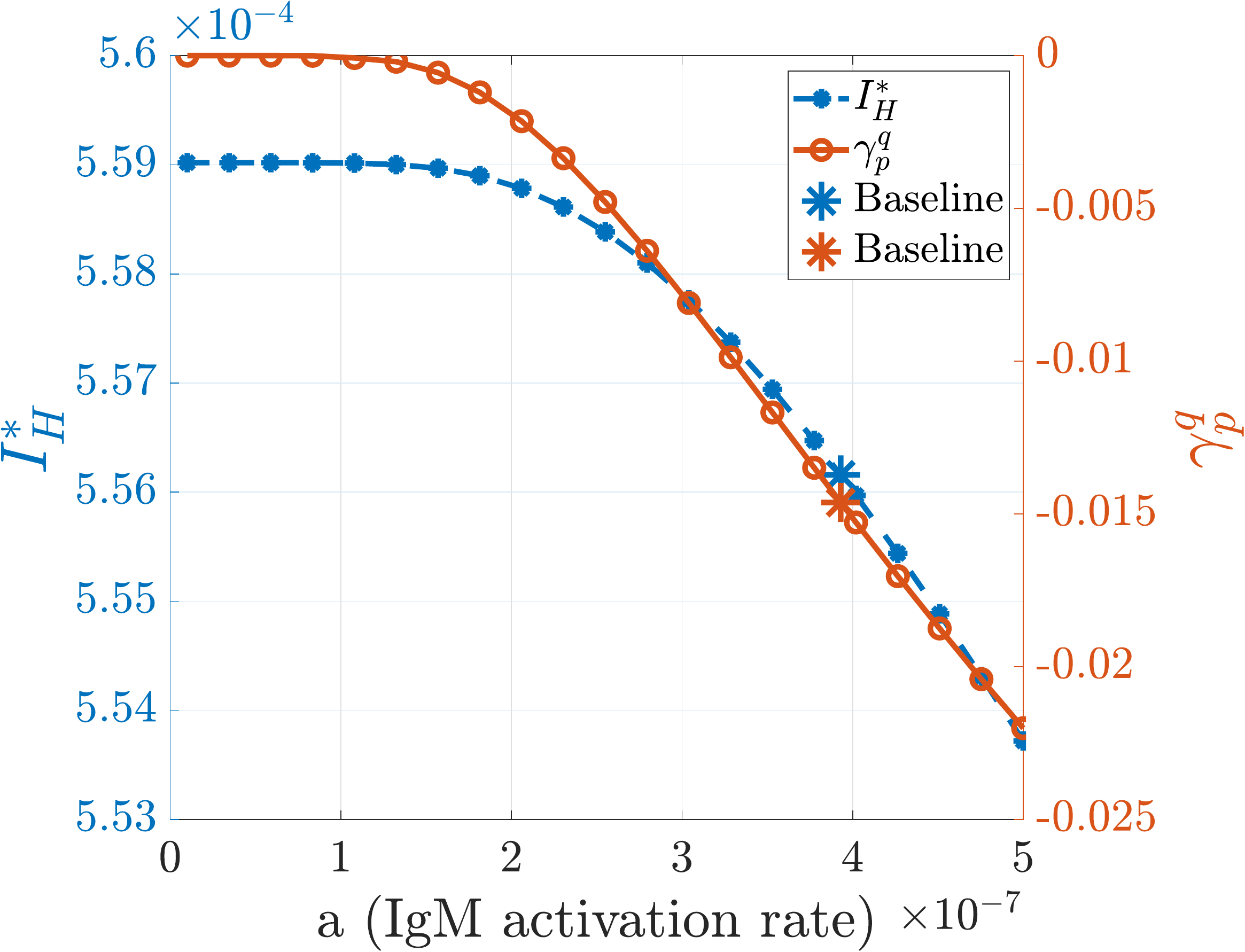} 
\includegraphics[width=0.49\textwidth]{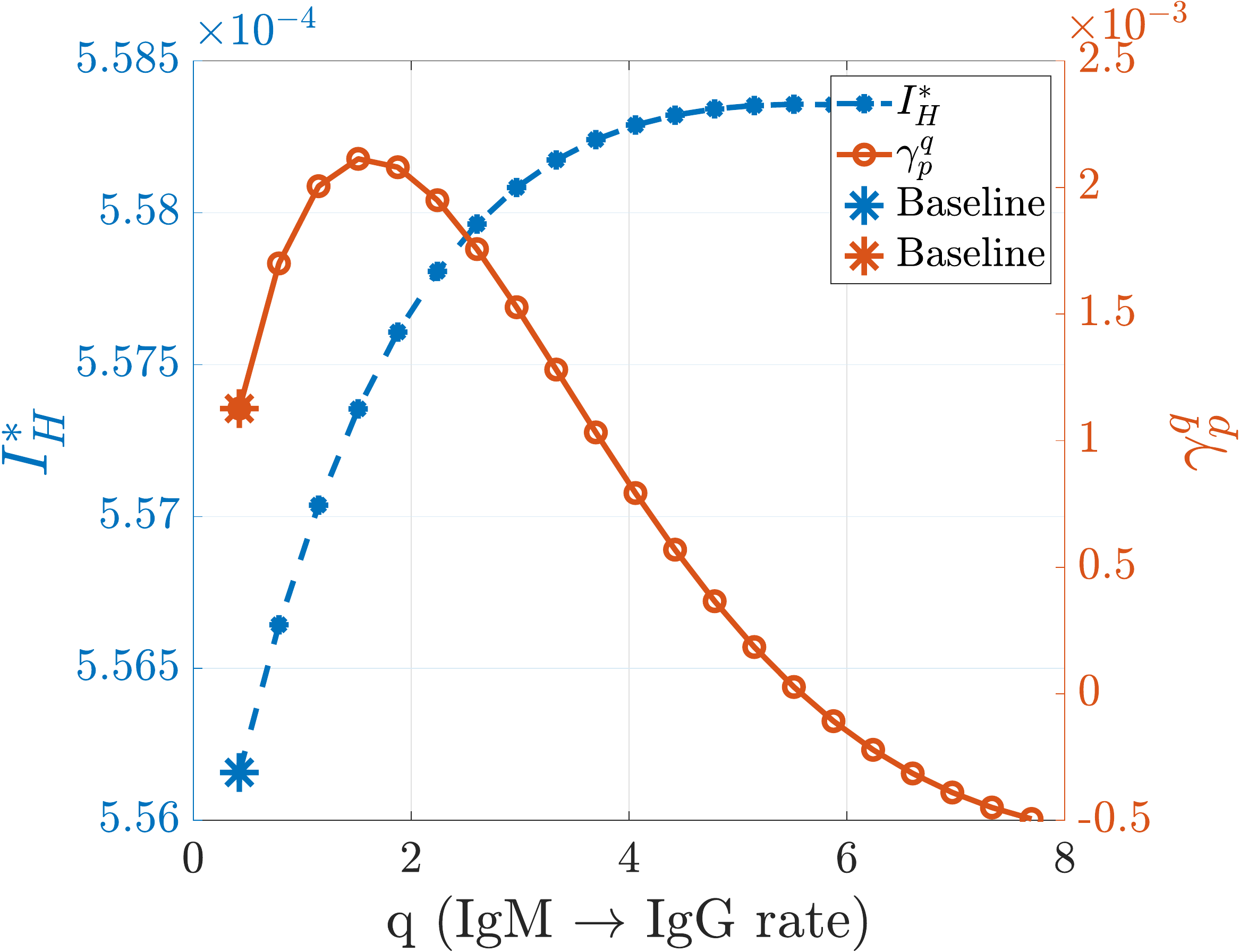}
\includegraphics[width=0.49\textwidth]{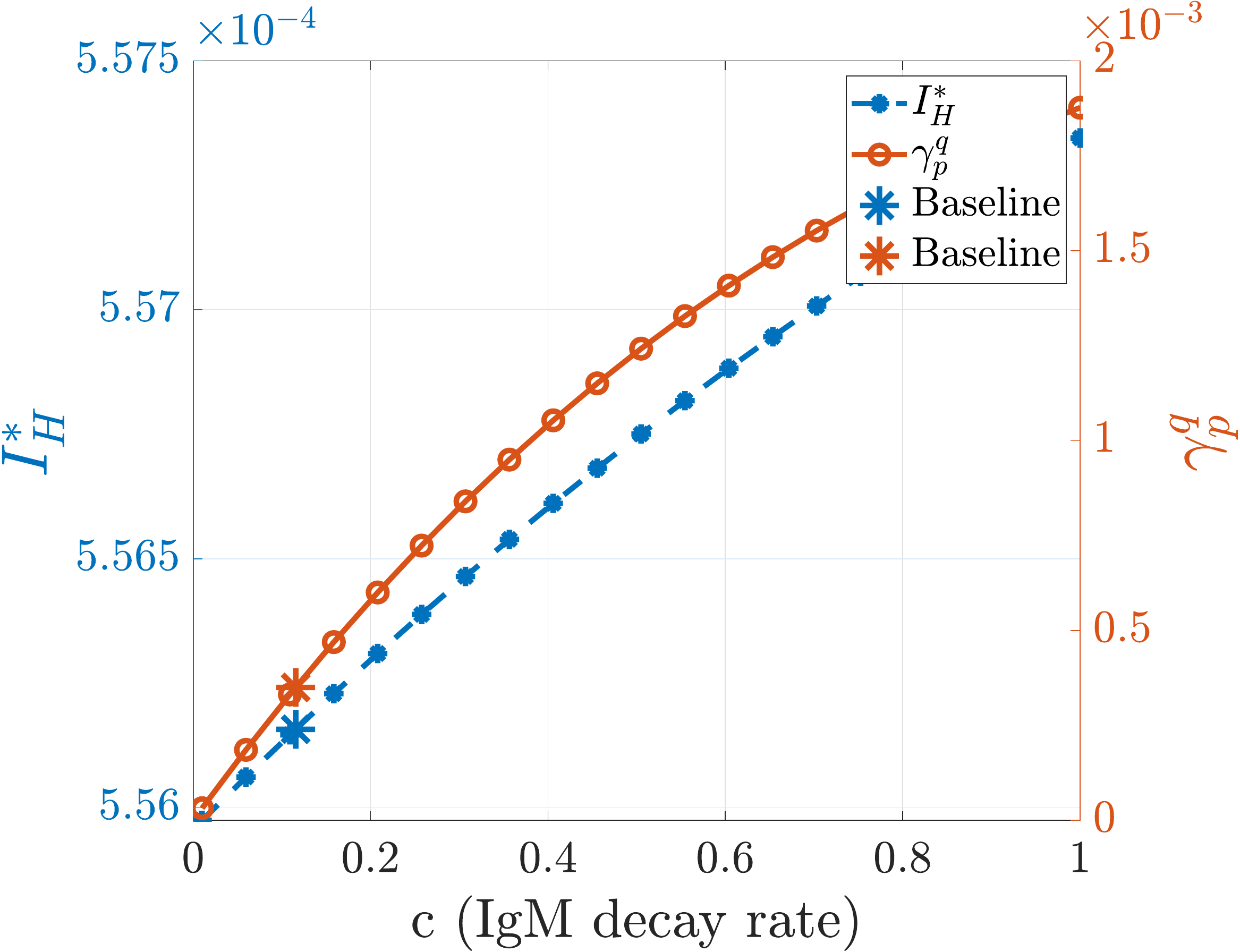}
\includegraphics[width=0.49\textwidth]{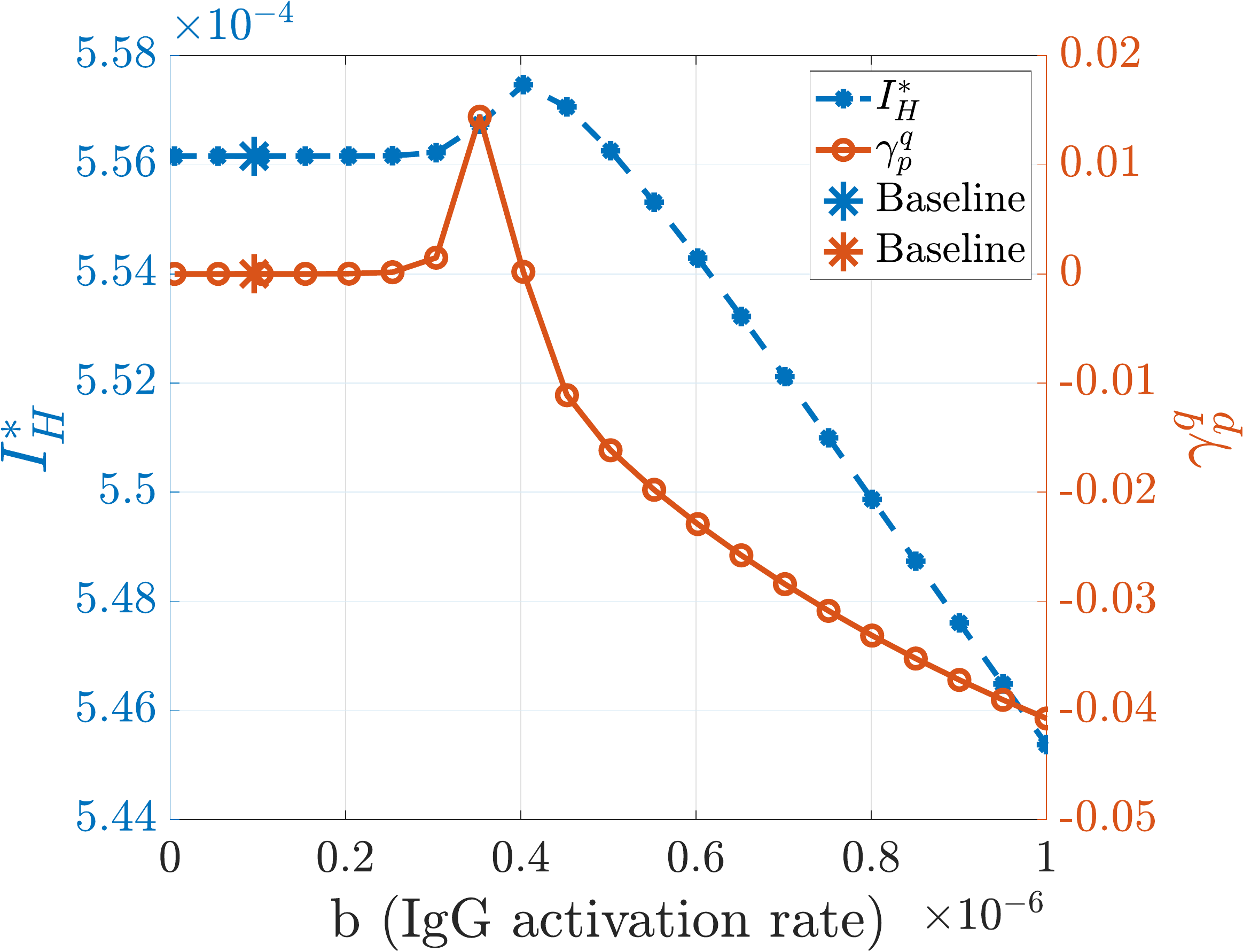}
\caption{\emph{Impact of immune parameters on the final epidemic size, $\mathcal I_H^*.$} The final infected host disease abundance, $\mathcal I^*_H,$ along with the response curves for the extended SA, $\gamma^{\R_0}_{\p},$ are plotted against the immune parameter values $p$ over a range of values around the baseline estimates.}  
\label{fig:SA_Ih}
\end{figure} 

%\noindent \textbf{Numerical results.} 
In Fig.\ref{fig:SA_R0}, the immune-response-dependent epidemiological reproduction number, $\R_0,$ along with the response curves for the extended SA, $\gamma^{\R_0}_{\p},$ are plotted against the immune parameter values $p$ over a range of values around the baseline estimates. In Fig.\ref{fig:SA_R0}(a),
the epidemiological threshold $\R_0$ is a non-monotone function of the within-host pathogen replication rate $r$ and there is a critical pathogen replication rate $r_c$ at which the reproduction number is maximal.  We present the resulting sensitivity indices at the baseline parameters in Table \ref{table:SA_R0}, where the indices are sorted by magnitude.  The numerical results suggest that:

\begin{itemize}
\item[i)] At the baseline (estimated) values of within-host immune parameters, the within-host viral growth rate $r$ has the largest impact on $\R_0:$ $1\%$ increase in $r$ causes $0.6\%$ reduction in $\R_0$, $0.8\%$ reduction in final infected host abundance $\mathcal I_H^*$ and $0.72\%$ reduction in final infected vector abundance $\mathcal I_V^*$.
\end{itemize}
%\blue{***Include the comments from the reviewer \#2, comment \#6 on the observation.***}

In Fig.\ref{fig:SA_R0}, the reduction in $\R_0$ with increasing within-host pathogen growth rate $r$ is counter-intuitive, but well known %biological assumption 
in the literature \cite{RMay}. The trade-off between pathogen virulence and infectiousness is first shown in \cite{Gilchrist2002}, articulated through a within-host model for directly transmitted diseases. Later it was extended to vector-borne diseases in \cite{gulbudak2017vector}. In particular, the reduction in $\R_0$ for large $r$ occurs due to  three mechanisms: \emph{(i) death due to ``aggressive'' immune response, (ii) death due to pathogen exploitation of target cells, and (iii) decreasing infectious period due to large viral clearance.} As the pathogen growth increases rapidly,  the immune response becomes more robust (See Fig. \ref{figure:WH_dyn}). Both cases can lead to rapid death of the host, shortening the infectious period, and subsequently leading to a reduction in the average number of secondary cases caused by these infectious individuals. In addition, note that  a $1\%$ increase in within-host pathogen growth rate $r$ in the parameter region $r \in [0.1, 2]$ leads to up to $8\%$ increase in $\R_0$ (Fig.\ref{fig:SA_R0}), $1\%$ increase in  $\mathcal I_H^*$ (Fig.\ref{fig:SA_Ih}), and  $1.5 \%$ increase in  $\mathcal I_V^*$ (Fig.\ref{fig:SA_Iv}), which are significant increases in population-scale disease quantities, suggesting that control strategies should be targeted toward reducing within-host pathogen growth rate to significantly change the disease outcomes in the long term. For example, within-host pathogen growth rate $r$ can be reduced by immunizing the host population, or through drug treatment during outbreaks, hampering viral growth within-hosts. 
%\end{remark}
\begin{itemize}
\item[ii)]  The within-host pathogen carrying capacity $K$ has the second largest impact on $\R_0$ at the baseline parameter value $K=7.57\times 10^{7}$: $1\%$ increase in $K$ causes  $0.079\%$ reduction in $\R_0$, $0.060\%$ reduction in  $\mathcal I_H^*$, and $0.095\%$ reduction in  $\mathcal I_V^*$.
\item[iii)]  The within-host IgM immune activation rate $a$ has the third largest impact on $\R_0$ at the baseline parameter value $a=1.1\times 10^{-7}$: $1\%$ increase in $a$ causes  $0.019\%$ reduction in $\R_0$, $0.015 \%$ reduction in  $\mathcal I_H^*,$ and $0.023 \%$ in  $\mathcal I_V^*.$
\end{itemize}
The faster the IgM antibodies activate, the faster they clear the pathogen, causing reduction in disease transmission (See Fig.\ref{fig:SA_R0}(c)).  A drastic decrease in $\R_0$ for large values of $a$ suggests that a  large improvement in vaccine efficacy is needed for efficient reduction of population-scale transmission. 

\begin{itemize}

\item [iv)] The immune response switching rate $q$ has the fourth largest impact on the epidemic quantities, albeit almost negligible: a $1\%$ increase in $q$ causes  $1.6 \times 10^{-3}\%$ increase in $\R_0,$ a $1.13 \times 10^{-3} \%$ increase in  $\mathcal I_H^*$ and $1.9 \times 10^{-3} \%$ increase in  $\mathcal I_V^*.$ 
\end{itemize}
These numbers suggest that it might be more favorable for the host population if the loss term due to B-cells switching production of IgM antibodies to IgG antibodies (with a per-capita rate $q$) is smaller.  In other words, because the IgM immune response antibodies are mainly responsible for \emph{rapid destruction of virus}, for better outcome in disease eradication,  rapid destruction of virus within hosts can be more crucial than the life-long immunity (memory) provided by IgG antibodies.

\begin{itemize}
\item[v)] The IgM immune response decay rate is the next most influential parameter  impacting epidemic quantities: a $1\%$ increase in $c$ causes a negligible $4.6\times 10^{-4}\%$ increase in $\R_0$, a $3.5\times 10^{-4} \%$ increase in  $\mathcal I_H^*$, and a $5.6 \times 10^{-4} \%$ increase in  $\mathcal I_V^*$. 
\item[vi)] Finally, the killing efficacy parameters $\theta, \delta$ for both immune responses appear to be least impactful immunological parameters. 
\end{itemize}
It is interesting to note that an increase in the killing efficacy parameters can lead to an increase in epidemic quantities, although just marginally. The biological insight behind this numerical observation is that: \emph{because the baseline parameters are in the high disease fatality parameter region, an increase in parameter killing efficacy slows down the viral density growth within-hosts, decreasing disease induced death rates; subsequently,  allowing an increase in infectious period, which leads an increase in epidemic quantities (see Fig.\ref{figure:WH_dyn} (bottom row subfigures)).}  In summary, comparing the %magnitude of the
impact of immune response parameters,  the IgG immune activation rate seems to be the least crucial, underscoring the importance of IgM immune parameters in disease eradication over IgG immune response. However this result might be an artifact of how the disease-induced death rate $\kappa$ is formulated and what the range of the parameter $\theta$ is. 
We observed that the value of the IgM killing efficacy parameter largely determines the peak viral load in such a way that, when the IgM killing efficacy is chosen sufficiently large, not only the viral relapse occurs but also there is a larger change in the virus density (see the Fig \ref{fig:theta}). Since the values of $\theta$ in the chosen range are small, we observe no viral relapse and a smaller impact on the viral dynamics, but a larger reduction in the antibody response dynamics.

\begin{table}
\caption{Different linking parameter formulations} 
\centering 
\begin{tabular}{llllll}
\toprule
Definition & Param. & LP$_{1}$ & Ref. &  LP$_{2}$ & Ref. \\
\midrule 
Host transmission rate &$\beta_H(\tau)$  & $\D\frac{C_\beta P(\tau)}{C_0+ P(\tau)}$ & \cite{tuncer2016structural} & $\D\frac{C_\beta^* P^2(\tau)}{(C_0^*)^2+ P^2(\tau)}$  & \cite{gulbudak2017vector}  \\ [0.4cm] 
Pathogen induced death rate & $\alpha(\tau)$  & $\zeta P(\tau)$ & \cite{tuncer2016structural} & $r \zeta^* P(\tau)$ &\cite{gulbudak2017vector,Gilchrist2002} \\[0.3cm] 
Antibody induced death rate & $\kappa(\tau)$   & $\xi M(\tau)$ & \cite{tuncer2016structural} & $ a\xi^* M(\tau) P(\tau)$  &\cite{gulbudak2017vector,Gilchrist2002} \\[0.2cm] 
Host recovery rate&$\gamma(\tau)$ & $C_\gamma\D\frac{G(\tau)}{P(\tau)+\epsilon_0}$ & \cite{tuncer2016structural} & $C_\gamma^*\D\frac{G(\tau)}{P(\tau)+\epsilon_0^*}e^{-P(\tau)/P(0)} $& \cite{gulbudak2017vector} \\   
\bottomrule
\end{tabular}
\label{table:distinct_linking} 
\end{table}

%%%%%%%%%%%%%%%%%%%%%%%%%%%%%%%%%%%%%%%%%%%%%%%%%%%%%%%%%%%%%%%%%%%%%

\begin{table}[htbp]
\caption{Re-calibrated parameters for the alternative linking functions ($LP_2$) in Table \ref{table:distinct_linking}. The parameters are estimated using nonlinear least squares to match the basic reproduction number $\R_0$.  \label{table:parameters_link2}}
\begin{tabular}{lllll}
\toprule
& Linking function parameters & Dimension & Estimated  & Range\\
\midrule
$C_{\beta}^*$ & transmission efficiency $(=C_{\beta})$ & (host$\times$ day)$^{-1}$ & $0.5365$ & - \\
$C_0^*$ &  half saturation in transmission rate & $\mbox{TCID}_{50}$ & $3.030 \times 10^3$ & $10^3 \sim 10^4$ \\
$\zeta^*$ & the pathogen cost coefficient &$(\mbox{TCID}_{50})^{-1}$ &$9.840\times 10^{-9}$ & $10^{-9} \sim 10^{-7} $\\
$\xi^*$ &  the immune response cost coefficient & $(\mbox{Elisa PP})^{-1}$ & $2.951\times 10^{-6}$ & $10^{-7}\sim 10^{-5}$\\ 
$C_{\gamma}^*$ &  recovery coefficient &  $\frac{\mbox{TCID}_{50}}{\mbox{Elisa PP}\times\mbox{day}}$ &  $4.860$ & $0\sim 10$\\
$\epsilon_0^*$ &  half saturation in recovery rate $(=\epsilon_0)$ & \mbox{TCID}$_{50}$ & $ 7.43\times 10^{-4}$ & - \\
\bottomrule& 
\end{tabular}
\end{table}

\begin{table}
\caption{Local sensitivity indices $\gamma^{QOI}_{\p}$ with respect to model parameters %\notes{I re-fit the parameters using only the $\R_0$ with very fine tolerance, now the order for the SI is identical to the $LP_1$. ZQ}
} 
\centering 
\begin{tabular}{llllllll}
\toprule
LP$_1$ & $\gamma^{\R_0}_{\p}$ & $\gamma^{\mathcal I_H^*}_{\p}$ & $\gamma^{\mathcal I_V^*}_{\p}$& LP$_2$ & $\gamma^{\R_0}_{\p}$& $\gamma^{\mathcal I_H^*}_{\p}$ & $\gamma^{\mathcal I_V^*}_{\p}$\\
\midrule
$r$ & $-0.598$ & $-0.817$ & $-0.720$ & $r$ & $-0.7520$  & $-0.925$ & $-0.906$\\
$K$ & $-0.0789$ & $-0.0595$& $-0.0950$ & $K$ & $-0.0801$  & $-0.0607$& $-0.0964$\\
$a$ & $-0.0194$ & $-0.0146$& $-0.0234$ & $a$ &$-0.0205$  & $-0.0155$ & $-0.0247$\\
$q$ & $1.56 \times 10^{-3}$ & $1.13 \times 10^{-3}$& $1.88 \times 10^{-3}$ & $q$  & $1.64 \times 10^{-3}$  & $1.22\times 10^{-3}$& $1.97 \times 10^{-3}$\\
$c$ & $4.64\times 10^{-4}$ & $3.50\times 10^{-4}$& $5.59 \times 10^{-4}$ & $c$ &  $4.61 \times 10^{-4}$   & $3.49\times 10^{-4}$ & $5.54\times 10^{-4}$\\
$\theta$ &  $3.19\times 10^{-6}$&  $2.40\times 10^{-6}$& $3.84\times 10^{-6} $ & $\theta$ &  $5.44\times 10^{-6}$ & $4.12\times 10^{-6} $ & $6.55\times 10^{-6}$\\
$\delta$ & $4.65 \times 10^{-7}$& $4.39 \times 10^{-7}$& $5.60\times 10^{-7}$ & $\delta$ &  $6.12\times 10^{-7}$  & $5.50\times 10^{-7}$& $7.37\times 10^{-7}$\\
$b$ &  $3.10 \times 10^{-7}$&  $2.31 \times 10^{-7}$ & $3.74\times 10^{-7}$ & $b$ &  $4.21 \times 10^{-7}$ & $3.18\times 10^{-7}$& $5.07\times 10^{-7}$\\   
\bottomrule& & & & 
\end{tabular}
\label{table:SA_R0} 
\end{table}

\begin{table}
\caption{Local epidemic sensitivity indices $\gamma^{\mathbf q}_{\p}$ with respect to within-host model parameters.} 
\centering 
\begin{tabular}{llll}
\toprule
\emph{Fatal} & $\gamma^{\R_0}_{\p}$ & \emph{Mild} & $\gamma^{\R_0}_{\p}$\\
\midrule
$r$ & $-0.598$ &$r$& $0.587$\\
$K$ & $-0.0789$ &$K$ & $-0.0137$   \\
$a$ & $-0.0194$ & $a$ &$-0.0029$   \\
$q$ & $1.56 \times 10^{-3}$ & $q$  & $-0.00193$   \\
$c$ & $4.64\times 10^{-4}$ & $c$ &  $5.54 \times 10^{-4}$   \\
$\theta$ &  $3.19\times 10^{-6}$&  $b$ &  $4.91\times 10^{-7}$   \\
$\delta$ & $4.65 \times 10^{-7}$& $\theta$ &  $3.88\times 10^{-7}$    \\
$b$ &  $3.10 \times 10^{-7}$&  $\delta$ &  $-3.65 \times 10^{-7}$  \\   
\bottomrule& & 
\end{tabular}
\label{table:SA_R0_mild_fatal} 
\end{table}

%%%%%%%%%%%%%%%%%%%%%%%%%%%%%%%%%%%%%%%%%%%%%%%%%%%%%%%%%%%%%%%%%%%%%

\subsection{The impact of the choice of linking parameters on the SA}
 An important question is, \textit{how does the choice of linking functions affect the outcome of the SA performed in Section \ref{sec:SA}?} To study this question, we consider different linking functions that have been implemented in the literature (Table \eqref{table:distinct_linking}), and we carry out the corresponding SA using a similar process to that described in Section \ref{sec:SA}. To do that we first re-calibrate the coefficients in the alternative linking functions ($LP_2$-columns in Table \ref{table:distinct_linking}) by solving the following optimization problem:
$$
\min_{C^*\in(C_L,C_R)}|\R_0(C^*)-\R_0|,
$$
to approximate the basic reproduction number $\R_0$ at the baseline setting. We have fit the four linking parameters, $C_0^*,\zeta^*,\xi^*, C_\gamma^*$, one at a time: we vary one linking parameter $C^*$ over its corresponding range, $(C_L,C_R)$, specified in Table \ref{table:parameters_link2}, while keeping all the other parameters in the  $LP_1$ forms at their baseline values. This single-variable optimization was repeated for each of the four linking parameters, and it was done using the \verb|fminbnd| algorithm in MATLAB with \verb|TolX| = $10^{-10}$. The resulting estimates are listed in Table \ref{table:parameters_link2}.\\

\noindent We present the resulting sensitivity indices for these new linking functions in Table \ref{table:SA_R0} ($LP_2$ columns). Comparing these with the indices for the original linking functions ($LP_1$ columns), we see the same ordering by magnitude and comparable values for different the paired linking functions. The numerical results suggest that the choice of linking functions, after fitting their linking parameters at the baseline values of the immunological and epidemiological parameters from Tuncer et. al.\cite{tuncer2016structural}, does not cause important differences in $\R_0$-values (Fig.\ref{fig:SA_link_R0}). We also observe similar patterns in the corresponding curves for final disease abundance $I^*_H$ and $I^*_V$ (Appendix \ref{sec:Appendix_SA_link}). 

\noindent Furthermore, notice that at the baseline (estimated) parameter values, we have  $\partial \R_0 / \partial r<0$ due to increasing disease induced fatality and decreasing infectious period. Therefore, we classify the diseases according to different ranges of of $r$ as: \emph{fatal}\/ when $\partial \R_0 / \partial r<0$; \emph{mild}\/ if $\partial \R_0 / \partial r>0$.
Then, independently of the choice of linking functions, we obtain that when the disease is fatal:
\begin{itemize} 
\item  The within-host pathogen growth rate $r$, and pathogen carrying capacity $K$ have the largest impact on the disease reproduction number and final disease abundance size, suggesting that the most efficient control measures are the ones that target reducing within-host viral growth rate, even if virus eradication within hosts cannot succeed. 
\item Compared to IgG immune response, the IgM immune response parameters have larger impact on the disease outcomes. 
\item Changes in the host immune response parameters have a larger impact on the final infected vector abundance ($I^*_v$) than on the final infected host abundance ($I^*_H$) because of the indirect transmission route. 
\end{itemize}

\noindent Next we consider the parameter region, where the disease is mild by fixing $r=1.$ For this case, we observe that SA of QOI to the within-host parameters $r, q$ and $\delta$ change signs, and the order of the sensitivity indices by magnitude is slightly different, suggesting that:
\begin{itemize}
\item Independently of disease fatality, the per capita parasite growth rate consistently appears to be the most important within-host parameter impacting disease spread. Yet in the mild case (as opposed to fatal case):$1\%$ increase in $r$ causes  $0.587\%$ increase in $\R_0$, i.e. increasing the final disease abundance.
\item  However $1\%$ increase in carrying capacity $K$ causes  $0.0137\%$ reduction in $\R_0.$ This reduction also holds when  $r$ is in the ``fatal" parameter region.
%which is counter intituitive since causing an increase in the in-host viral logistic growth rate, $f(P)$.
The biological insight behind this finding is that an increase in within-host pathogen carrying capacity causes the host to harbor more pathogens, resulting in increased fatality due to pathogen resource use.
\item An increase in the immune-response activation rates decreases the epidemic quantities. Yet, IgM parameters, compared to IgG,  still appear to be more impactful. This suggests that vaccine efficacy can be tested by measuring the increase in IgM and IgG antibody densities comparing data pre- and post-vaccination.
\end{itemize}
However, in the mild case, other parameters may also see a change in their baseline values corresponding to changing values of $r$ because the parameters in the model may not be independent from each other. For a more rigorous comparison of the mild and the fatal cases,  one needs to have another data set representing the mild disease, and fit other parameters as well, which we suggest as valuable future work.  

\begin{figure}[t]
\center
a)\includegraphics[width=12cm,height=9cm]{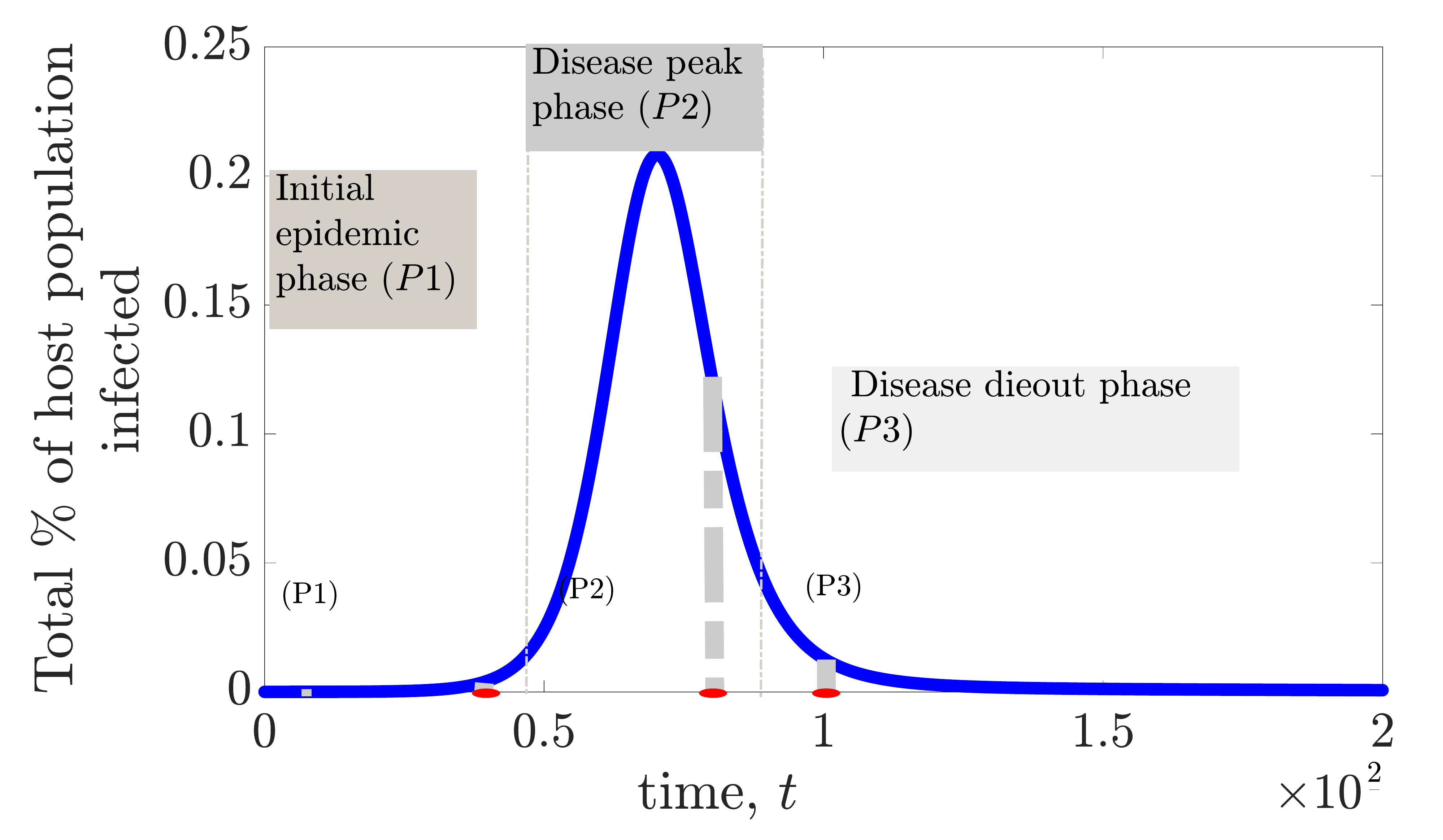}
b)\includegraphics[width=12cm,height=9cm]{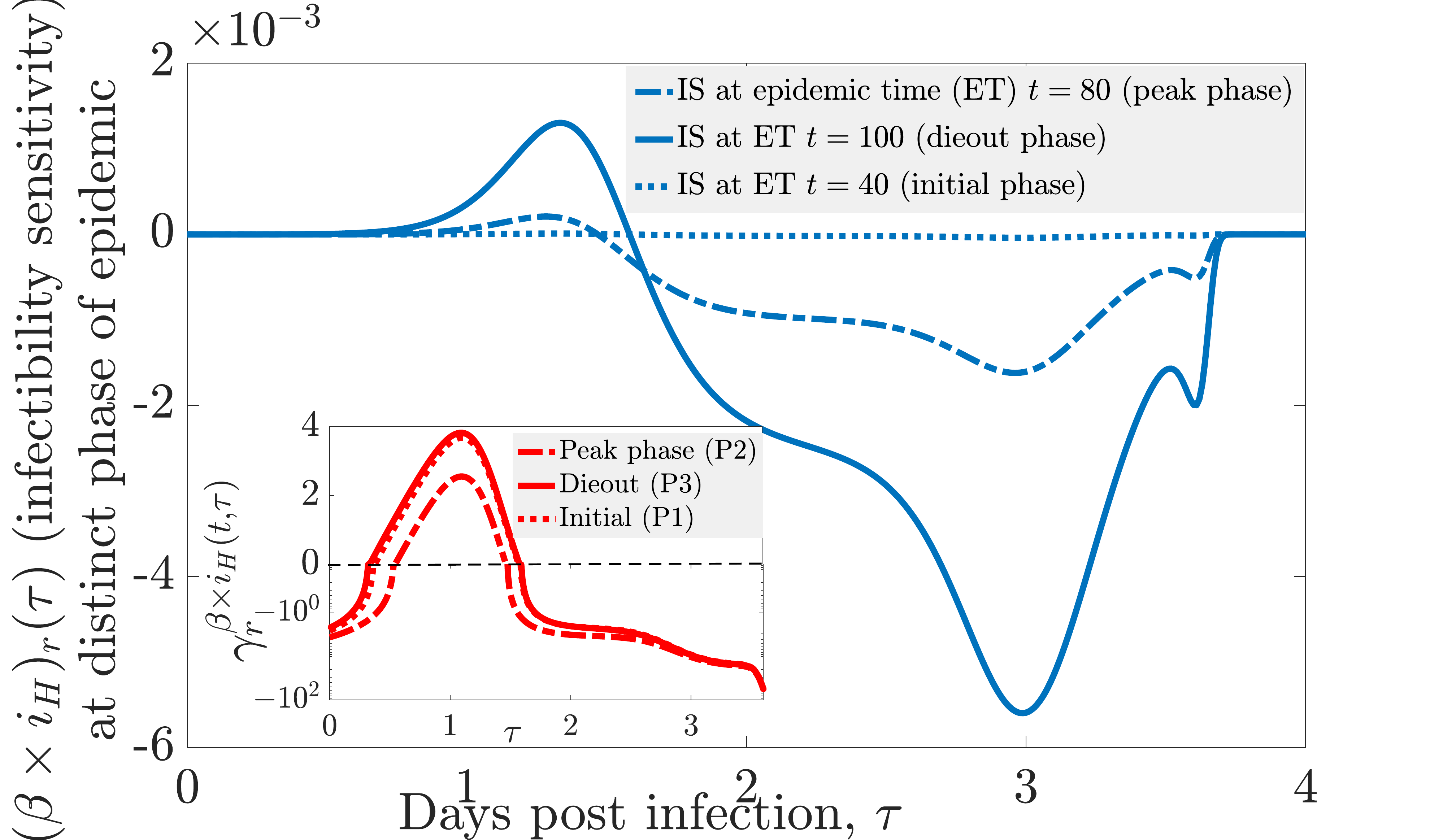}
%c)\includegraphics[width=7cm,height=6cm]{beta_i_H_a_at_distinct-eps-converted-to.pdf}
%d)\includegraphics[width=7cm,height=6cm]{beta_i_H_theta_at_disctinct-eps-converted-to.pdf}
\caption{\emph{ Sensitivity of infectability of host (at distinct epidemic phase at time $t$) to viral replication rate $r$.}
} 
\label{epidemic_phases}
\end{figure}

\subsection{Impact of viral replication rate on infectability during distinct phases of infection and epidemic}\label{SA_outbreaks}
So far we focused on the impact of immune parameters on the basic reproduction number $\mathcal R_0$ and the steady state infected host, $I_H^*$, and infected vector abundance $I_v^*$. 
Next, we will show how \textit{infectability} sensitivity with respect to key immune parameters, $\gamma^{\beta(\tau)i_H(\tau,t)}_{\textbf{p}}$ (derived in Appendix \ref{Inf_sensitivity}), during different outbreak phases and (within-host) infection-times shows heterogeneous impact of virulence-transmission trade-off. Recall that the \textit{normalized forward sensitivity index} $\gamma^{\beta(\tau)i_H(\tau,t)}_{\textbf{p}}$ quantifies  \textit{how much the infectability of cohort of individuals $i_H(\tau,t)$ changes when the parameter $\textbf{p}$} is increased by $1\%.$  
In contrast to the reproduction number and steady states, infectability, defined with the epidemic quantity $\beta(\tau)i_H(\tau,t)$, is a function of times since outbreak and individual infection start, $t$ and $\tau$, respectively, allowing us to see within-host effect on the main population level determinant of force of infection which varies with $t$ and $\tau$.  Although sensitivities are calculated for all immune parameters, because the viral replication rate is the fastest evolving parameter determining virulence \cite{Gilchrist2002,gulbudak2017vector}, we highlight here the influence of $r$ on infectability, dependent on both infection-age of cohorts of infected individuals and time since outbreak began on population scale (see Appendix \ref{Inf_sensitivity} for other parameter sensitivities and additional figures).  In this way we can understand how, in general, a virus might evolve in response to control strategies acting at different epidemic phase and different times during infection course (e.g. isolation of infected cases), or conversely how interventions may be optimally timed given the viral replication rate.

We observe in Fig.\ref{epidemic_phases} and Fig.\ref{epidemic_phase1}(a) that the raw (non-normalized) sensitivity of infectability to viral replication rate, $\left(\beta\times i_H\right)_{r}(\tau,t)$, is positive during early infection (small $\tau$) and negative after a sufficient amount of time-since infection passes, and the magnitude amplifies at later phases of the epidemic.  Indeed, increasing $r$ causes the within-host infection to be more severe and acute, with larger and earlier peaks of both virus and immune response, thereby tending to shift the transmissions caused by an infected individual to occur earlier during infectious period.  Certain control strategies, such as contact tracing, self-isolation, case finding and hospitalization, require some time $\tau$ after infection to act due to delays in symptom onset or tracking, and thus the increased transmissibility during early infection of viral strains with higher $r$ values may make these interventions harder to implement \cite{browne2015modeling,browne2022differential,macdonald2021modelling}. In this way, a virus may tend to evolve larger replication rates in response to these control strategies, which may increase disease virulence.  With this knowledge, public health authorities may prioritize rapid response for active and passive case finding in order to combat evolution of and existent virulent strains.

Furthermore, Fig.\ref{epidemic_phases} suggests that the change in viral replication rate can impact the infection transmissibility of individuals differently depending on epidemic phase. In particular, in the inset figure in Fig.\ref{epidemic_phases}(b), we observe that when within-host parasite growth rate is increased by $1\%,$  infection transmissibility increases up to $4\%$  at epidemic time $t=40$ (initial epidemic phase) and at $t=100$ (epidemic die out phase), but only up to $2.5\%$ during epidemic peak phase ($t=80$), with the maximum increase occurring around 1 day post infection and almost 100\% decrease in transmissibility during the end of infection because of the more acute infection. This normalized sensitivity and the raw sensitivities show that varying $r$ may dramatically shift infectability to earlier infection ages at the later declining phase of epidemic, making the implications of virulence and control strategies discussed in previous paragraph even more relevant.  
%  A $1\%$ increase in the IgM immune activation rate $a$ decreases the infection transmissibility up to $8\%,$ suggesting that drug treatment or vaccination can help curbing the new incidences significantly. Compare to viral growth parameter $r,$ an increase in the immune parameters such as immune activation rate $a$ and killing efficacy $\theta$ lead the same relative impact on the infection transmissibility at host infection age $\tau$ at all phases of epidemic (see inset figures in the Fig.\ref{epidemic_phases_other_param}(c)-(d) in Appendix \ref{Inf_sensitivity}). 
 Note that these results might be due to estimated parameters values for RFVD. 
% However, even though not a vector-host disease, we may extrapolate implications to the COVID-19 pandemic.  In the presence of unprecedented control efforts which included intensive case finding, COVID-19 evolved more virulent variants, such as the Delta variant, which rapidly spread in some countries that were in declining phases of epidemic waves. This evolution of increased viral replication rate contrasts some previous theory that viruses might evolve to be less virulent, and suggest that closer consideration of multi-scale infection-age models may be important to understand virulence-transmission trade-offs and evolution.
 More rigorous assessment of immune parameters on the infection transmissibility requires further developed techniques in fitting data from within-host and between-host scales to immuno-epidemiological models.

\section{Discussion}
 There are some good reasons for the use of SA in immuno-epidemiological models. First, infectious diseases operates at both immunological (individual) and epidemiological (population) scales. Therefore, quantifying how the underlying immunological processes affect the dynamics of the epidemic can be naturally approached by SA of multi-scale systems. Secondly, SA of epidemic models is frequently used to assess the impact of control strategies that impact the within-host (or in-vector) virus-and-immune-response dynamics, such as vaccination, drug treatment or bio-control strategies (e.g. Wolbachia-based).  A rigorous approach of SA requires the assessment of the sensitivity of epidemic quantities to immunological parameters. Finally, since obtaining population scale data is a challenging task and requires costly surveillance efforts, the  parameters estimated from within-host viral and immune-response data that can obtained less costly from laboratory experiments, can be helpful for control strategy development and its assessment. Therefore, to investigate how within-host immune parameters affect disease spread among host population, the SA of the multi-scale models, as the one introduced here, can be informative and give crucial insights on the expected impact of control strategies and their efficacy.

 %Here we first introduced a structured hybrid disease model, where epidemic parameters are governed by the state variables of a coupled ODE system describing in-host virus-and-immune-response dynamics. We summarized the stability and persistence conditions for the disease, showing that the \textit{immune-response dependent epidemiological basic reproduction number}  $\R_0$ \eqref{R0} is both a threshold between extinction and persistence and a measure of fitness for the pathogen, making it a strict threshold for disease eradication. 
%Furthermore, we developed a numerical algorithm for the SA of three population-scale disease quantities ---$\R_0$, and the steady-state infected host and infected vector population densities $\mathcal I_H^*$ and $\mathcal I_v^*$ to immunological parameters. 
Our results suggest that the within-host per capita parasite growth rate $r$ has the largest impact on disease spread: \emph{$1\%$ increase in $r$ in the parameter region $r \in [0.1, 2]$ leads to up to $8\%$ increase in $\R_0,$ and up to $1 \%$ increase in  $\mathcal I_H^*,$ and $1.5 \%$ increase in $\mathcal I_V^*$, which are significant numbers concerning disease outcomes.} Among the rest of the within-host immune parameters none has a significant impact on the epidemic at the population scale. We may still note that, IgM immune response (mainly responsible for the initial rapid destruction of virus within the host) parameters have larger impact compared to those of IgG immune response (responsible for life-long immunity). When the disease is fatal, a $1\%$ increase  in $q$ (IgM$\rightarrow$ IgG switching rate) leads to an increase in $\mathcal R_0$ and when it is mild, it decreases the epidemic quantities. This suggests that when the disease is mild, it is more favorable for the host population to have more IgG antibodies that provide life-long immunity. On the other hand, when the disease is fatal, it emphasizes the benefit of a smaller switching rate. Interestingly, increases in the IgM or IgG virus killing efficacy parameters $\theta,$ and $ \sigma$ (respectively), both lead to increases in $\mathcal R_0$, albeit insignificant ones. 
 
In previous studies \cite{gulbudak2017vector,Gilchrist2002,tuncer2016structural}, the choice of linking epidemic parameters as functions of immune variables were different. Therefore to address how sensitivity indices, $\gamma^{.}_{\textbf{p}},$ change with respect to linking functions chosen,  we first non-dimensionalize all linking functions, and then fit the unknown parameters in them using data from prior studies. By doing so, we show that the different linking functions considered do not cause differences in the main outcomes of SA of immunological parameters on epidemic quantities.
An interesting numerical result is that even though the unknown parameters in linking functions are fitted at the baseline (estimated) parameters, the consistency on the overlapping parameter values continues over a range of parameter values given as nontrivial intervals around those baseline values.

\noindent SA is a common methodological approach to determine  the impact of model parameters on the within- and between-host population dynamics quantities.  However, these two scales are typically treated separately. Since infectious diseases operate at both scales, in order to understand the crucial mechanisms behind disease dynamics and the impact of targeted control strategies, we need the formulation and analysis of unified models describing disease progression on both scales. One of the biggest challenges for using immuno-epidemiological models is the lack of multi-scale parameter estimations. The sensitivity analysis approach described here is local in nature. Generally, the sensitivity near one choice of parameter values maybe very different from that near a different choice of parameter values, so that a separate sensitivity analysis would need to be performed for every different choice of baseline parameters. However, for the system at hand, we based our parameter estimates on \cite{tuncer2016structural}, where it was shown that the model is globally structurally identifiable; that is, a unique set of parameters produces the best fit to the data. Therefore, the fitted parameter estimates we use are reliable. 
%For our future work, we plan to address the parameter estimates and their sensitivities for distinct arbovirus diseases.}

%\blue{Clearly, local sensitivity analysis is local by nature and, in general, the sensitivity for one set of parameters may be very different from that for other parameter choices. In another words, we cannot predict global sensitivities from local information. A similar analysis would need to be performed for different baseline parameters.
%However, for the system at hand, the identifiability of the immunological model was studied in [15]. It is shown there that the original model is not identifiable (A model is unidentifiable when more than one parameter combination fit the data).  The model was re-scaled to obtain a structurally (globally) identifiable model. That is, only a unique set of parameters produces the same output (data). Then, the parameter fitting was performed for the identifiable model. Thus, the parameter estimates we use in the present article are reliable.}
Finally, the model introduced here can be reformulated to incorporate seasonality and temperature-dependent vector growth cycle and vertical transmission in the mosquito population. We plan to extend the present model in the future by incorporating these mechanisms to investigate the impact of seasonality and temperature-dependent vector growth cycle along together with within-host immunological parameters on the disease dynamics at population scale. This might enable us to forecast the impact of implemented control strategies and inform changes to increase their efficacy to help reduce or eradicate the disease. Even though not a vector-host disease, we may also extrapolate implications to the COVID-19 pandemic.  In the presence of unprecedented control efforts which included intensive case finding, COVID-19 evolved more virulent variants, such as the Delta variant, which rapidly spread in some countries that were in declining phases of epidemic waves. This evolution of increased viral replication rate contrasts some previous theory that viruses might evolve to be less virulent, and suggest that closer consideration of multi-scale infection-age models may be important to understand virulence-transmission trade-offs and evolution.
\newpage

\section{Appendix}
\renewcommand{\theequation}{A.\arabic{equation}}
\setcounter{equation}{0}
\renewcommand{\thefigure}{A.\arabic{figure}}
\setcounter{figure}{0}
\renewcommand{\thesection}{A}
\setcounter{section}{0}
\setcounter{theorem}{0}
\setcounter{proposition}{0}

\begin{figure}[htbp]
\centering
\includegraphics[width=0.325\textwidth]{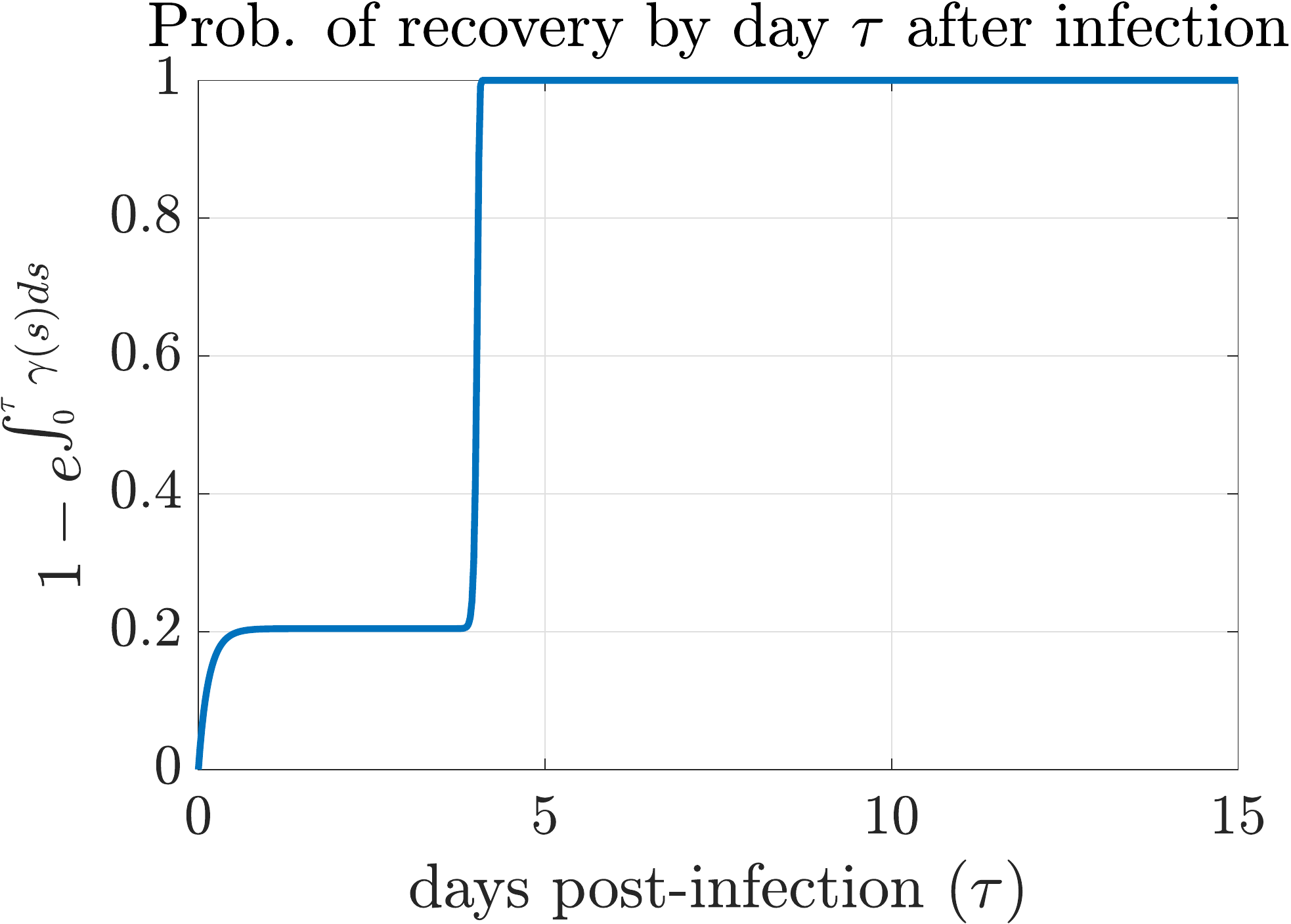}
\includegraphics[width=0.325\textwidth]{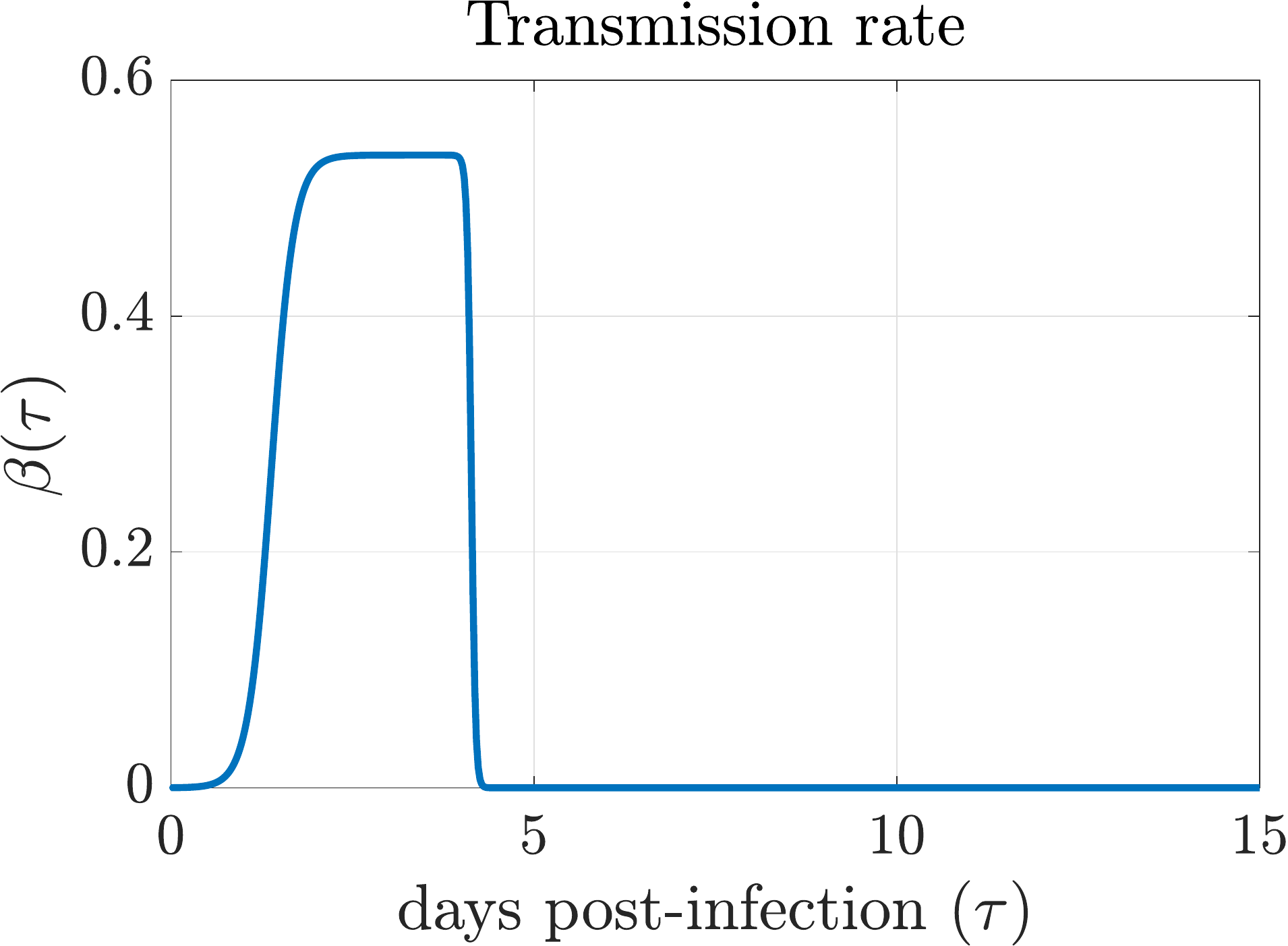}
\includegraphics[width=0.325\textwidth]{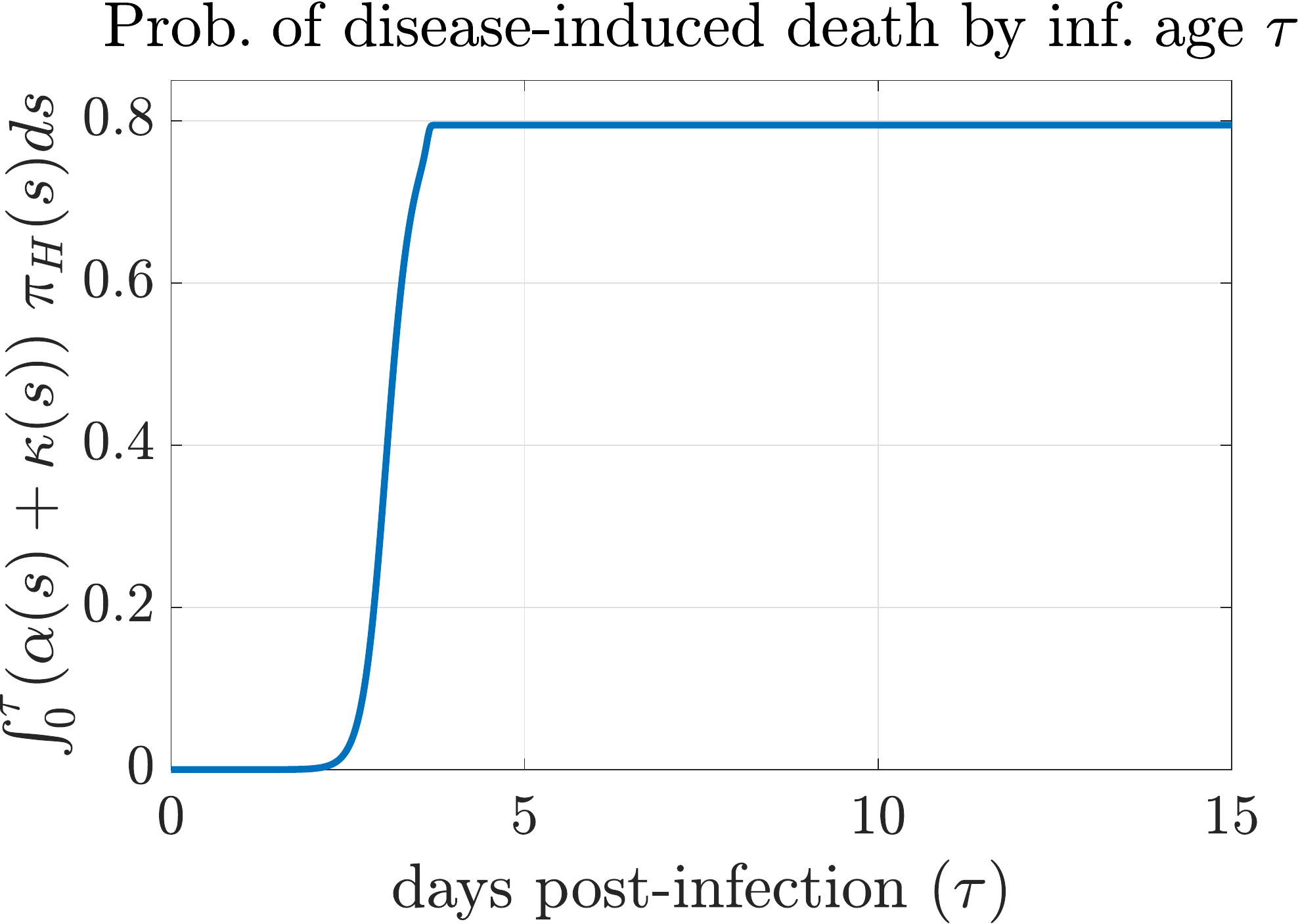}
\caption{\emph{The evolution of epidemiological parameters over the course of infection w.r.t. days post infection $\tau.$} 
a) Probability of the host being recovered at infection age $\tau,$ \emph{given that infectious hosts can exit the infected compartment only through recovery.} b) Transmission rate w.r.t. the infection age $\tau.$ %c) Disease induced death rate at time $\tau$ due to pathogen resource use and aggressive immune response. 
c) Probability of death occurring due to disease by infection age $\tau.$
Parameter values are given in Table \ref{table:parameters} and the corresponding within-host dynamics is displayed in the right subfigure of Fig.\ref{fig:within-hostdynamics}.}
\label{fig:linking_baseline}
\end{figure}

%%%%%%%%%%%%%%%%%%%%%%%%%%%%%%%%%%
\begin{figure}[htbp]
\centering
\includegraphics[width=0.24\textwidth]{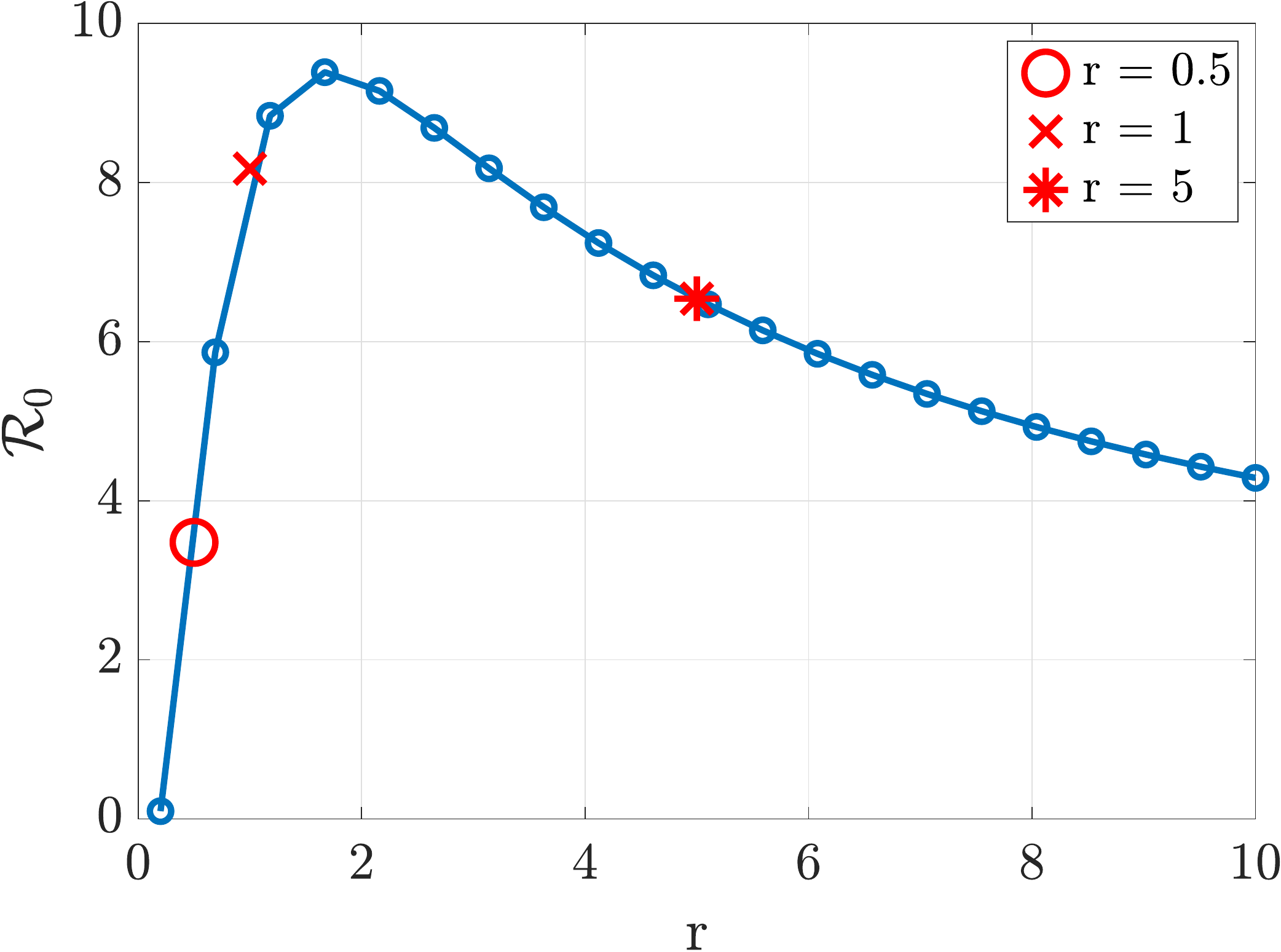}\hfill\includegraphics[width=0.24\textwidth]{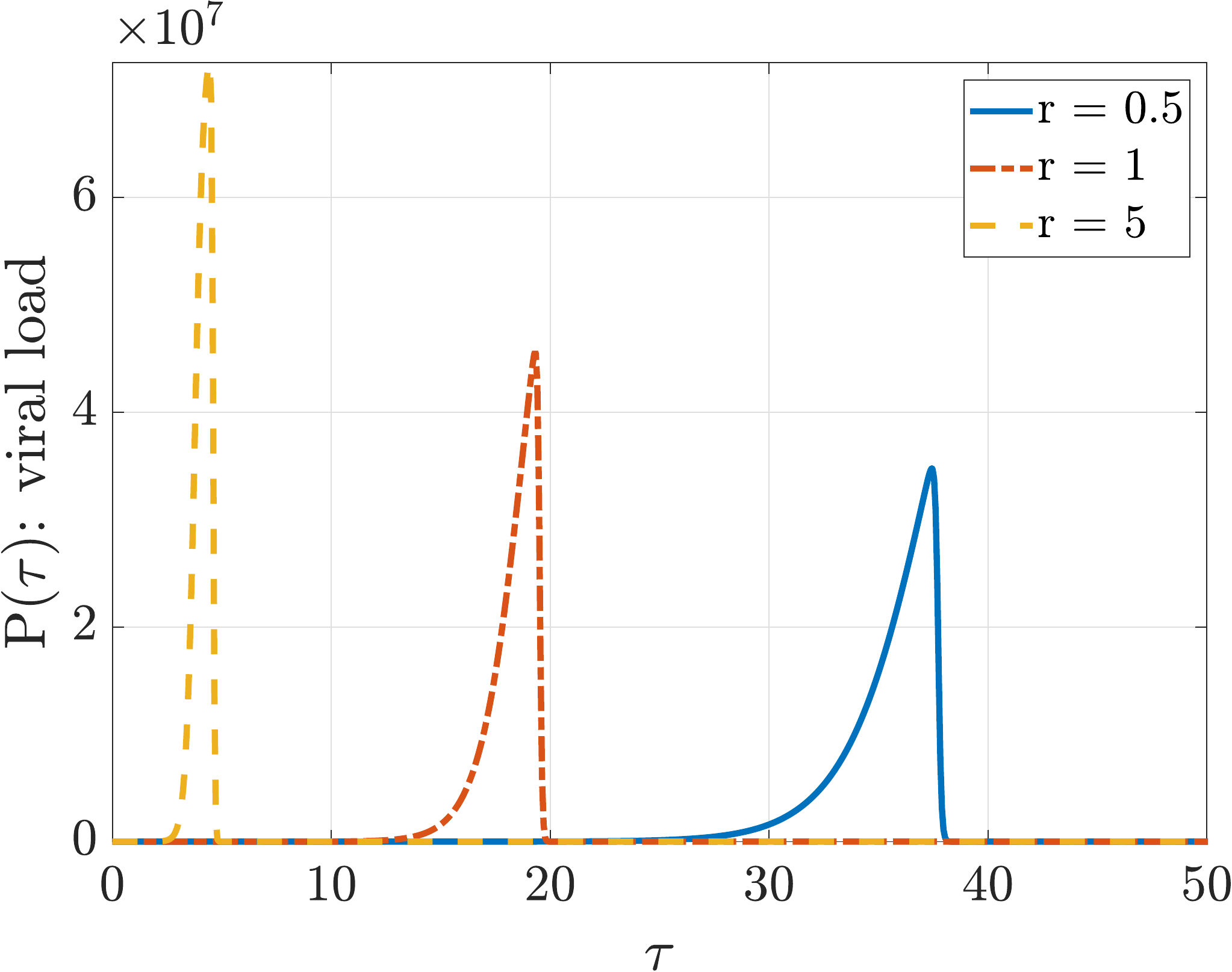}\hfill\includegraphics[width=0.24\textwidth]{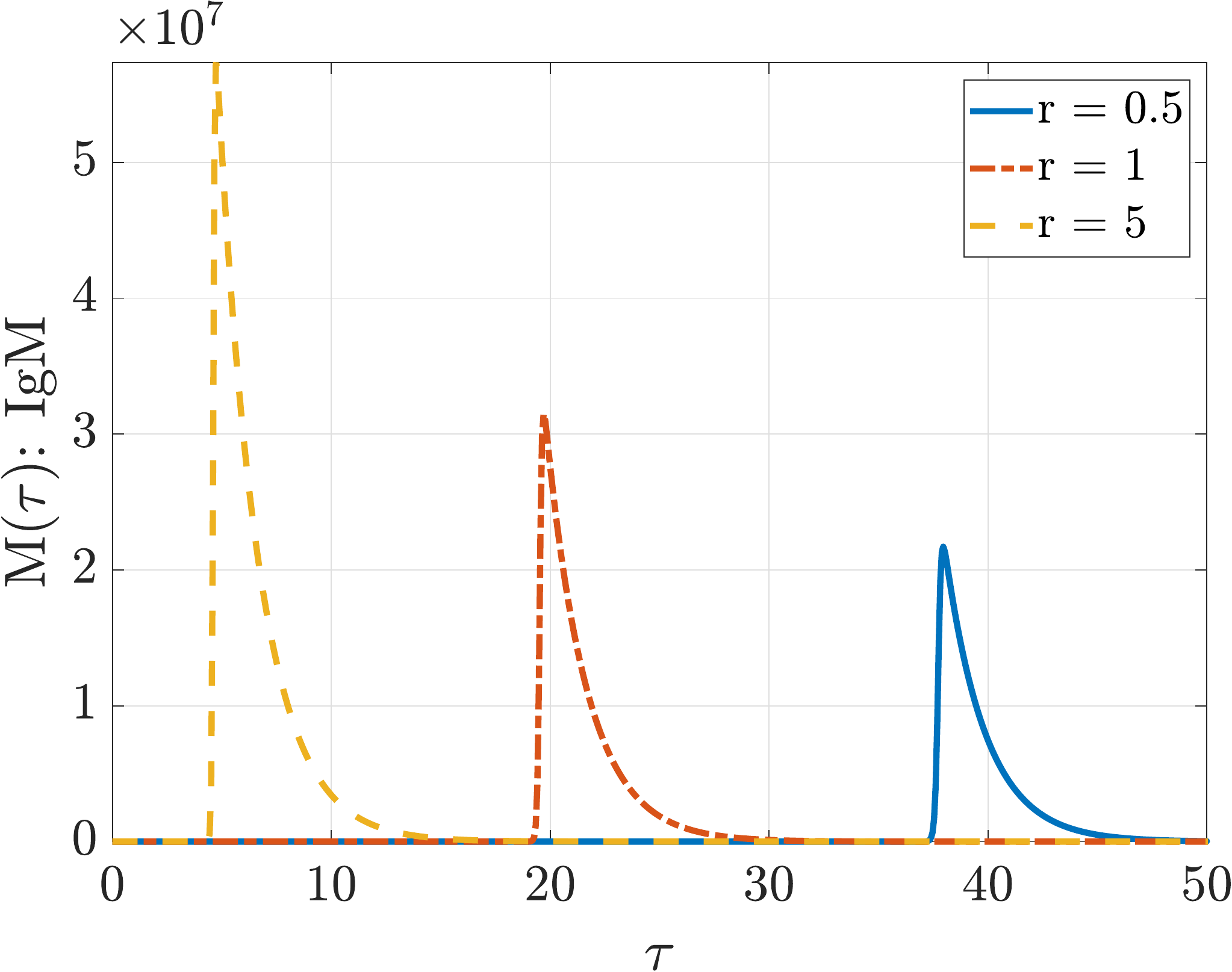}\hfill\includegraphics[width=0.24\textwidth]{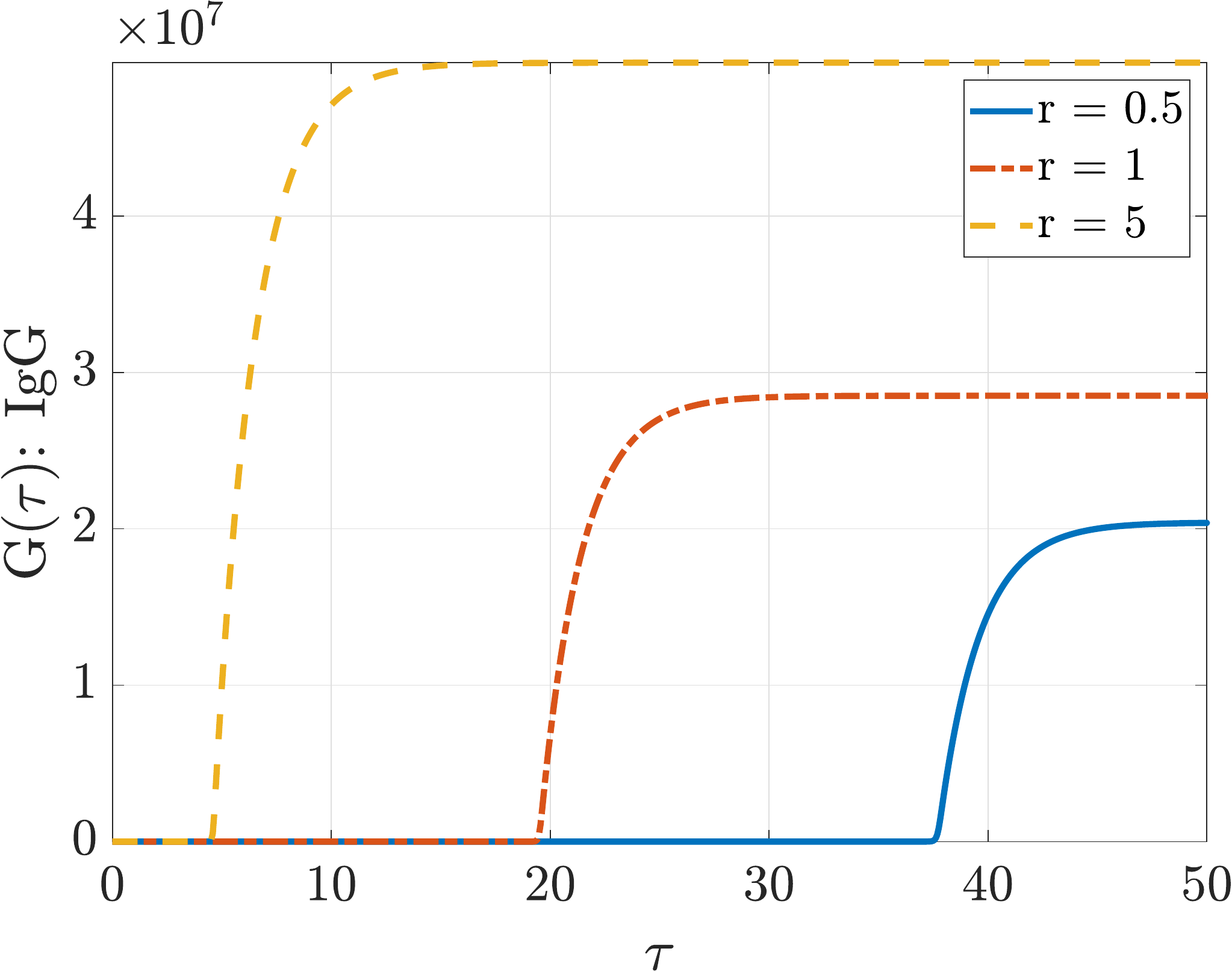}\\
\includegraphics[width=0.24\textwidth]{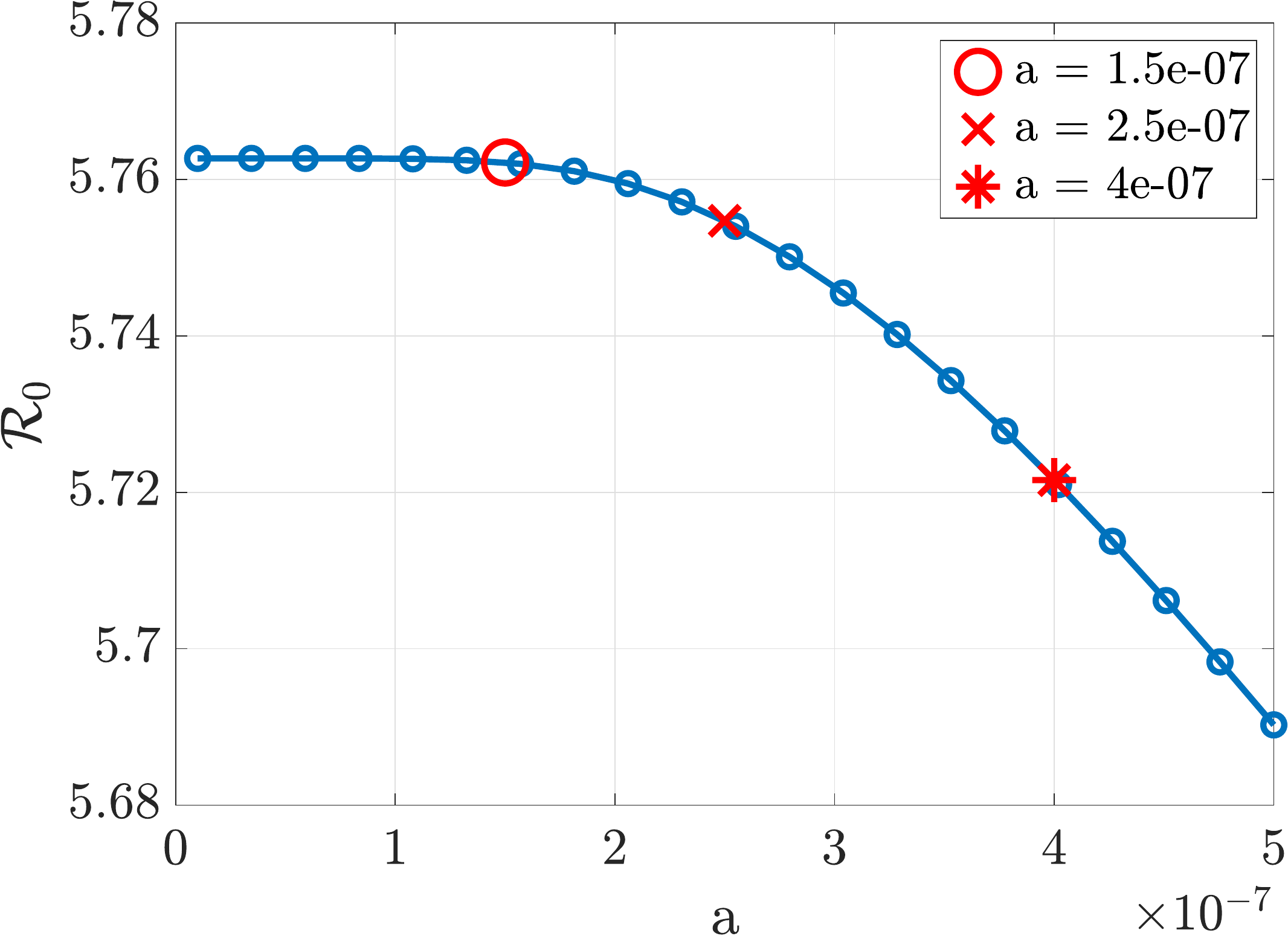}\hfill\includegraphics[width=0.24\textwidth]{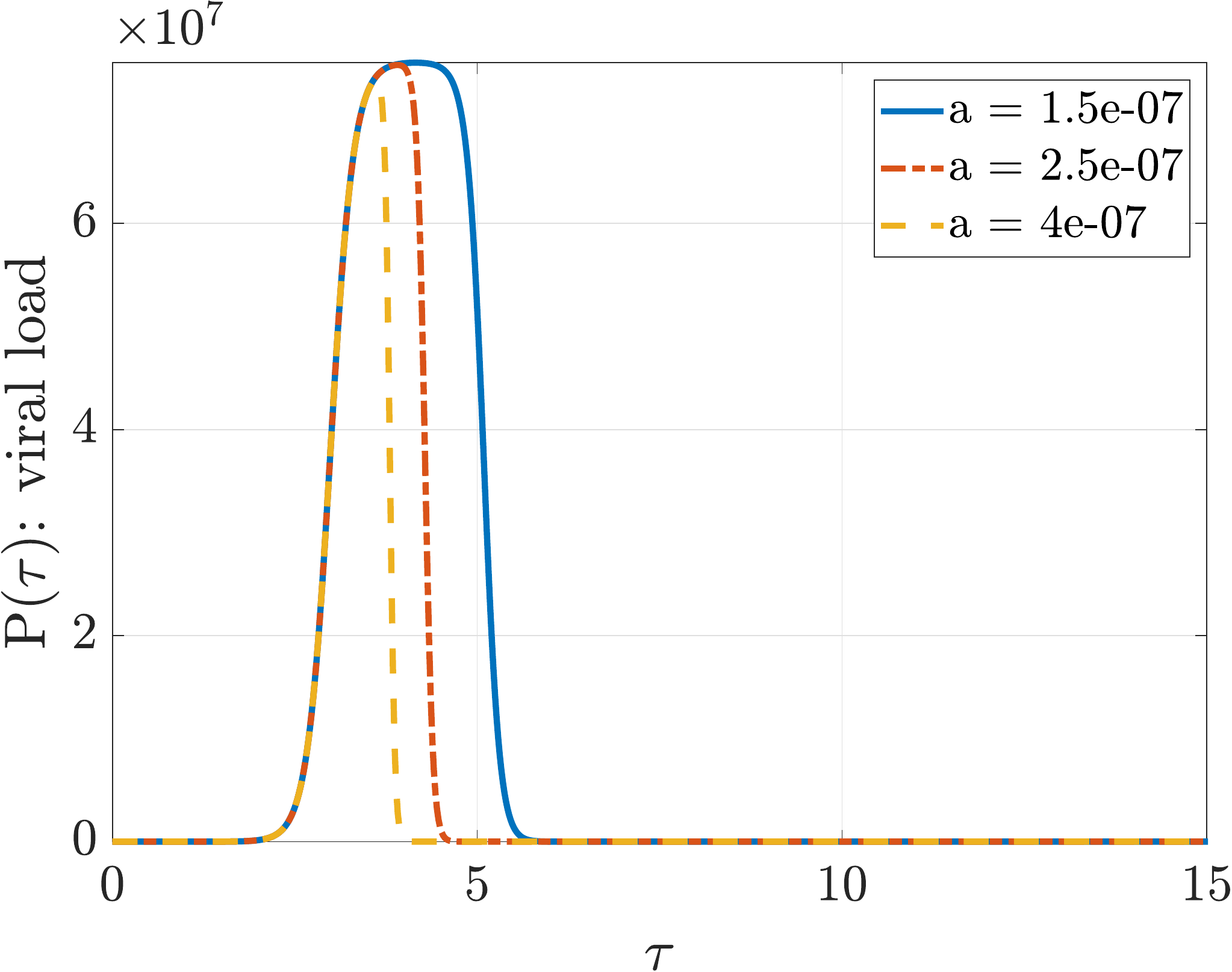}\hfill\includegraphics[width=0.24\textwidth]{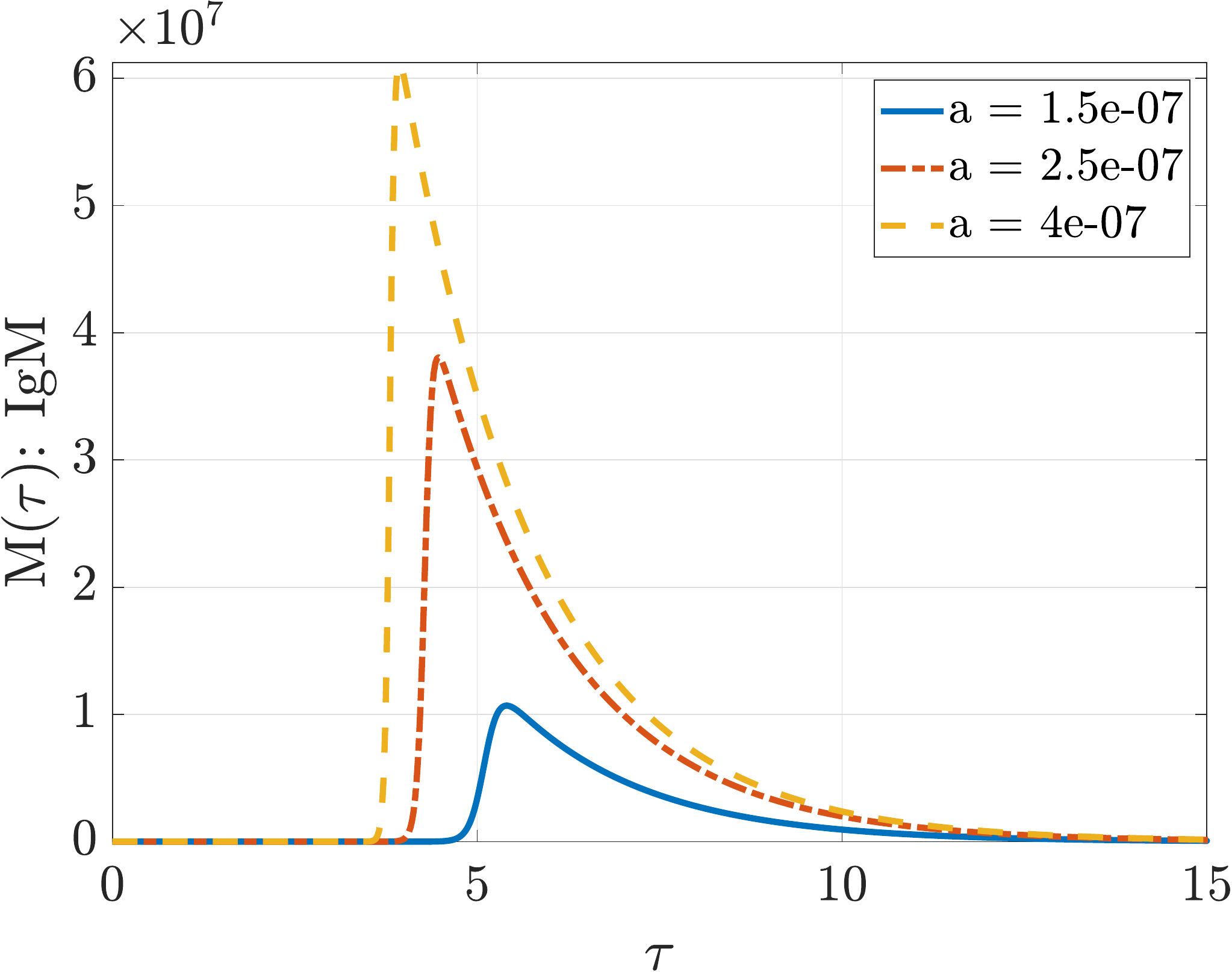}\hfill\includegraphics[width=0.24\textwidth]{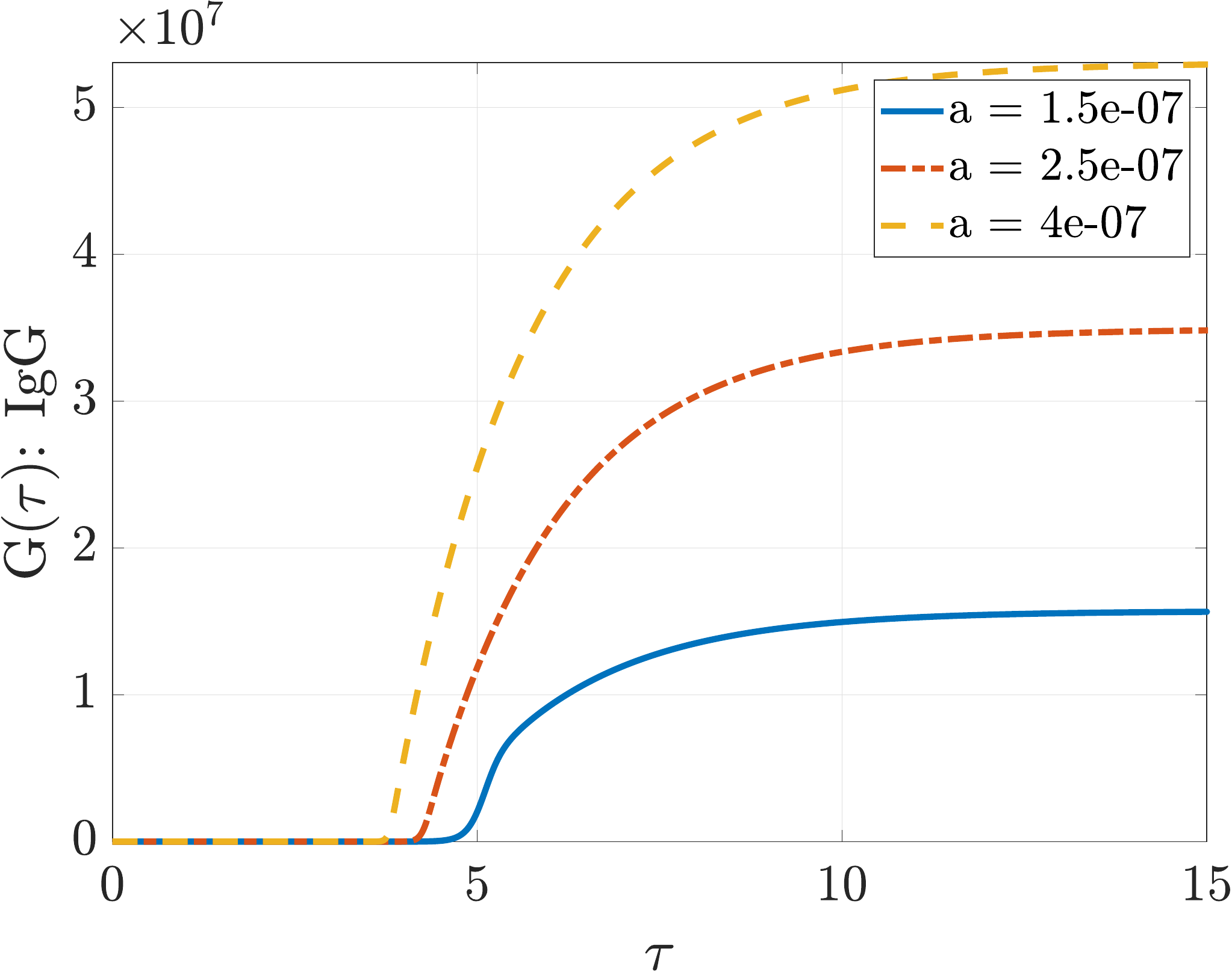}\\
\includegraphics[width=0.24\textwidth]{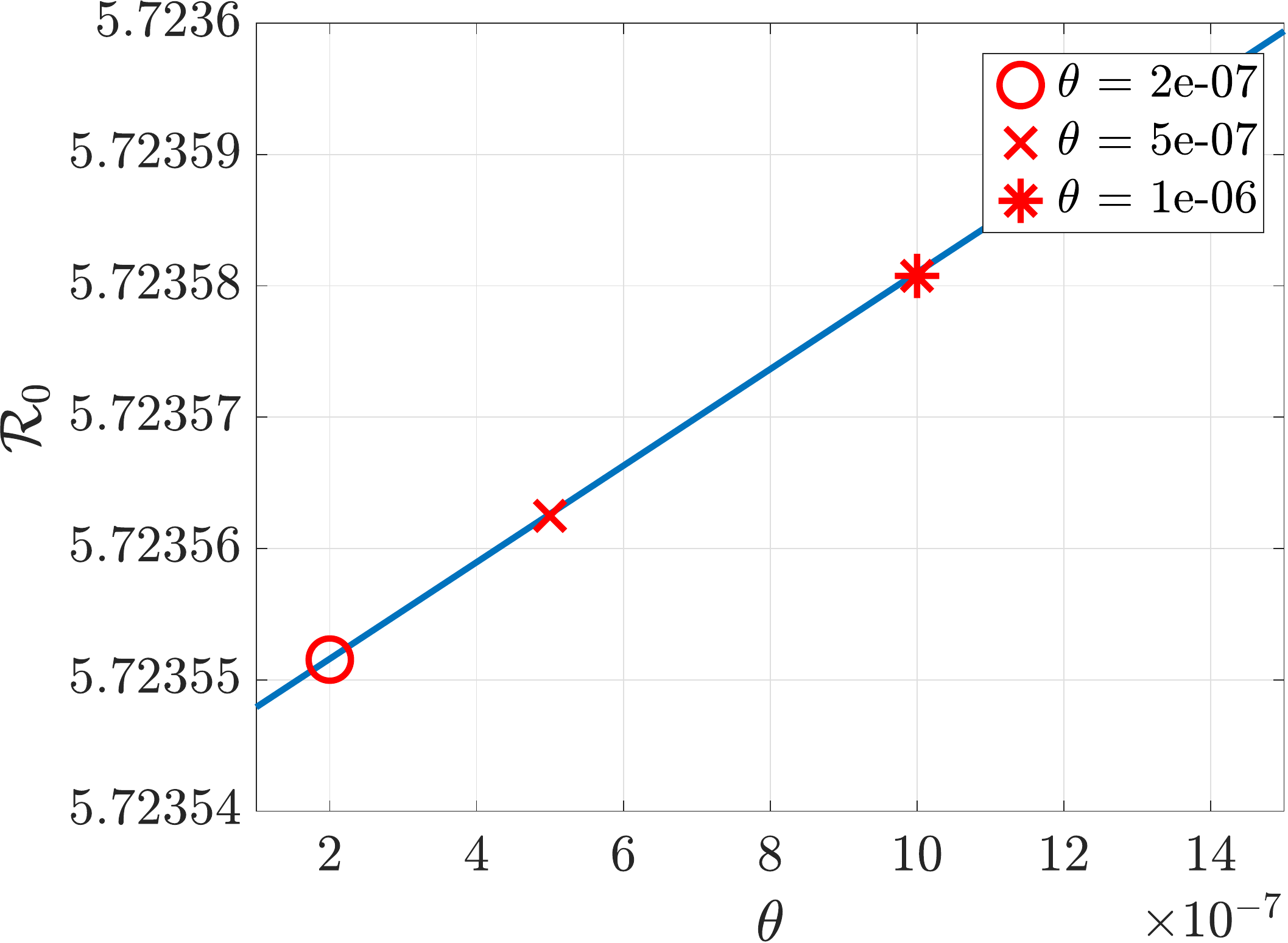}\hfill\includegraphics[width=0.24\textwidth]{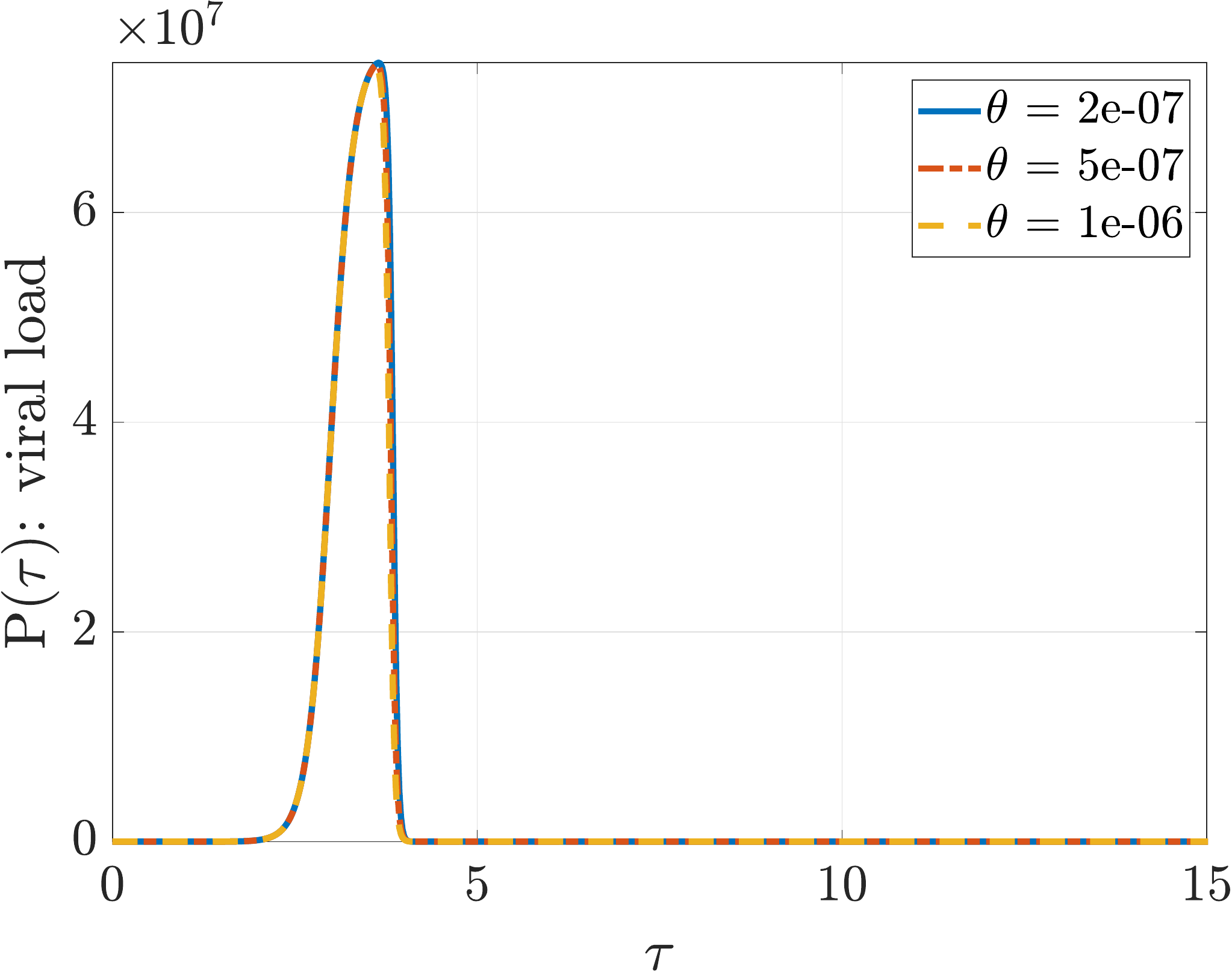}\hfill\includegraphics[width=0.24\textwidth]{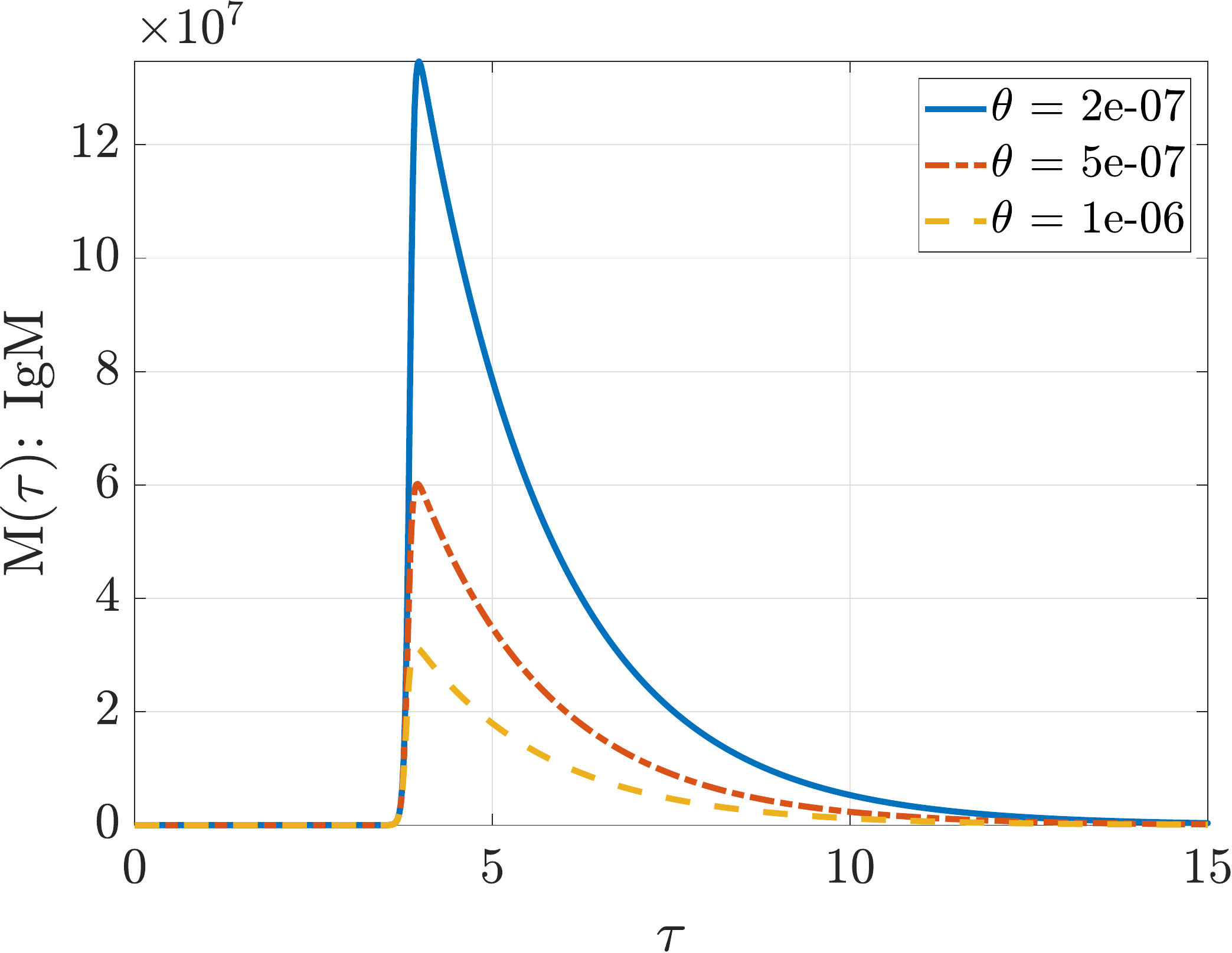}\hfill\includegraphics[width=0.24\textwidth]{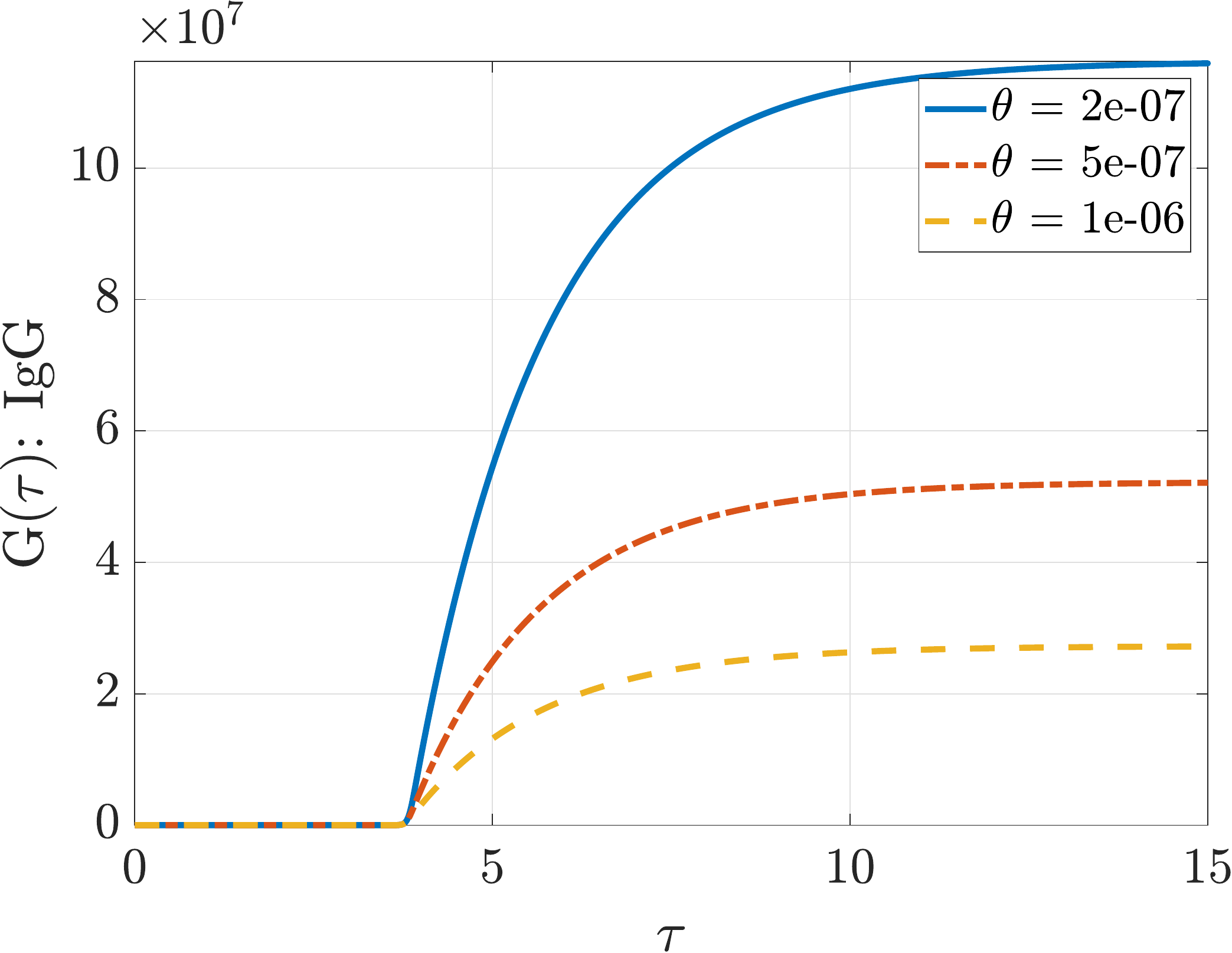}
\caption{\label{figure:WH_dyn}\textbf{Top row:} $\R_0$ vs. pathogen growth rate $r.$ As pathogen growth rate $r$ increases, the infectious period decreases with larger peak load and the immune response antibodies become more prevalent.\textbf{Middle row:} $\R_0$ vs. IgM activation rate $a$. The faster the IgM antibodies activate, the faster they clear the pathogen with smaller peak load and decreasing infectious period. \textbf{Bottom row:} $\R_0$ vs. killing efficiency of IgM $\theta$. As the immune response triggered by IgM antibodies become more efficient at killing pathogens, the pathogen load and subsequent immune response get smaller. }
\end{figure}
%%%%%%%%%%%%%%

\begin{figure}[htbp]
\centering
\includegraphics[width=0.5\textwidth]{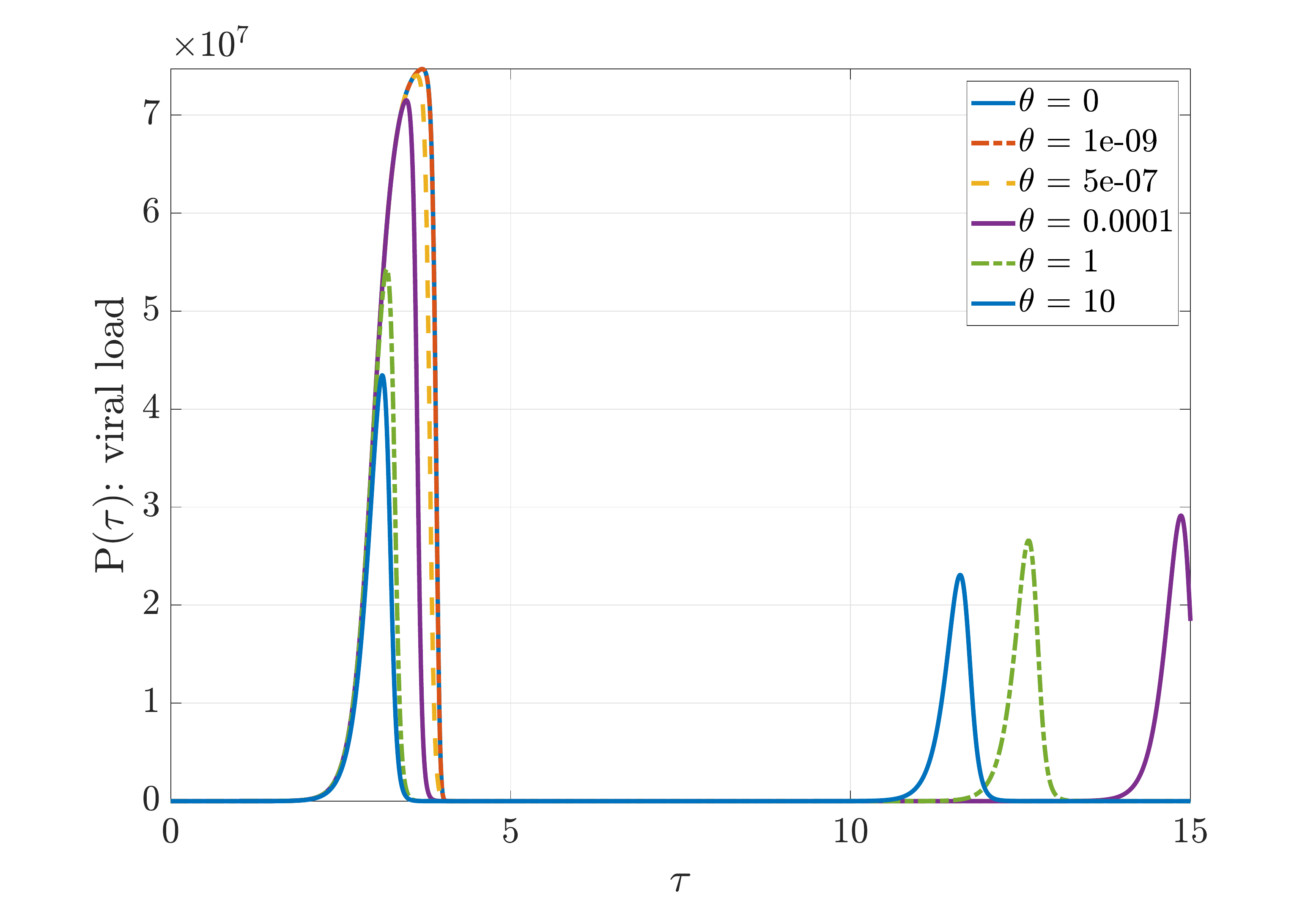}
\caption{Impact of killing efficacy $\theta$ on the viral dynamics. \label{fig:theta}}
\end{figure}

\subsection{SA curves for final infected vector abundance, $\mathcal I_V^*$ \label{sec:Appendix_SA}}
\begin{figure}[htbp]
\centering
\includegraphics[width=0.327\textwidth]{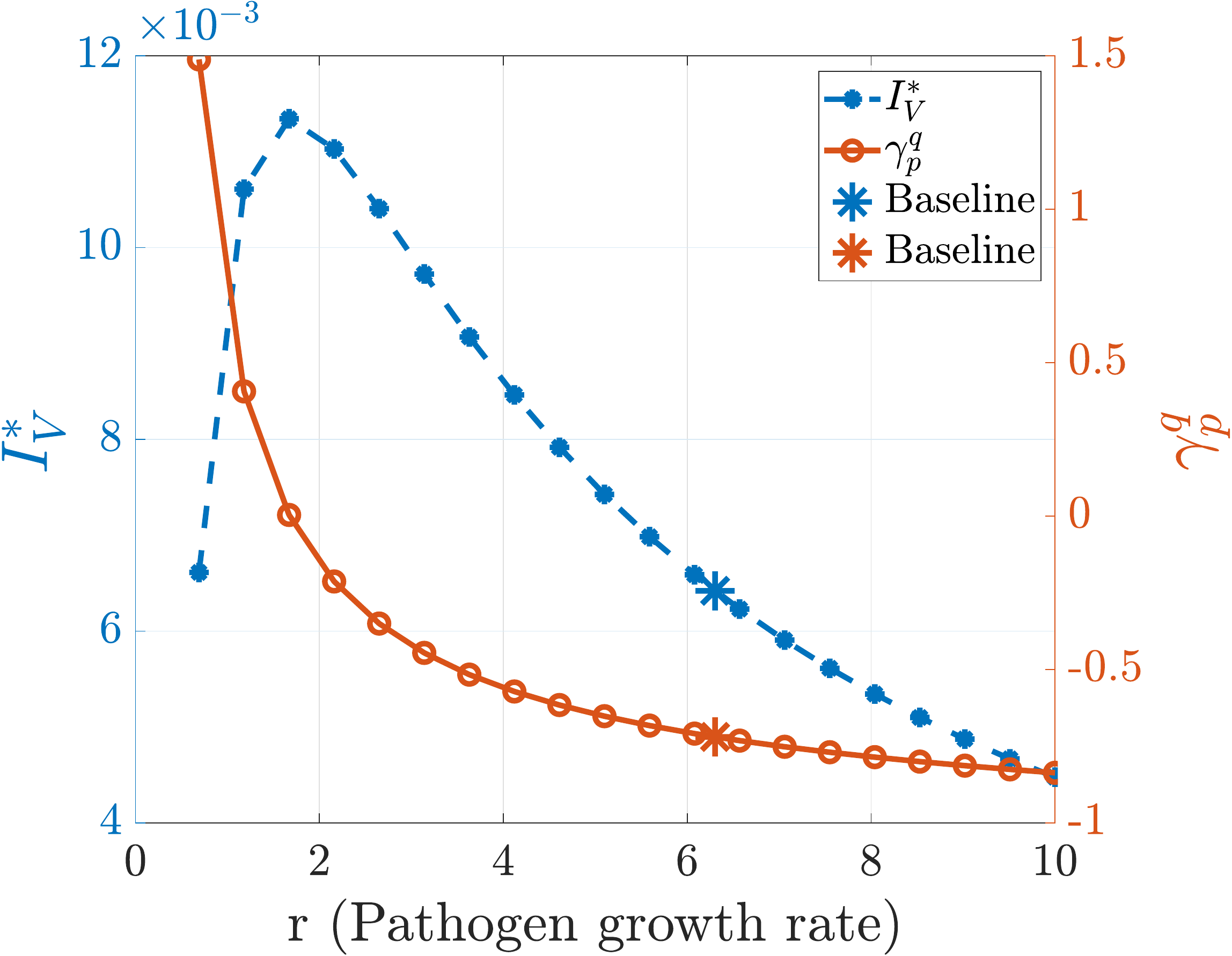} 
\includegraphics[width=0.327\textwidth]{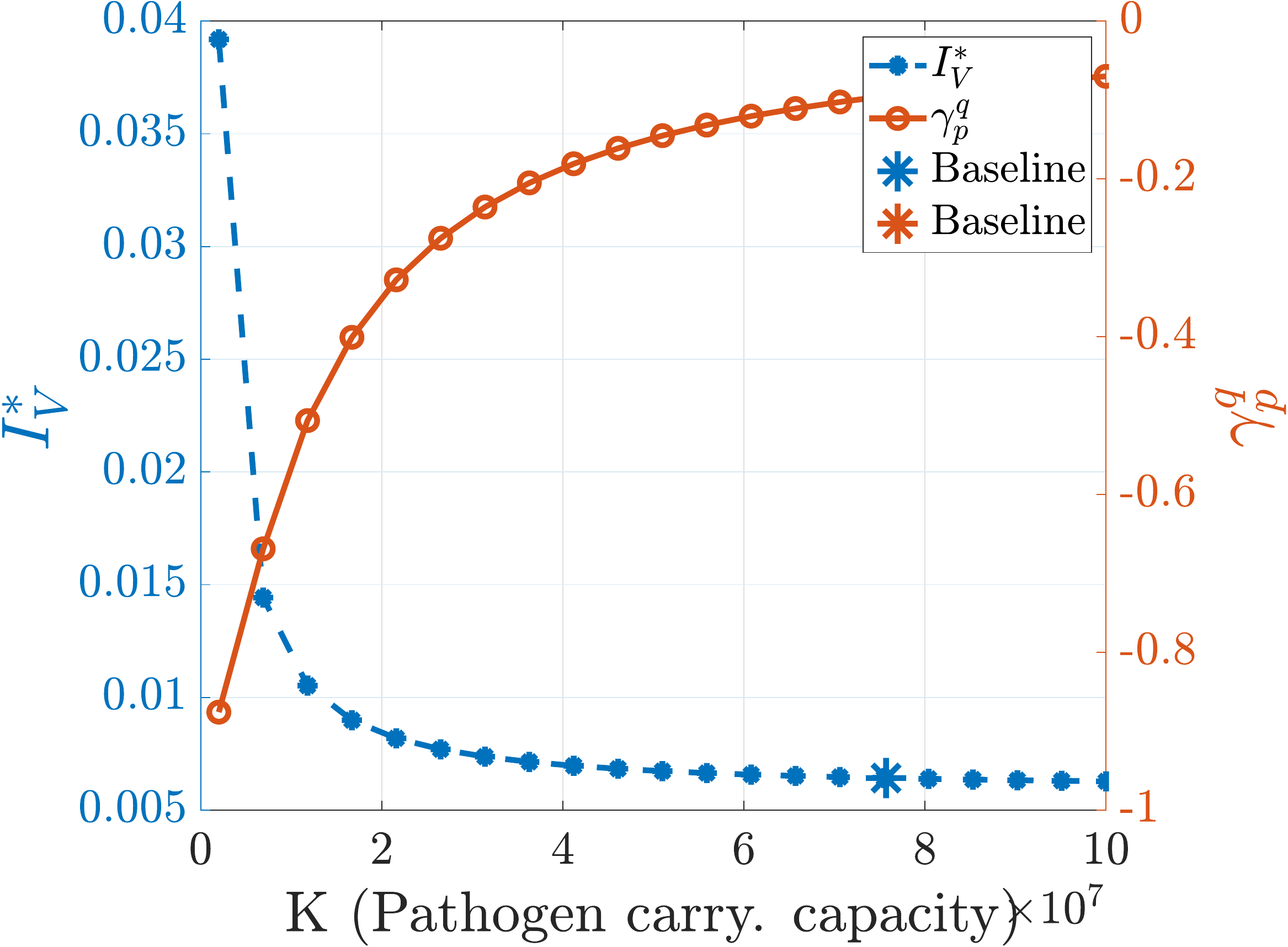}
\includegraphics[width=0.327\textwidth]{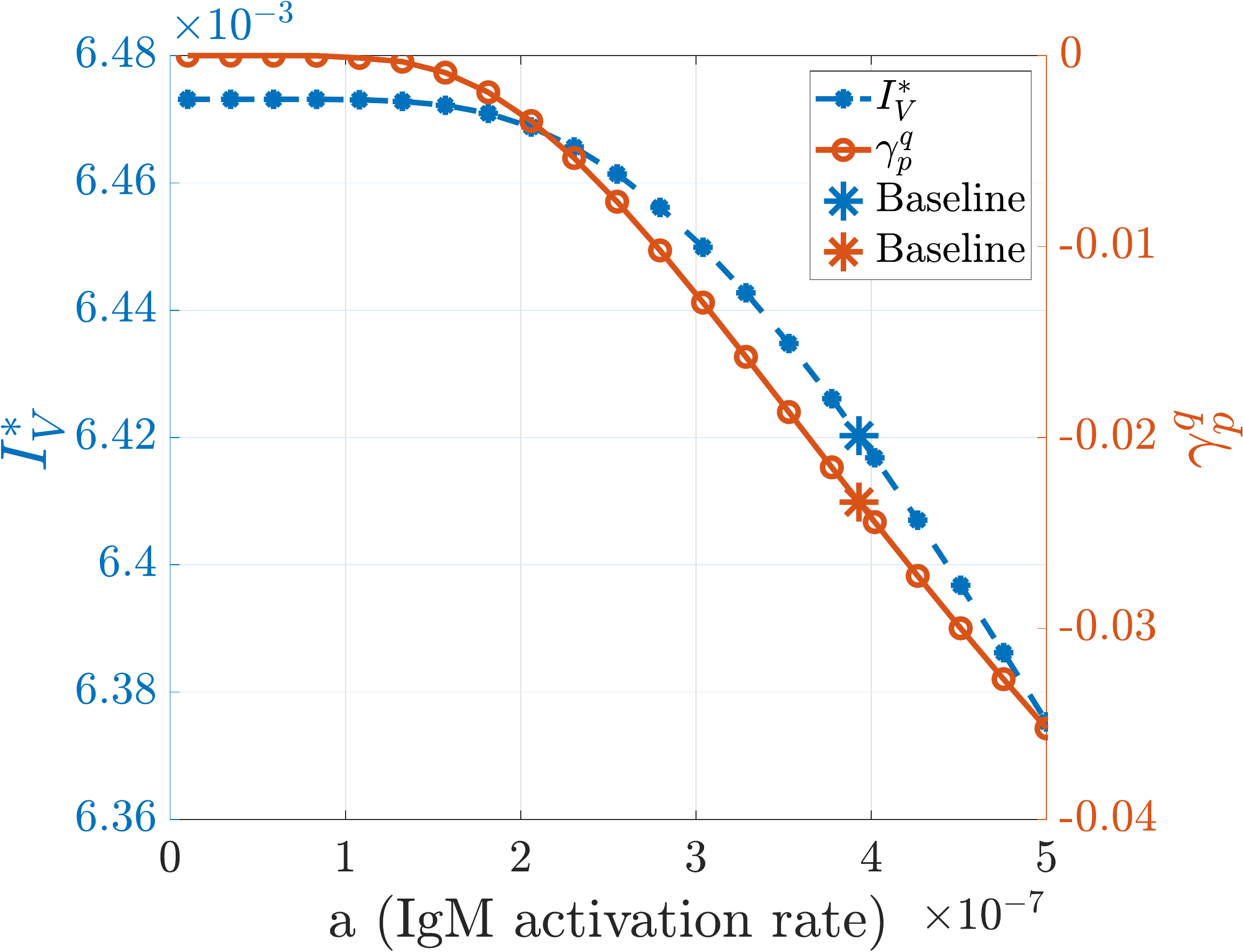} 
\includegraphics[width=0.327\textwidth]{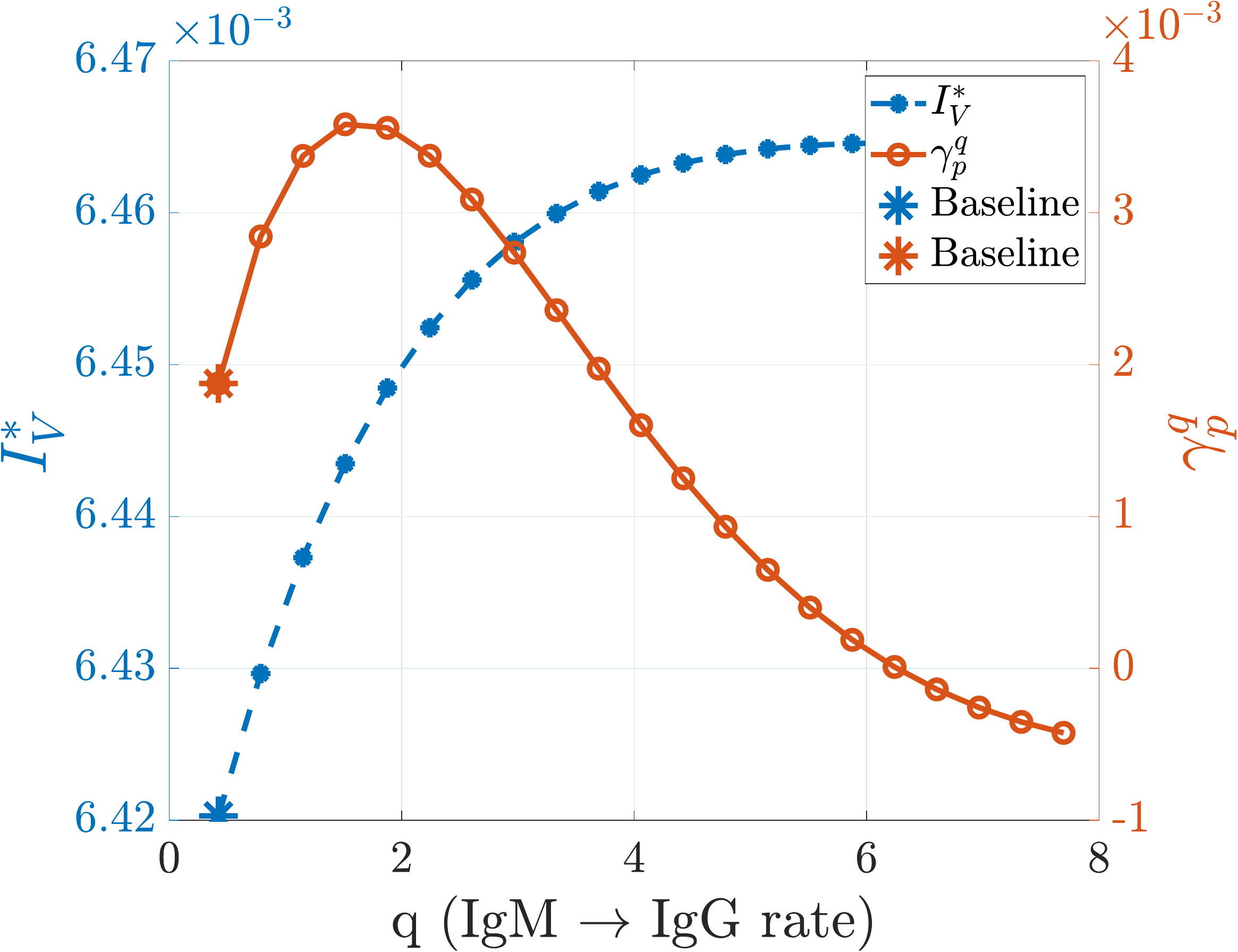}
\includegraphics[width=0.327\textwidth]{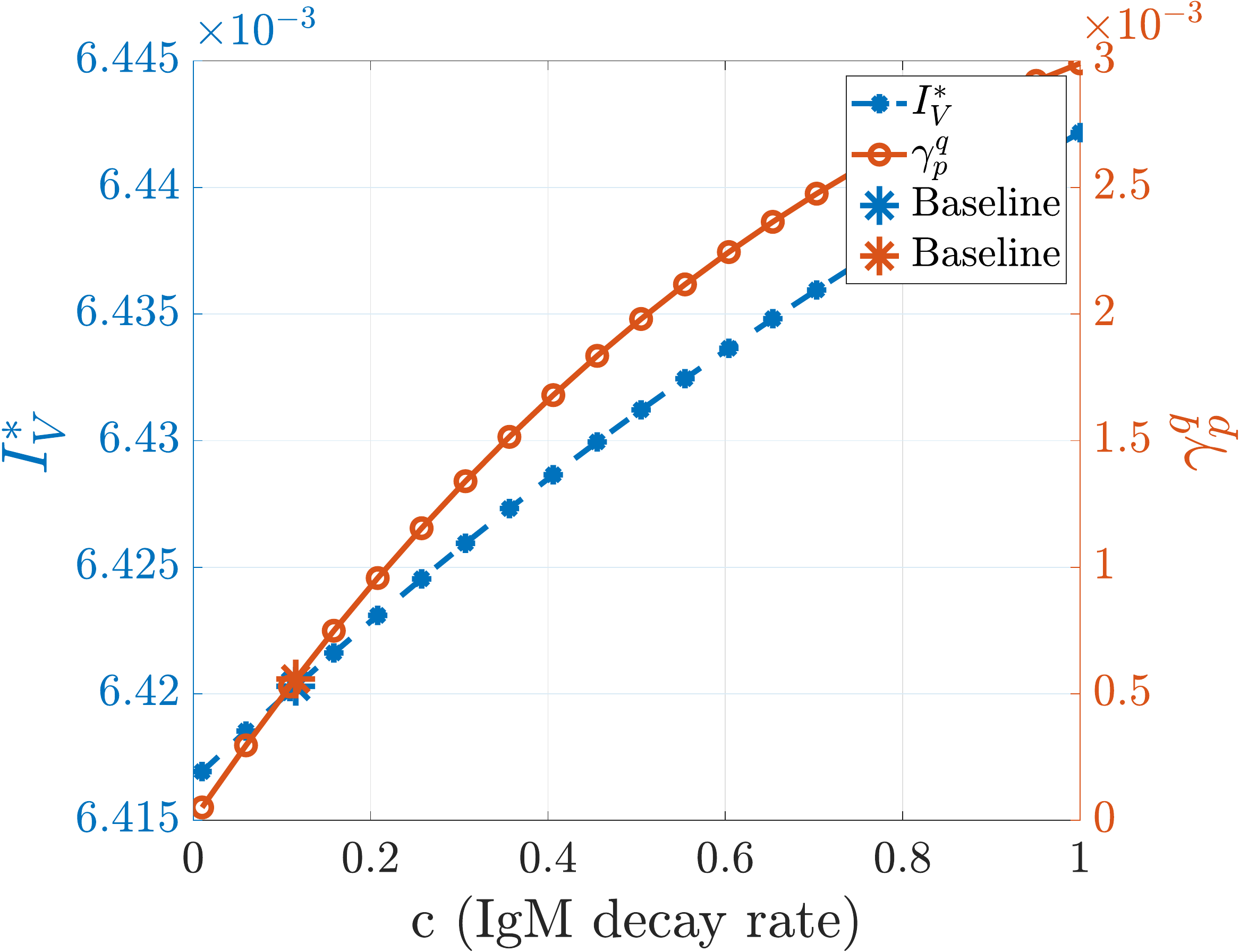}
\includegraphics[width=0.327\textwidth]{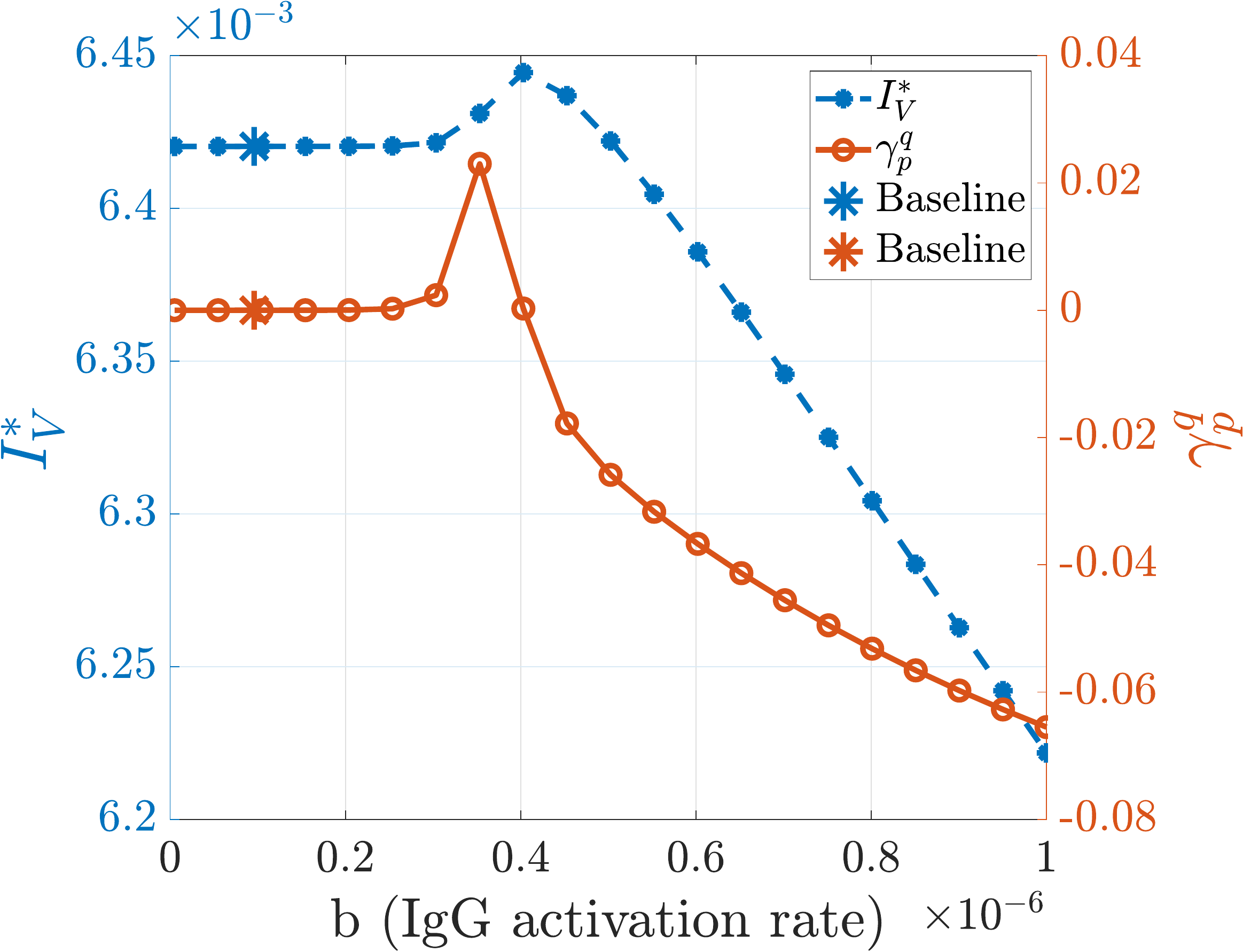}
\caption{Impact of immune parameters on the steady-state infected vector abundance, $\mathcal I_V^*.$ Notice that numerical results for SA for vector component are  similar to SA of $\mathcal R_0,$ and the steady-state host endemic size $\mathcal I^*_H.$}  
\label{fig:SA_Iv}
\end{figure}

See Fig. \ref{fig:SA_Iv}
%%%%%%%%%%%%%%%%%%%%%%%%%%%%%%%%%%
\subsection{Impact of using different linking functions on SA: $\mathcal I_H^*$ and $\mathcal I_V^*$ \label{sec:Appendix_SA_link}}
\begin{figure}[htbp]
\centering
\includegraphics[width=0.47\textwidth]{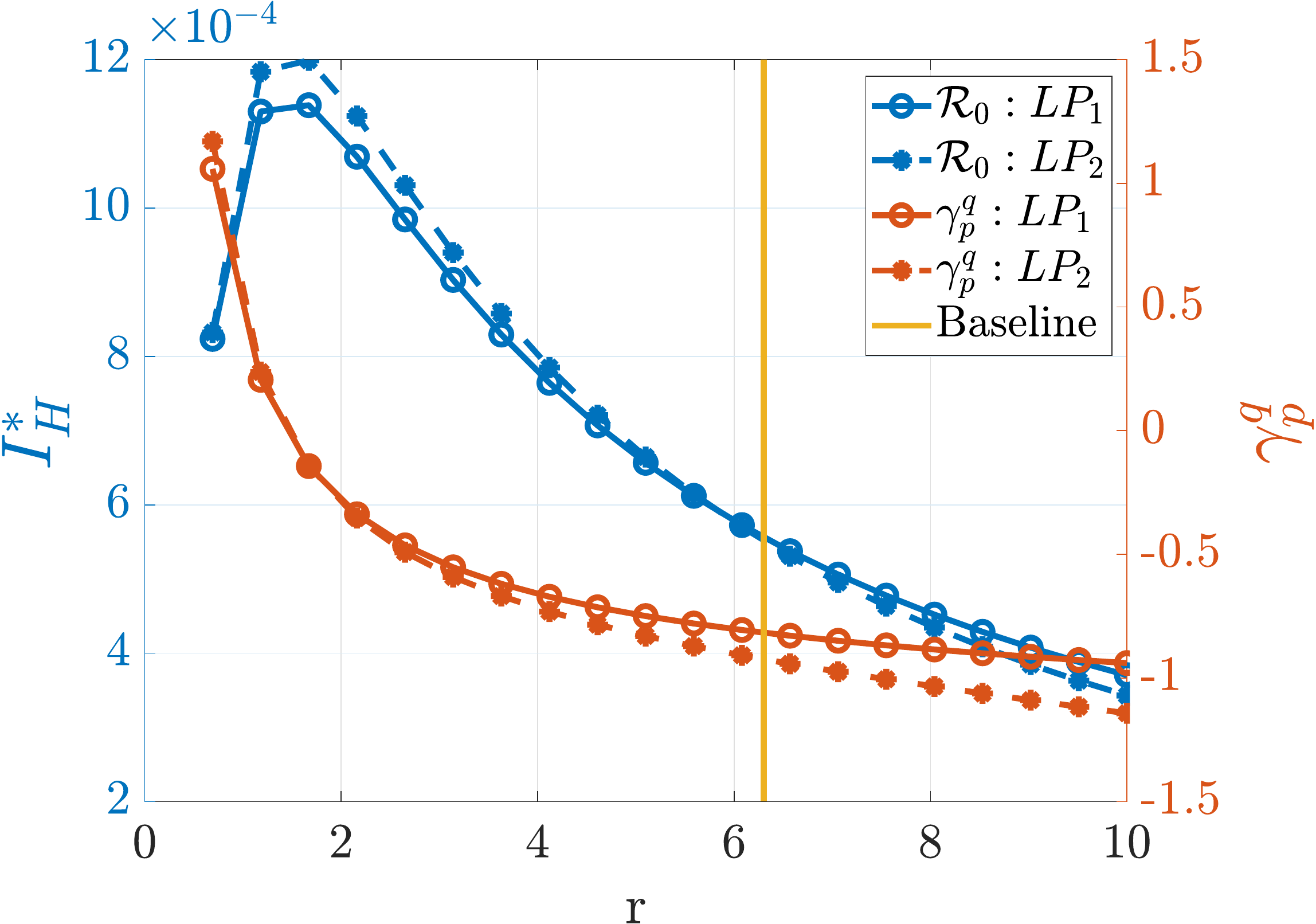}\hfill
\includegraphics[width=0.47\textwidth]{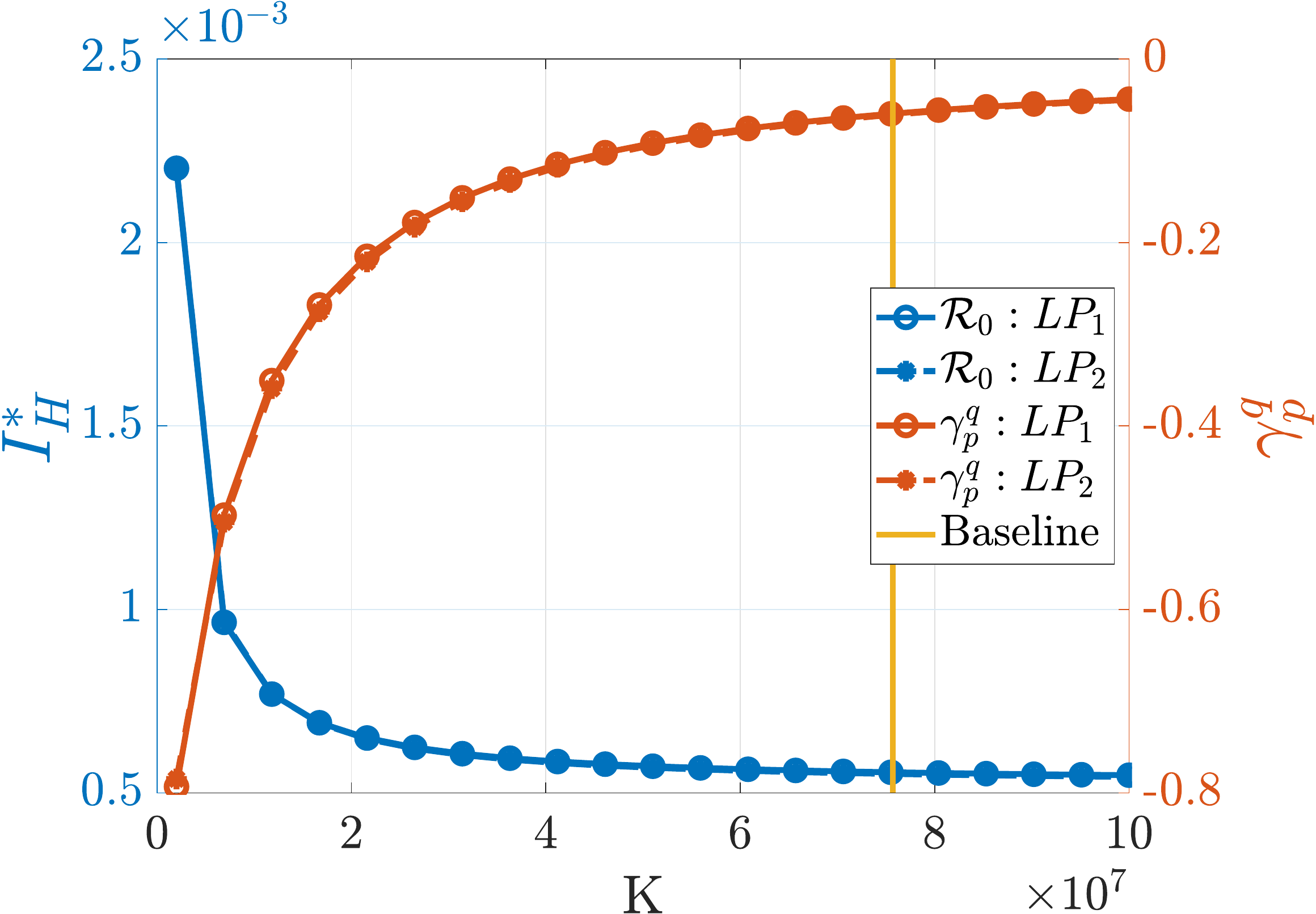}
\includegraphics[width=0.47\textwidth]{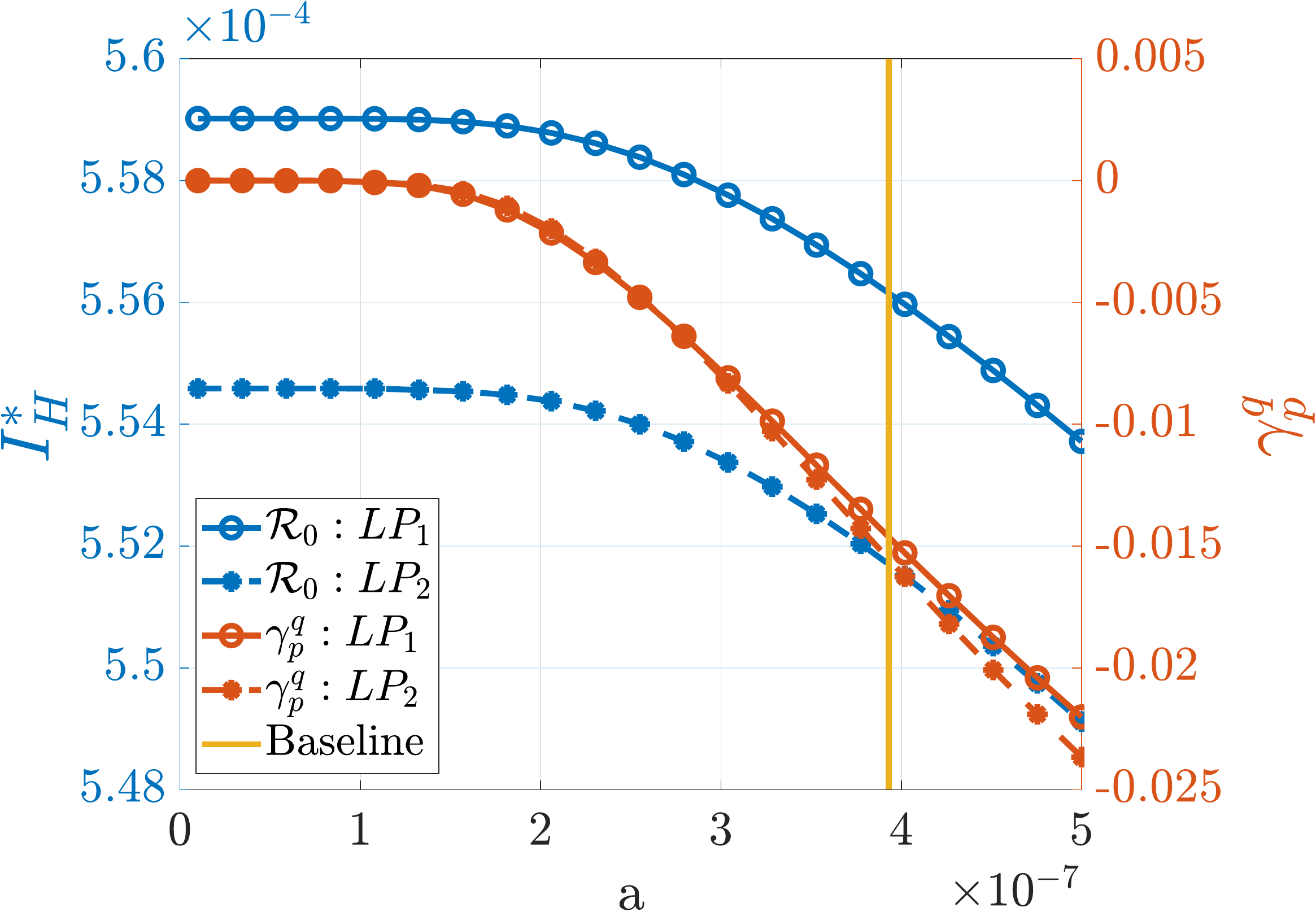} \hfill
\includegraphics[width=0.47\textwidth]{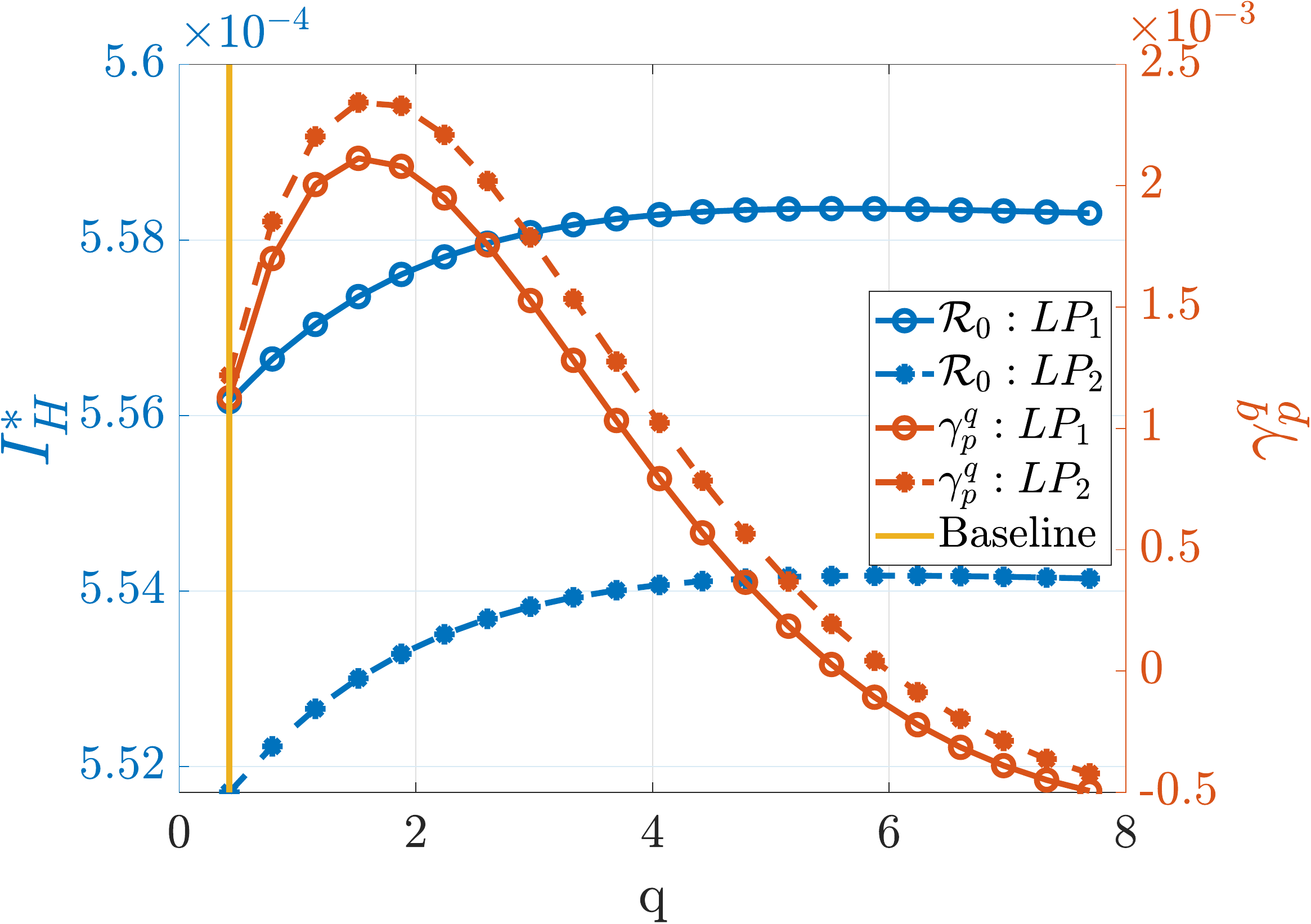}
\caption{Impact of the choice of linking functions on the SA. \label{fig:SA_link_Ih}}  
\end{figure}

\begin{figure}[htbp]
\centering
\includegraphics[width=0.47\textwidth]{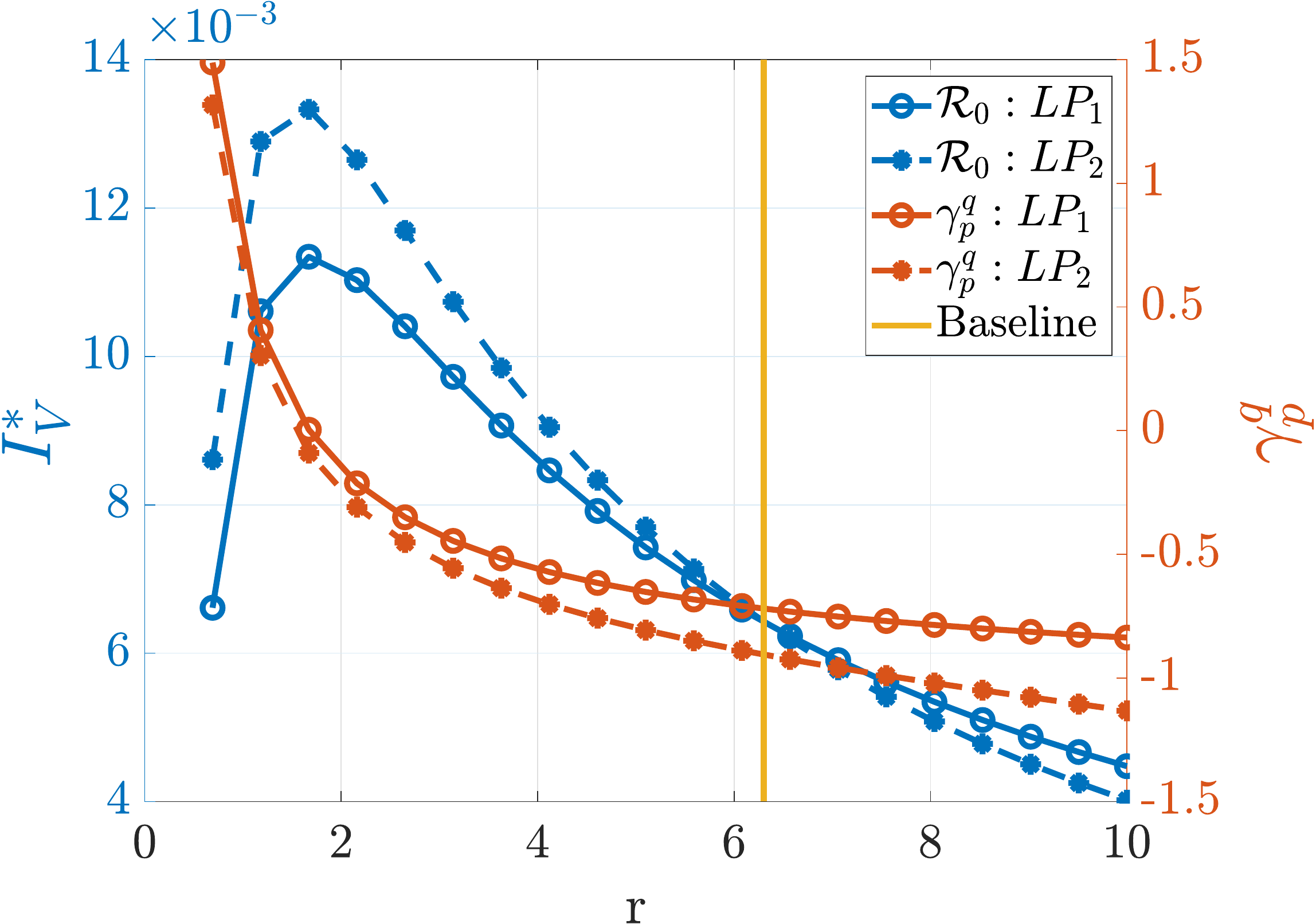}\hfill
\includegraphics[width=0.47\textwidth]{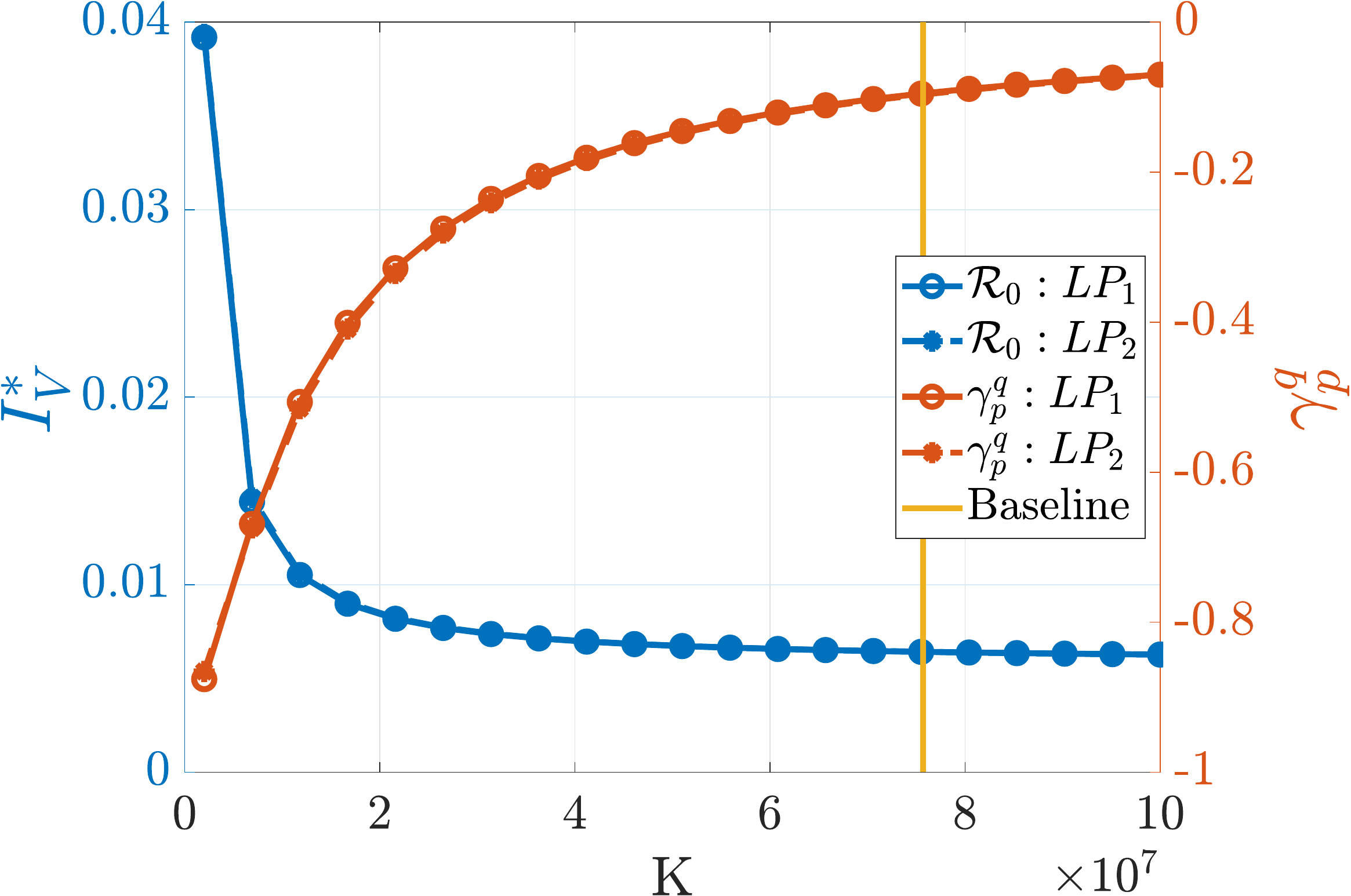}
\includegraphics[width=0.47\textwidth]{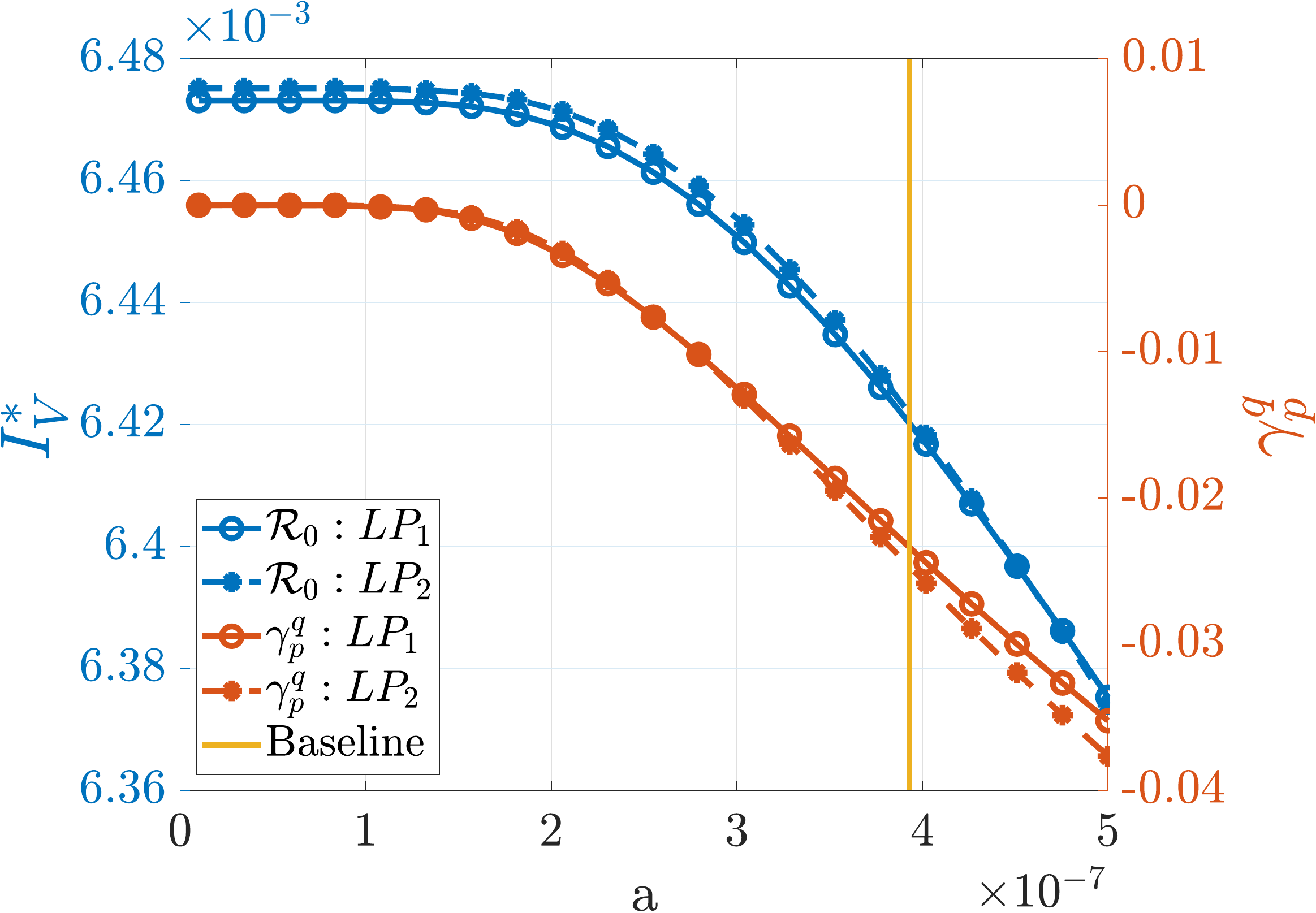} \hfill
\includegraphics[width=0.47\textwidth]{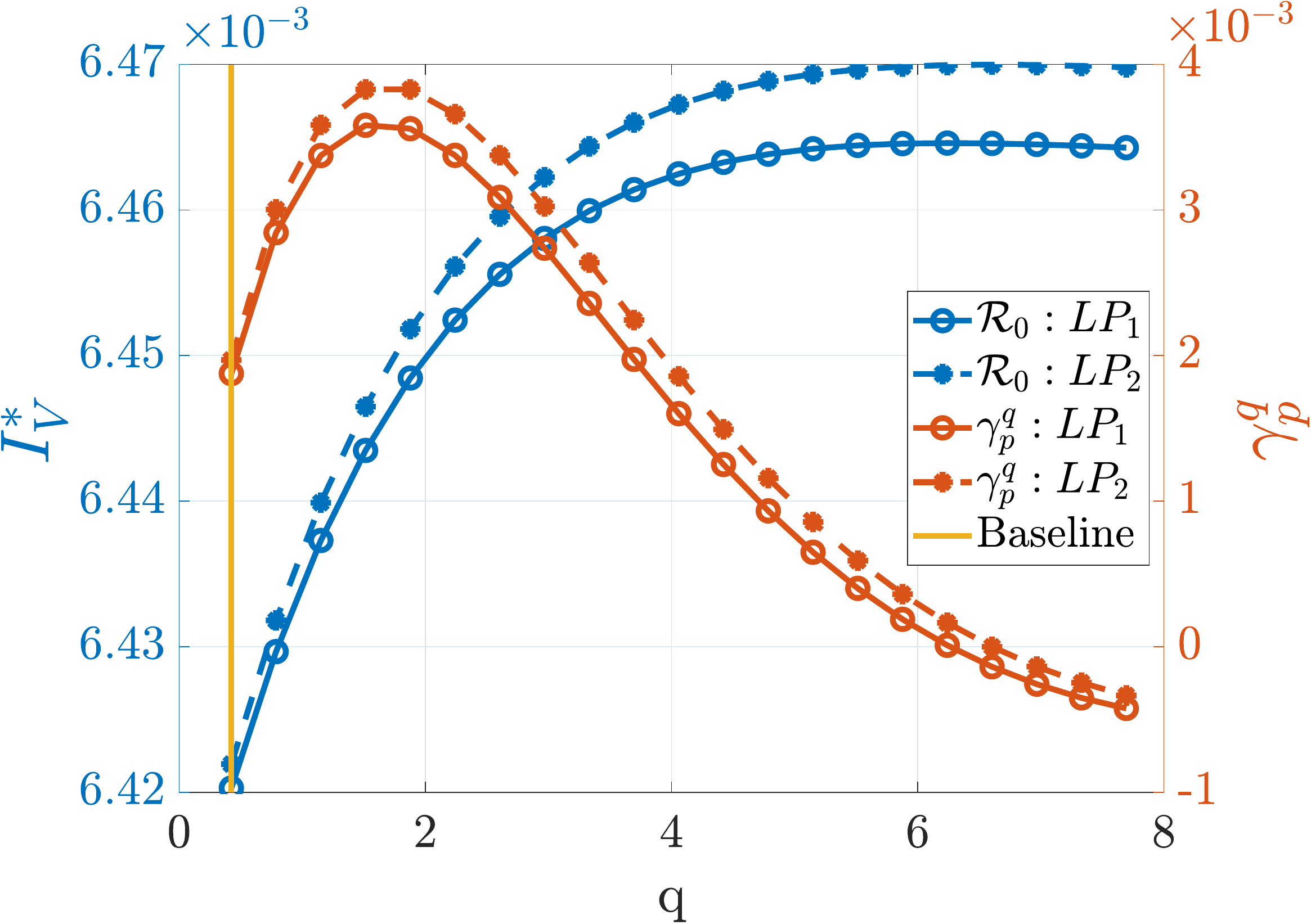}
\caption{Impact of the choice of linking functions on the SA. \label{fig:SA_link_Iv}}  
\end{figure}

See Fig. \ref{fig:SA_link_Ih} and Fig. \ref{fig:SA_link_Iv}.
\newpage

\begin{figure}[htbp]
\centering
\includegraphics[width=0.47\textwidth]{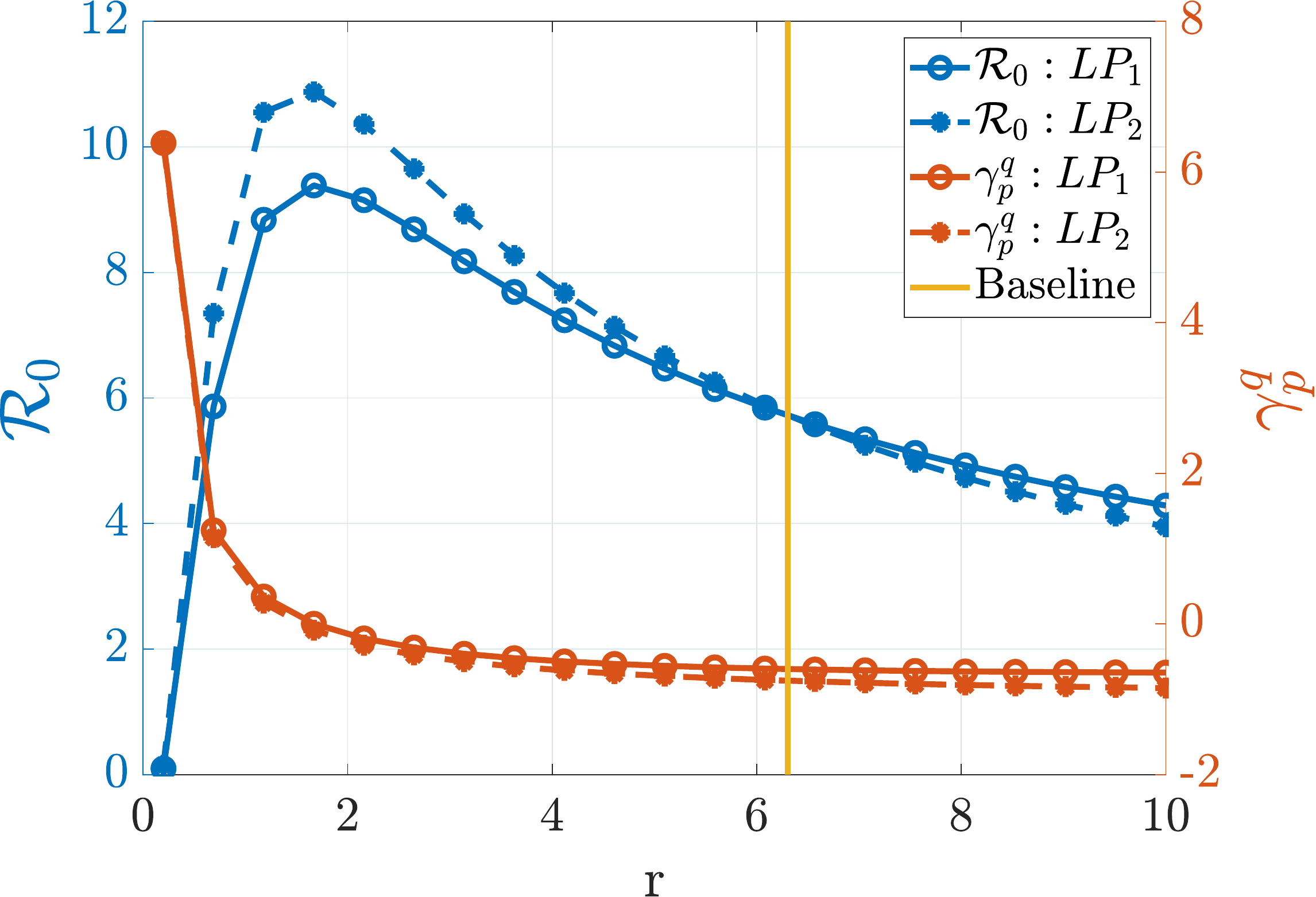}\hfill
\includegraphics[width=0.47\textwidth]{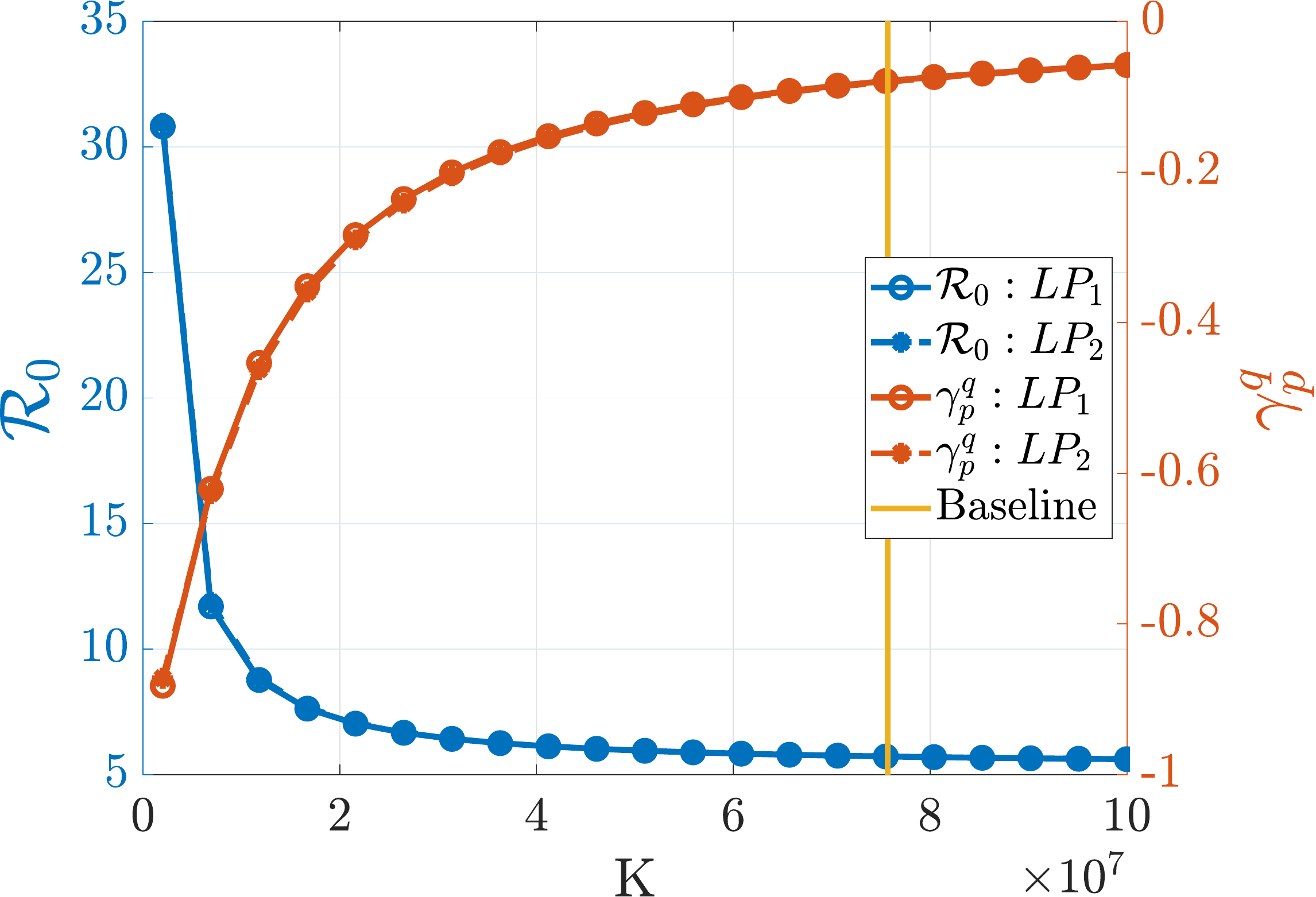}
\includegraphics[width=0.47\textwidth]{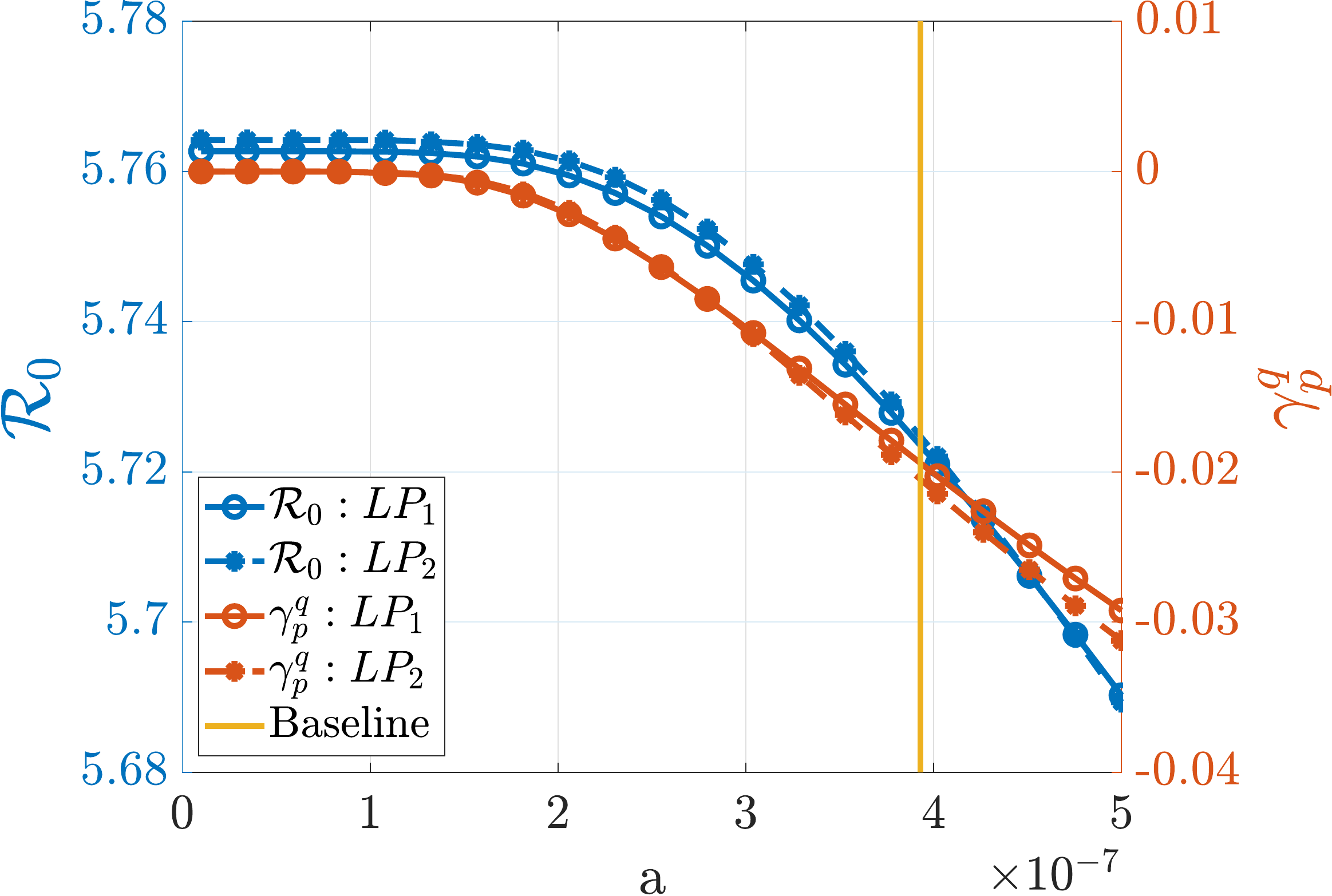} \hfill
\includegraphics[width=0.47\textwidth]{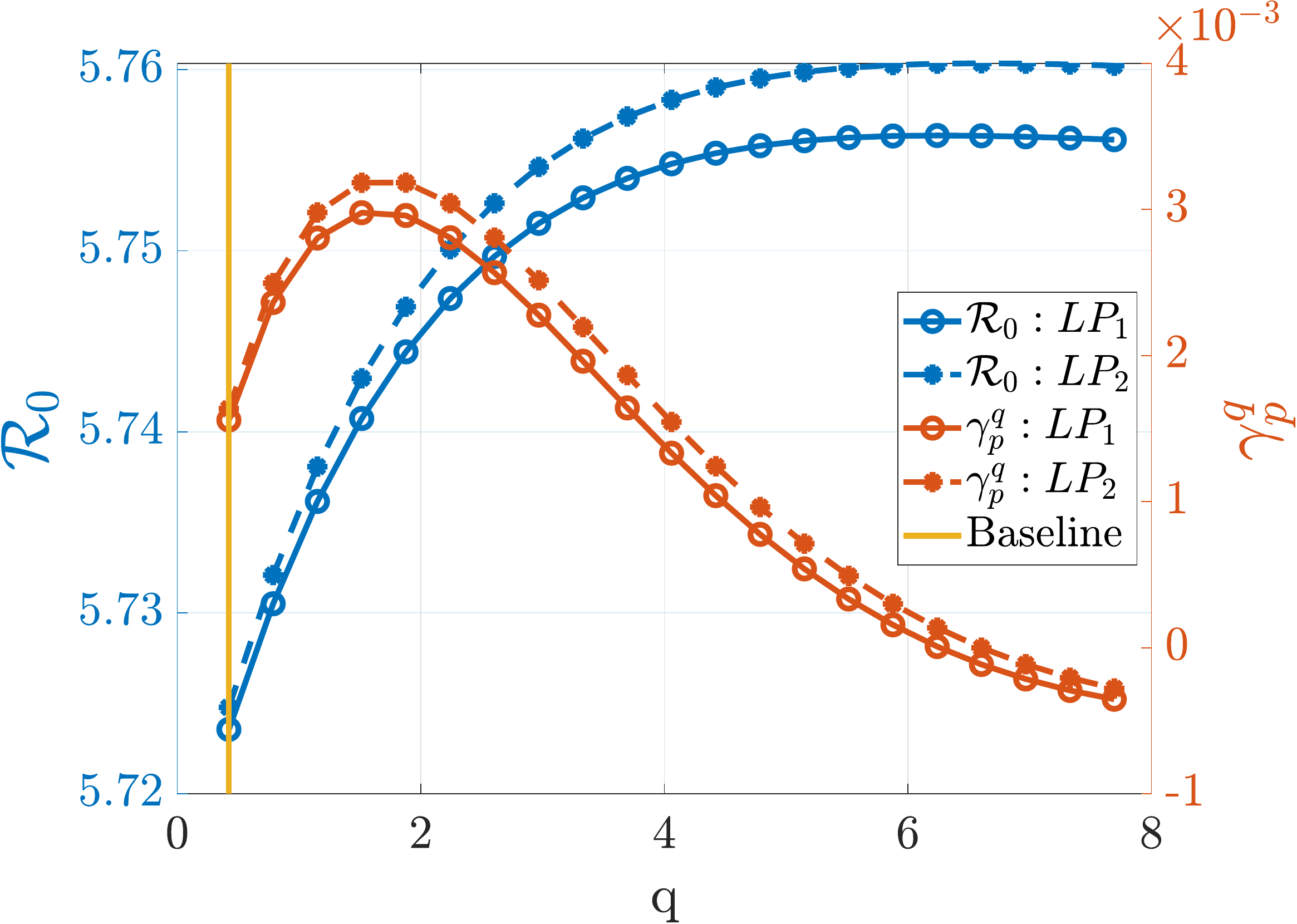}
\caption{Impact of the choice of linking functions on the SA. \label{fig:SA_link_R0}}  
\end{figure}


\begin{figure}[t]
\center
(a)\includegraphics[width=7cm,height=5.5cm]{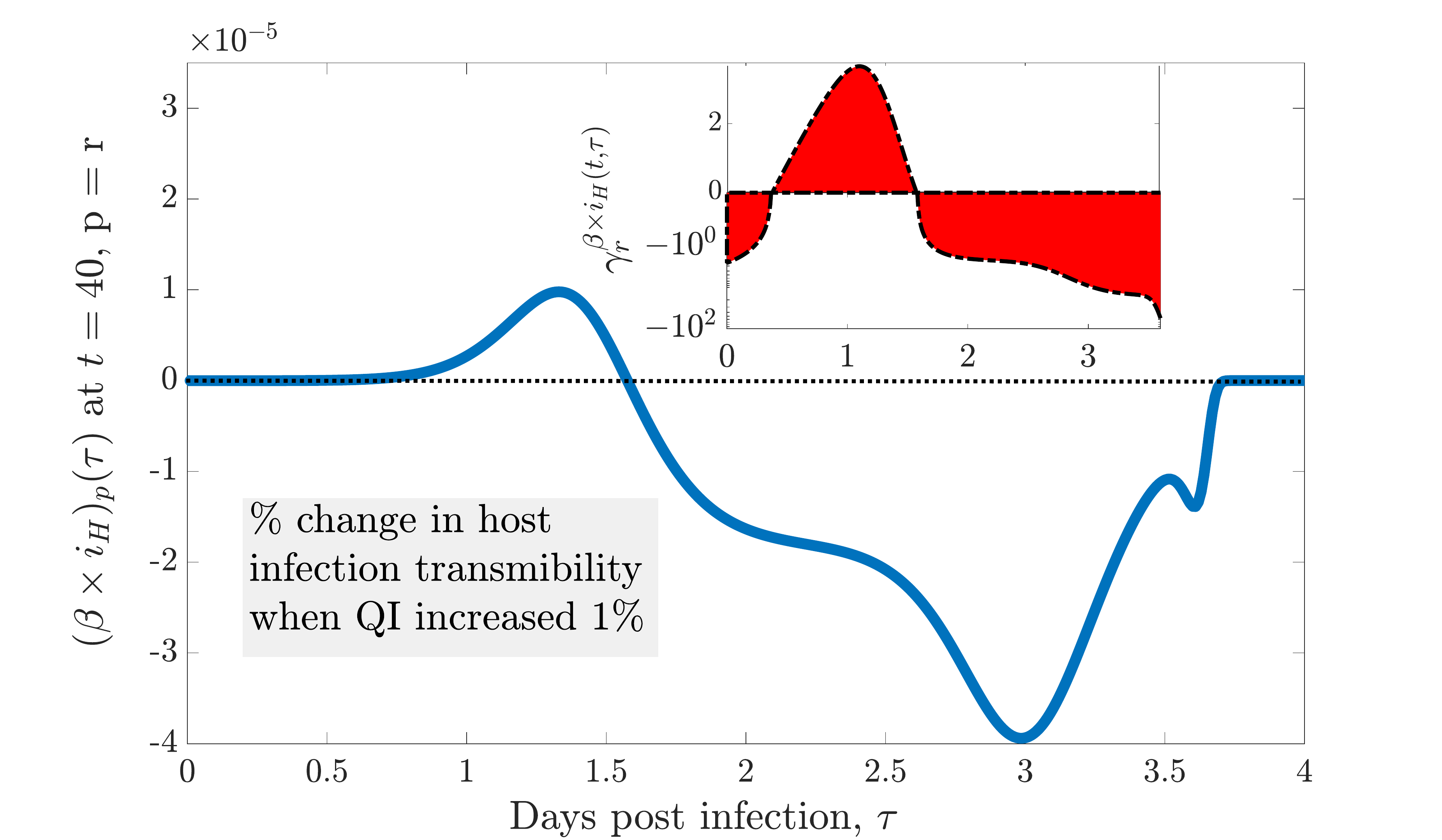}
(b) \includegraphics[width=7cm,height=5.5cm]{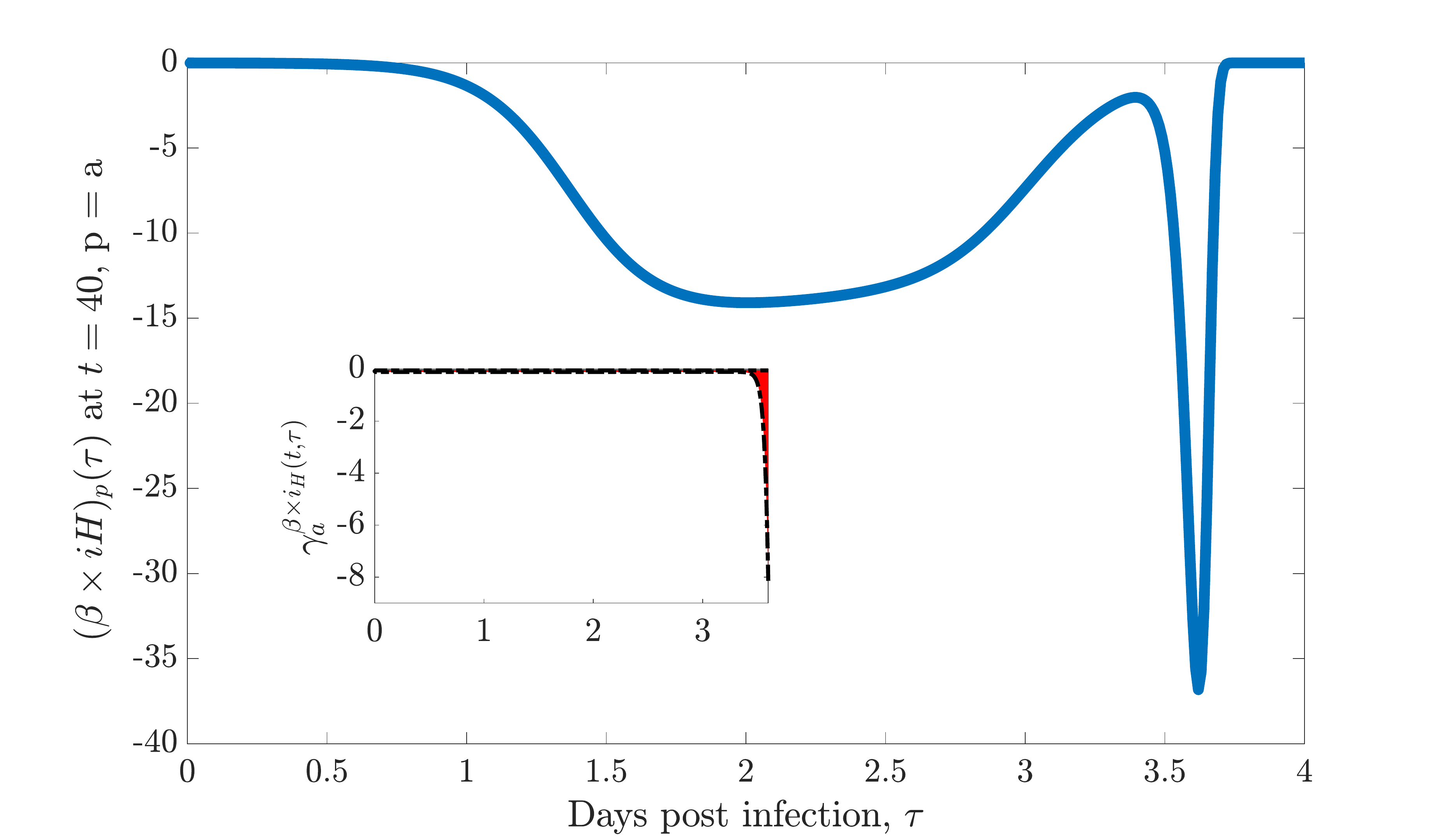}
(c) \includegraphics[width=7cm,height=5.5cm]{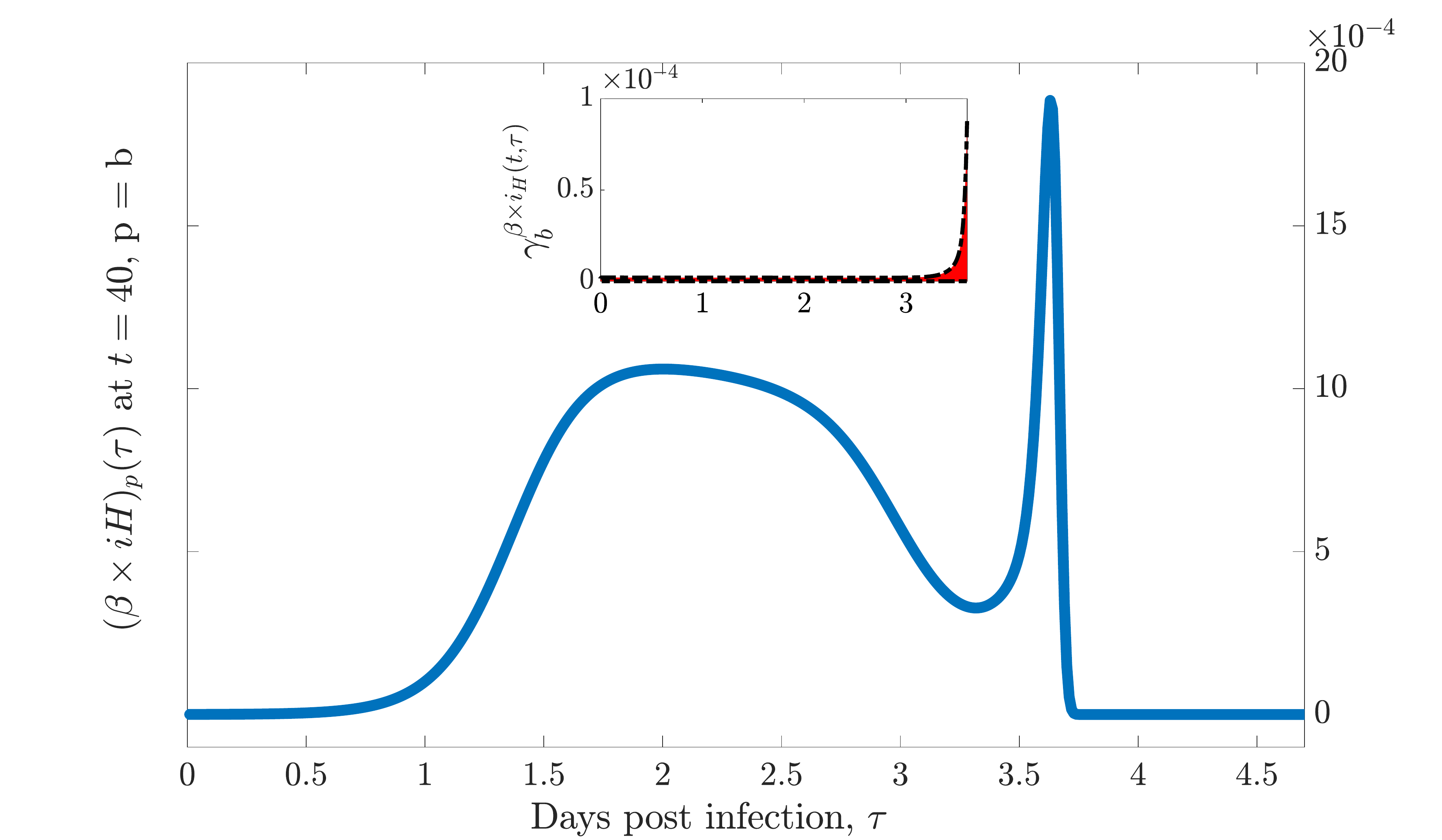}
(d) \includegraphics[width=7cm,height=5.5cm]{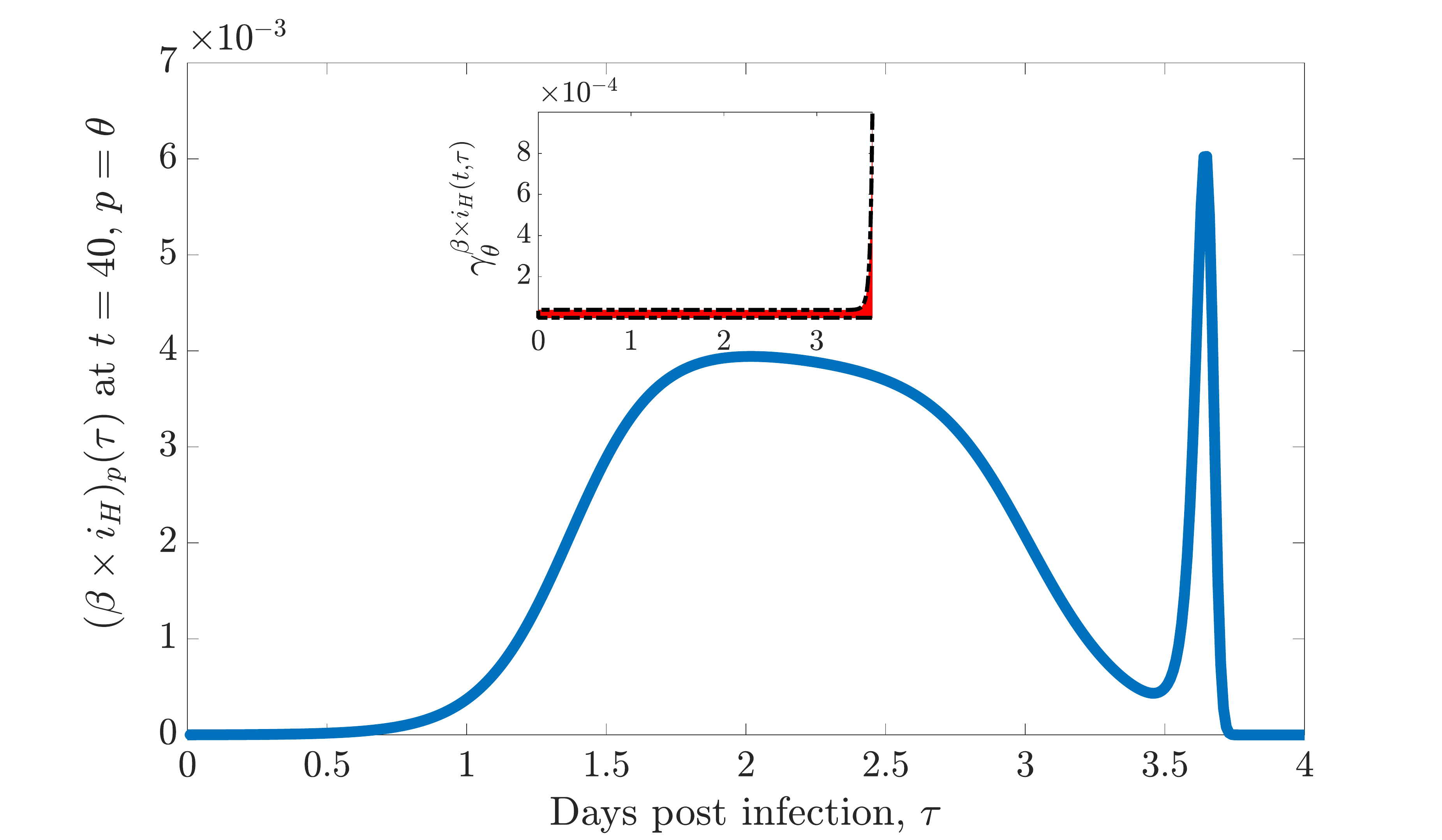}
(e) \includegraphics[width=7cm,height=5.5cm]{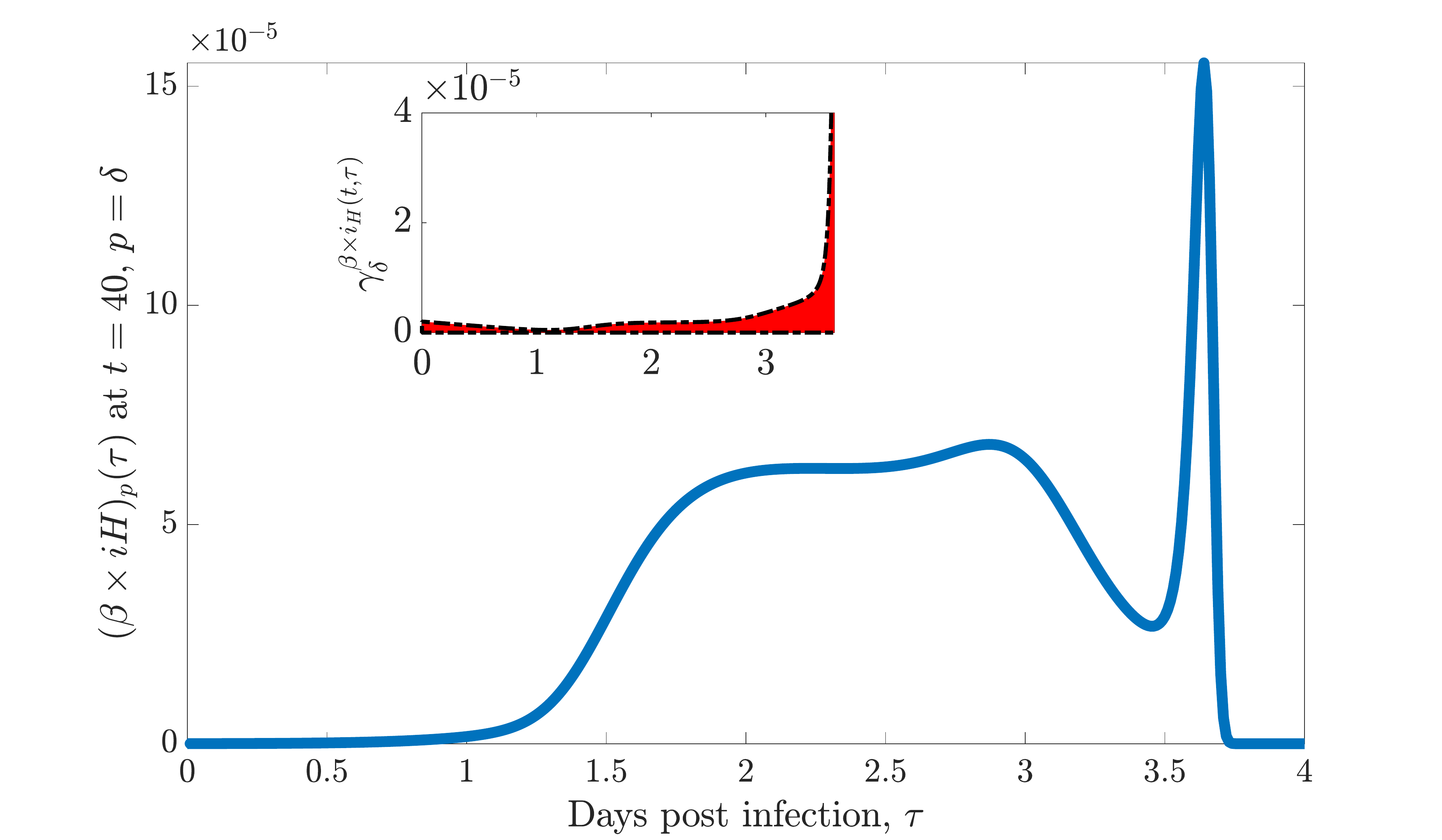}
(f) \includegraphics[width=7cm,height=5.5cm]{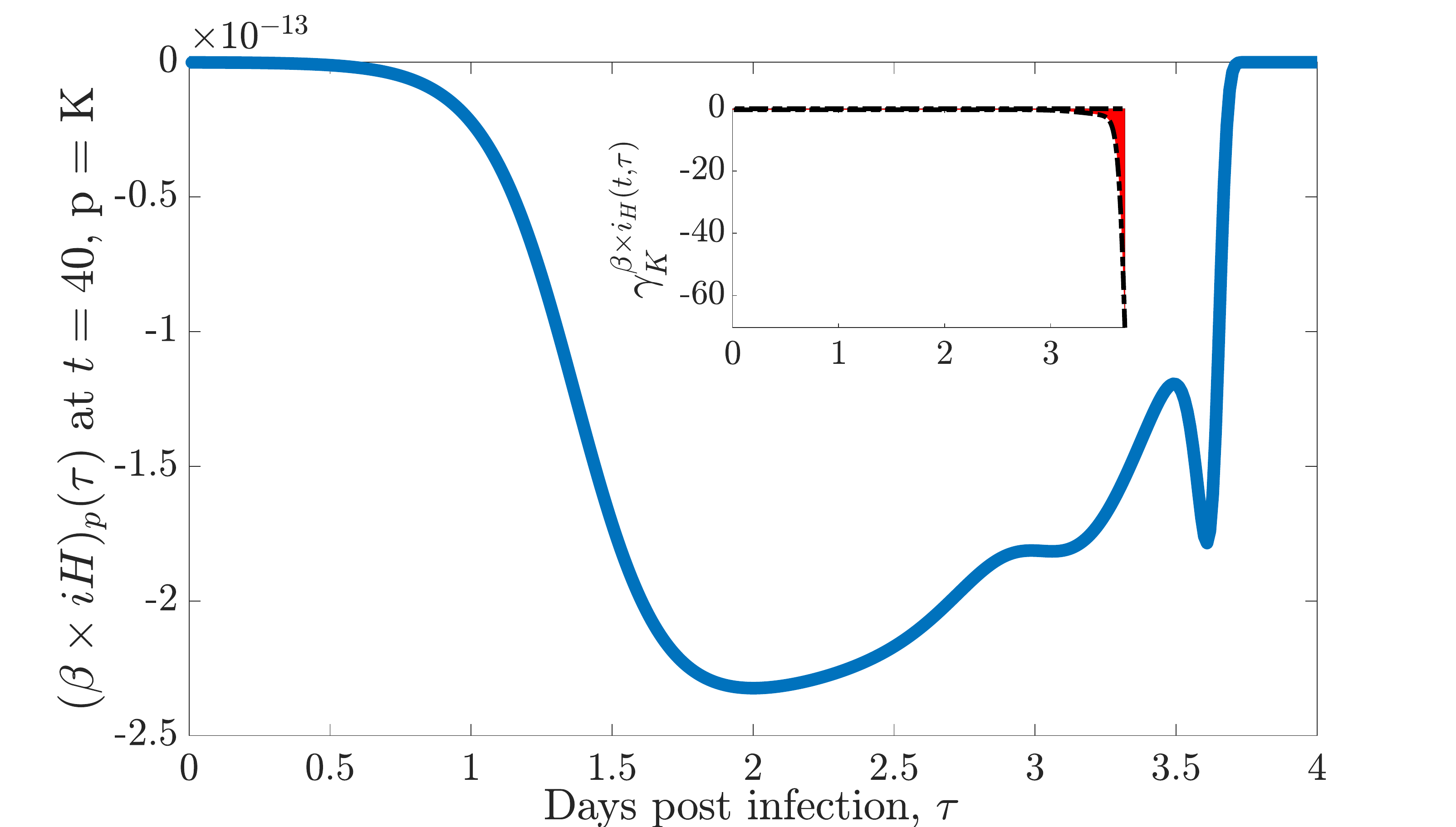}
\caption{ \emph{ How \textit{infection transmissibility of hosts (with different infection group $\tau$}) changes  when quantity of interest (QI) such as an immune parameter $p$ is increased by $1\%.$}  At initial phase of epidemic ($t=40$ days after epidemic started), the percentage change in infection transmissibility among infected host groups is the largest when the change is occurred in parasite growth rate $r,$ comparing to the rest of the immune parameters. 
}  
\label{epidemic_phase1}
\end{figure}

\newpage
\subsection{Derivation of infectability Sensitivity Matrix}\label{Inf_sensitivity}
By utilizing Clairaut's theorem, we first obtain the derivative of the system \eqref{Epi_model} with respect to a model parameter $p$:
\begin{equation}\label{SA_Epi_model}
%\left\{
\begin{aligned}
\frac{\partial}{\partial t}\left(\frac{\partial S_H}{\partial p}\right) & = - \beta_V \frac{\partial S_H}{\partial p}I_V(t) -\beta_VS_H(t) \frac{\partial I_V}{\partial p} -d \frac{\partial S_H}{\partial p}\\
\D\frac{\partial}{\partial t}\left(\frac{\partial i_H}{\partial p}\right)&+\frac{\partial}{\partial \tau}\left(\D\frac{\partial i_H}{\partial p}\right)  =  
 -\D\frac{\partial \left((\alpha (\tau)+ \kappa (\tau)+ \gamma(\tau)+ d)\right)}{\partial p}i_H(\tau,t) - (\alpha (\tau)+ \kappa (\tau)+ \gamma(\tau)+ d)\left(\D\frac{\partial i_H}{\partial p}\right),\\
\D\frac{\partial i_H(0,t)}{\partial p} & = \beta_V \frac{\partial S_H}{\partial p}I_V(t) +\beta_VS_H(t) \frac{\partial I_V}{\partial p},\\
\D\frac{\partial}{\partial t}\left(\frac{\partial R_H}{\partial p}\right) &= \D\int _0^\infty \left(\D\frac {\partial \gamma(\tau)}{\partial p}i_H(\tau,t) + \gamma(\tau) \D\frac {\partial i_H(\tau,t)}{\partial p}\right)d\tau- d\frac{\partial R_H}{\partial p}, \\
\D\frac{\partial}{\partial t}\left(\frac{\partial S_V}{\partial p}\right) &=  -  \frac{\partial S_V}{\partial p}\D \int_0^\infty {\beta_{H}(\tau)i_H(\tau,t) d \tau} - S_V(t)\D \int_0^\infty \left(\frac{\partial \beta_{H}}{\partial p}i_H(\tau,t)  + \beta_H(\tau)\frac{\partial i_H(\tau,t)}{\partial p} \right)d \tau-\mu  \frac{\partial S_V}{\partial p},\\ 
\D\frac{\partial}{\partial t}\left(\frac{\partial I_V}{\partial p}\right) &= \frac{\partial S_V}{\partial p}\D \int_0^\infty {\beta_{H}(\tau)i_H(\tau,t) d \tau} + S_V(t)\D \int_0^\infty \left(\frac{\partial \beta_{H}}{\partial p}i_H(\tau,t)  + \beta_H(\tau)\frac{\partial i_H(\tau,t)}{\partial p} \right)d \tau-\mu \frac{\partial I_V}{\partial p}.
\end{aligned}
%\right.
\end{equation}
Following the numerical scheme in \cite{tuncer2016structural}, we discretize the system using the Backward Euler difference quotient along the characteristic, and obtain the following fully-discretized finite difference scheme:
\begin{equation}
%{\mbox{\textbf{Finite Difference Algorithm:}}\hspace{1cm}}
\begin{cases}
\D\frac{S_H^{n+1}-S_H^{n}}{\Delta t} &\hspace{-4mm}= \Lambda-\beta_vS_H^{n+1}I_v^n-dS_H^{n+1}, \
\D\frac{i_H^{k+1,n+1}-i_H^{k,n} }{\Delta t}=-\left(\alpha_{k+1}+\gamma_{k+1}+d  \right)i_H^{k+1,n+1}, \vspace{1.5mm}\\
i_H^{0,n+1}&\hspace{-4mm}= \beta_v S_H^{n+1} I_v^{n+1}, \vspace{1.5mm}\\
\D\frac{S_v^{n+1}-S_v^{n}}{\Delta t}&\hspace{-4mm}=\eta-S_v^{n+1}\D\sum_{k=1}^M \beta_H^k i_H^{k,n+1} -\mu S_v^{n+1}, \
\D\frac{I_v^{n+1}-I_v^{n}}{\Delta t}=S_v^{n+1}\D\sum_{k=1}^M \beta_H^k i_H^{k,n+1}-\mu I_v^{n+1} \vspace{1.5mm}\\
\D\frac{S_{H_p}^{n+1}-S_{H_p}^{n}}{\Delta t} &\hspace{-4mm}= -\beta_vS_{H_p}^{n+1}I_v^n-\beta_v S_{H}^{n+1}I_{v_p}^n-dS_{H_p}^{n+1},\vspace{1.5mm}\\
\D\frac{i_{H_p}^{k+1,n+1}-i_{H_p}^{k,n} }{\Delta t}&\hspace{-4mm}=-\D\frac{\partial}{\partial p}\left(\alpha_{k+1}+\gamma_{k+1}+d  \right)i_H^{k+1,n+1}-\left(\alpha_{k+1}+\gamma_{k+1}+d  \right)i_{H_p}^{k+1,n+1},\vspace{1.5mm}\\
i_{H_p}^{0,n+1}&\hspace{-4mm}= \beta_v S_{H_p}^{n+1} I_v^{n}+  \beta_v S_{H}^{n+1} I_{v_p}^{n}, \vspace{1.5mm}\\
\D\frac{S_{v_p}^{n+1}-S_{v_p}^{n}}{\Delta t}&\hspace{-4mm}=
-S_{v_p}^{n+1}\D\sum_{k=1}^M \beta_H^k i_H^{k,n+1}-S_{v}^{n+1}\D\sum_{k=1}^M \D\frac{\partial \beta_H^k}{\partial p} i_H^{k,n+1}-S_{v}^{n+1}\D\sum_{k=1}^M \beta_H^ki_{H_p}^{k,n+1}-\mu S_{v_p}^{n+1},\vspace{1.5mm}\\
\D\frac{I_{v_p}^{n+1}-I_{v_p}^{n}}{\Delta t}&\hspace{-4mm}=S_{v_p}^{n+1}\D\sum_{k=1}^M \beta_H^k i_H^{k,n+1}+S_{v}^{n+1}\D\sum_{k=1}^M \D\frac{ \partial \beta_H^k}{\partial p} i_H^{k,n+1}
+S_{v}^{n+1}\D\sum_{k=1}^M  \beta_H^k i_{H_p}^{k,n+1}-\mu I_{v_p}^{n+1}, \vspace{1.5mm}\\
\end{cases}
\label{nummethod}
\end{equation}
where the system is discretized on the domain  $D = \{(\tau,t): 0\leq \tau \leq A,\; 0\leq t \leq T \},$ the grid sizes for time ($\Delta t$) and age ($\Delta \tau $) are chosen to be the same $\Delta t = \Delta \tau$, and $T = N\Delta t$ and $A = M\Delta \tau$. We approximate the value at each grid point $(\tau_k,t_n)$, $\tau_k = k\Delta t$, $t_n = n\Delta t\,, k=1\ldots,M,\; n=1,\ldots,N\,$, which are denoted by $S(t_n) \approx S^n,\; i_H(\tau_k,t_n) \approx i_H^{k,n},\;\beta_H(\tau_k) \approx \beta_H^k\,$ along with derivatives $\D\frac{\partial S_H(t_n)}{\partial p} \approx S_{H_p}^n,\; \D\frac{\partial i_H(\tau_k,t_n)}{\partial p} \approx i_{H_p}^{k,n},\; \D\frac{\partial S_v(t_n)}{\partial p} \approx S_{v_p}^n,\; \D\frac{\partial I_v (\tau_k,t_n)}{\partial p} \approx I_{v_p}^{k,n}.$ Lastly, the linked parameters $\beta_H(\tau_k),\; \alpha(\tau_k),\; \gamma(\tau_k)$ and their derivatives with respect to the model parameters $p$ are pre-calculated following the numerical method as described in Section \ref{sec:SA}.

Furthermore recall that whenever $\mathcal R_0>1,$ independent from (non-zero) initial conditions, the solutions converge to the EE $\mathcal E^*.$ In addition to that, in the previous section, we analytically and numerically derive the sensitivity index, $\gamma^{ i_H^*}_{\p}$ from the explicit expression of  $ i_H^*(\tau),$ given in \eqref{eq_host_e}.
%Therefore to test the accuracy of numerical simulations for the SA, $\gamma^{\beta(\tau) i_H(\tau,t)}_{\textbf{p}}$, we first run the time-dependent infected host density solutions $i_H(\tau,t)$ (with baseline parameter values: $\mathcal R_0>1$) long enough so that at $t=t_{final}$, $i_H^*(\tau) \approx i_H(\tau,t_{final})$. Then we test the accuracy of SA results via the approximation $\gamma^{i_H^*(\tau)}_{\p} \approx \gamma^{ i_H(\tau, t_{final})}_{\p},\  \forall \tau \in [0, \tau_{max})$ \textit{ with } $i_H(\tau_{max}, .) \not  \approx 0.$  Note that to obtain $\gamma^{ i_H(\tau, t_{final})}_{\p}=\frac{\partial  i_H(\tau, t_{final})}{\partial p}\times \frac{p}{ i_H(\tau, t_{final})},$ we utilize the finite difference algorithm, given by \eqref{nummethod}, for simulating both $i_H(\tau, t_{final})$ and $\frac{\partial  i_H(\tau, t_{final})}{\partial p}$.

The subfigures (a)-(d) in Fig.\ref{epidemic_phase1} display the sensitivity, $\dfrac{d \beta \times i_H(\tau)}{d \mathbf{p}}$ of the infection transmissibility (of infected individuals) with respect to immunological parameters $\mathbf{p} \in \{r,a,b, \theta, \delta, K\}$ (at varying host infection age $\tau$) $t=40$ days after epidemic starts, which we called initial phase of epidemic (see Fig.\ref{epidemic_phases}(a)). The inserted subfigures in each figure display $\%$ of change in host infection transmissibility when QI increased $ 1 \%$ at infection age $\tau;$ i.e. $\gamma^{\beta \times i_H}_{\mathbf p}(\tau).$  In the inserted subfigure of Fig.\ref{epidemic_phase1}(a), we observe that the changes in the immune parameters can modulate the infection transmissibility of distinct infected host groups significantly. For example, at the initial phase of the epidemic, in particular $t=40$ days after epidemic started,  $1\%$ increase in-host parasite growth rate $r$ leads up to $4\%$ increase in the infectivity of individuals whom infected approximately at least $1.6$ and at most $3.6$ days ago. Notice that the infectious period for hosts is approximately $4$ days. Significant reduction in the host infection transmissibility when host infected more than $3.6$ days ago (before virus clearance) is due to increase in disease death rate induced by larger parasite growth, and instant large immune response in this infectious time interval. On the other hand, the decrease in the transmission infectivity of individuals whom got infected up to $1.6$ days ago, might be due to stronger immune response leading containment of virus in-host, or causing death. At initial phase of epidemic, the percentage change in infection transmissibility among infected host groups is the largest when the change is occurred in parasite growth rate $r,$ comparing to the rest of the immune parameters $a$ in Fig.\ref{epidemic_phase1}(b), $b$ in Fig.\ref{epidemic_phase1}(c), $\theta$ Fig.\ref{epidemic_phase1}(d), $\delta$ in Fig.\ref{epidemic_phase1}(e), $K$ Fig.\ref{epidemic_phase1}(f).

\begin{figure}[t]
\center
%a)\includegraphics[width=7cm,height=6cm]{RVF_Simulation_total_infected-eps-converted-to.pdf}
%b)\includegraphics[width=7cm,height=6cm]{beta_i_H_r_at_dictinct-eps-converted-to.pdf}
c)\includegraphics[width=7cm,height=6cm]{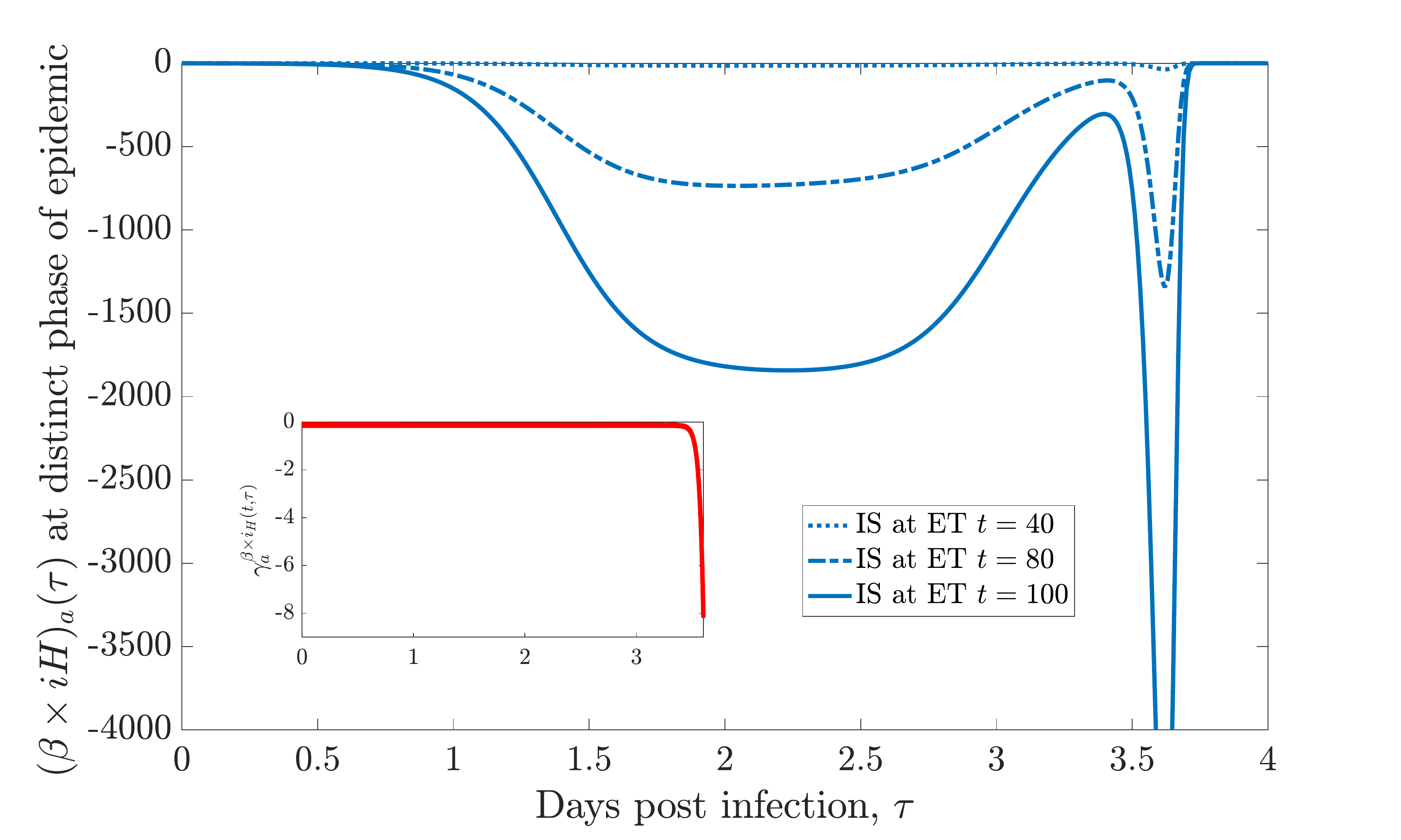}
d)\includegraphics[width=7cm,height=6cm]{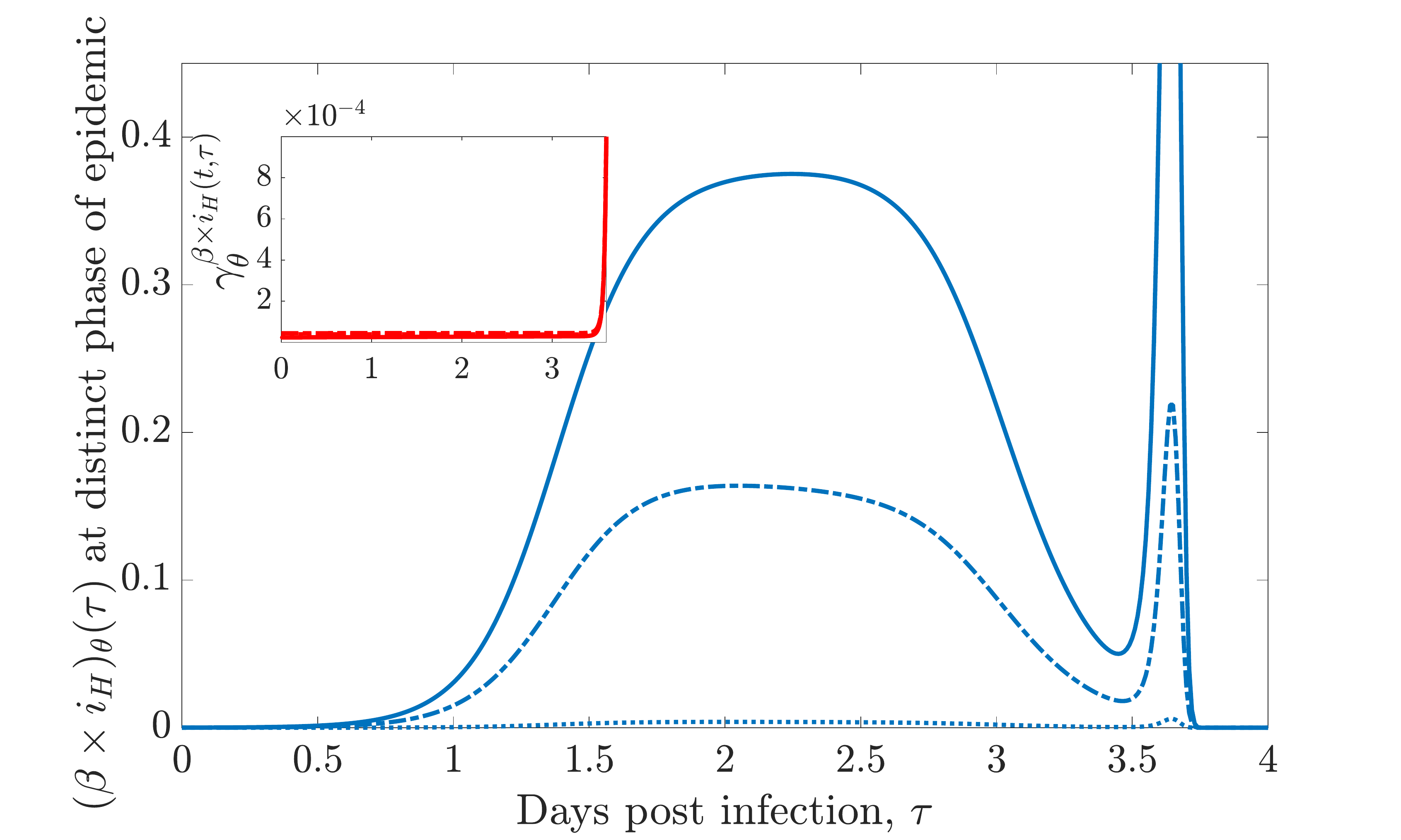}
\caption{\emph{ Sensitivity of infection transmissibility at distinct epidemic time $t$ to immune parameters $a$, and $\theta.$}
} 
\label{epidemic_phases_other_param}
\end{figure}
\newpage
\section*{Acknowledgement}
The authors thank two anonymous reviewers for their helpful comments and feedback on the manuscript, and Mac Hyman of Tulane University for his helpful discussions.  
Hayriye Gulbudak was supported by NSF grant (DMS-1951759) and Simons Foundation/SFARI(638193). The research of Necibe Tuncer was partially supported by NSF grant (DMS-1951626). 
\bibliography{immunoepi_sensitivity} 
\end{document}